\documentclass[preprint]{aastex6}
\usepackage[section] {placeins}
\shorttitle{CARMA 1 CM SURVEY OF ORION-KL}
\shortauthors{Friedel \& Looney}

\begin{document}
\newcommand\acetone{(CH$_3$)$_2$CO}
\newcommand\dme{CH$_3$OCH$_3$}
\newcommand\mef{CH$_3$OCHO}
\newcommand\fa{HCOOH}
\newcommand\fmal{H$_2$CO}
\newcommand\mtoh{CH$_3$OH}
\newcommand\kms{km s$^{-1}$}
\newcommand\jbm{Jy/beam}
\newcommand\vycn{C$_2$H$_3$CN}
\newcommand\mtcn{CH$_3$CN}
\newcommand\etcn{CH$_3$CH$_2$CN}
\newcommand\etoh{C$_2$H$_5$OH}
\newcommand\acal{CH$_3$CHO}
\newcommand\vlsr{$v_{\rm LSR}$}
\newcommand{\icm}{cm$^{-2}$}
\newcommand\e[1]{$\times10^{#1}$}
\newcommand{\mae}{CH$_3$CCH}
\newcommand{\tfmal}{H$_2$CS}
\newcommand{\sot}{SO$_2$}
\newcommand{\cms}{cm$^{-2}$}

% mtoh
\newcommand{\mtohNa}{$3.7(5)\times10^{18}$ cm$^{-2}$}
\newcommand{\mtohNb}{$2.8(9)\times10^{18}$ cm$^{-2}$}
\newcommand{\mtohNc}{$4.0(1)\times10^{18}$ cm$^{-2}$}
\newcommand{\mtohTa}{$248(102)$ K}
\newcommand{\mtohTb}{$138(26)$ K}
\newcommand{\mtohTc}{$117(1)$ K}
% 13mtoh
\newcommand{\thmtohN}{$2.8(2)\times10^{17}$ cm$^{-2}$}
%12/13
\newcommand{\ttth}{9.7(2.8)}
\title{CARMA $\lambda$ = 1 CM SPECTRAL LINE SURVEY OF ORION-KL}

\author{Douglas N. Friedel}
\affil{Department of Astronomy, 1002 W. Green St., University of Illinois, Urbana IL 61801\\
National Center for Supercomputing Applications, 1205 W. Clark St., Urbana IL 61801}
\email{friedel@astro.illinois.edu}
\and
\author{Leslie W. Looney}
\affil{Department of Astronomy, 1002 W. Green St., University of Illinois, Urbana IL 61801}

\begin{abstract}
Orion-KL is a well known high mass star forming region that has long been the target of spectral line surveys and 
searches for complex molecules. One spectral window where the region had never been surveyed is around wavelengths of 
$\lambda$=1 cm. This is an important window to observe due to the fundamental and low energy transitions of numerous 
complex molecules that indicate the maximum spatial extent of the molecular species;
knowing the spatial distribution of a molecule aids in determining the formation mechanism(s) of that molecule. 
Additionally, there are fewer 
transitions in this window, reducing confusion caused by blended lines that can be very problematic at shorter 
wavelengths ($\lambda<$3 mm). In this work, we present the first spectral line survey at $\lambda$=1 cm of the Orion-KL 
region. A total of 89 transitions were detected from 14 molecular species and isotopologues and two atomic species. The 
observations were conducted with the Combined Array for Research in Millimeter-wave Astronomy in both interferometric 
and single dish modes.
\end{abstract}


\section{Introduction}

The Orion-KL region, the closest high-mass star-forming region, is very complex, both chemically and spatially. 
The region was originally brought to the attention of astronomers by the discovery of an extremely bring infrared object (BN) by \citet{becklin67}. Future studies have determined that BN is an embedded B type star moving away from the Trapezium region \citep[e.g.][]{goddi11,plambeck13}. Additional infrared studies of the region have found a plethora of compact source within 20$\arcsec$ of BN \citep[e.g.][]{beuther04}. Radio studies of Orion-KL have found both a compact rotating disk and bipolar massive outflow, traced by SiO maser emission and centered on Source I \citep[e.g.][]{wright83,wright95,plambeck09}, water and methanol masers \citep[e.g.][]{plambeck88,horiuchi00}, emission from CS, DCN, and many complex molecules \citep[e.g.][]{mundy86,mangum91,blake87}, and many compact continuum sources \citep[e.g.][]{eisner06,friedel11}. Studies of BN and Source I over time have indicated that they likely interacted $\sim$560 years ago \citep{goddi11}. Whether this interaction triggered the outflow and subsequent molecular release is not known at this point \citep{friedel12,slww12}.

Orion-KL has been extensively studied due to its rich chemical 
inventory.  There have been dozens of molecular searches and spectral line surveys of the region spanning wavelengths 
of $\lambda$ = 1.3 cm to 188$\mu$m \citep[e.g.,][]{turner89,turner91,sutton85,friedel12,gong15}. Several studies have 
indicated that the complex chemistry of the Orion-KL region displays so-called nitrogen-oxygen "chemical 
differentiation," where the emission from complex nitrogen-bearing molecules such as ethyl cyanide [\etcn] trace the 
Orion Hot Core, and emission from complex oxygen-bearing molecules such as methyl formate [\mef] trace the Orion 
Compact Ridge, thus being near totally spatially distinct. This concept comes from the original millimeter line survey 
of this source \citep{blake87}, where the spectral lines from these two classes of molecules were found to have 
different rest velocities and were therefore assumed to occupy different parcels of gas. Follow-up studies over the 
last three decades confirmed this differentiation through imaging of molecular emission at $>$1 arcsecond spatial 
resolution.  However, the concept of these spatially distinct regions in Orion-KL was called into question by 
\cite{friedel08}, who showed that acetone [\acetone] traces the gas at the intersection of the regions shown to contain 
spatially-distinct N-bearing and O-bearing molecules.  Additionally, \citet{friedel11} showed that the $\lambda$=3 mm 
continuum of Orion-KL is comprised of numerous, compact, bright continuum sources; and \citet{friedel12} and 
\citet{slww12} suggested that much of the molecular emission may actually trace these compact regions, rather than the 
more extended structures traditionally associated with the Hot Core and Compact Ridge.

Spectral line surveys are a particularly powerful method in investigating the coupled dynamical and chemical evolution 
of molecular clouds. They provide a uniformly calibrated set of data that diminishes the errors associated with derived 
quantities. This survey studies a frequency range (26.938 - 34.938 GHz) that has not been surveyed before in Orion-KL. 
In this range there are many low energy transitions of complex molecules that indicate the true spatial distribution 
of the individual species. This in turn can give insight into the formation mechanisms of the different species. 
Additionally, this frequency range is less cluttered than those at 90 GHz and above, where many transitions are 
present, and at $\sim$30 GHz the lines are less likely to be blended with others, leading to lower uncertainties in 
derived quantities.

Figure~2 of \citet{friedel11} and the associated text give an overview of the major sources in this region. Due to the limited spatial resolution ($>5\arcsec$) of our observations we do not focus on many of the individual sources, but instead focus on the larger scale structures (CS1, IRc6, Compact Ridge), and those compact structures which have strong molecular emission (Hot Core/I). 
Table~\ref{tab:desc} gives an
overview of the different sources used in this work, including
coordinates and typical \vlsr\ and FWHM values.

In this paper, we use our observations to constrain the physical conditions of the region by
comparing/contrasting the emission of the low and high energy
transitions and dense gas tracers, and  place the molecular formation mechanisms 
into context with the observed structures. In \S2 we discuss the observations and in \S3 we present the results for the individual atomic and molecular species observed, including comparisons to previous works. In \S4 we describe the temperature and molecular structure of the Orion-KL region as we currently understand it.

\floattable
\begin{deluxetable}{lrrcclr}
\rotate
\tablecolumns{7}
\tabletypesize{\scriptsize}
\tablewidth{0pt}
\tablecaption{Description of Sources in Orion-KL\label{tab:desc}}
\tablehead{\colhead{} & \colhead{} & \colhead{} & \colhead{\vlsr} & \colhead{FWHM} & \colhead{} & \colhead{}\\
\colhead{Name} & \colhead{$\alpha$(J2000)} & \colhead{$\delta$J2000} & \colhead{(\kms)} & \colhead{(\kms)} & \colhead{Description} & \colhead{Ref.}}
\startdata
BN & $05^h35^m14^s.106$  & $-05{\degr}22{\arcmin}22{\arcsec}.528$ & \nodata & \nodata & Embedded B type star & 1\\
CS1 & $05^h35^m15^s.179$  & $-05{\degr}22{\arcmin}03{\arcsec}.76$ & 10.5 & 1.5 & Extended structure containing two compact sources MM5 \& MM6 & 2,3\\
I/Hot Core & $05^h35^m14^s.511$  & $-05{\degr}22{\arcmin}32{\arcsec}.101$ & 2.5-7.5 & 5.5 & Compact source and a significant region of complex molecular emission & 4,5\\
IRc6 & $05^h35^m14^s.155$  & $-05{\degr}22{\arcmin}27{\arcsec}.19$ & 7 & 1.5-6 & Compact source with a mix of O and N bearing molecules & 4,6\\
Compact Ridge & $05^h35^m14^s.15$  & $-05{\degr}22{\arcmin}37{\arcsec}.40$ & 7-8 & 1-3 & Extended structure, source of complex molecular emission & 6,7\\
HC-NE & $05^h35^m14^s.78$  & $-05{\degr}22{\arcmin}20{\arcsec}.86$ & 9-21 & 4-8 & An extension of the Hot Core to the northeast & 8 \\
CR-SW & $05^h35^m13^s.55$  & $-05{\degr}22{\arcmin}53{\arcsec}.80$ & 9 & 1.1 & An extension of the Compact Ridge to the southwest & 8 \\
Orion Bar & \tablenotemark{a} & & \tablenotemark{b} &  & Highly irradiated region dominated by atomic lines & 9,10\\
\enddata
\tablenotetext{a}{The Orion Bar is a very extended structure, no single coordinate set can describe its position.}
\tablenotetext{b}{The structure is very large and has a wide variety of velocities, FWHM, and structure sizes.}
\tablerefs{\footnotesize{(1) \citet{plambeck13}; (2) \citet{mundy86}; (3) \citet{eisner06}; (4) \citet{slww12}; (5)\citet{wright96b}; (6) \citet{friedel08}}; (7) \citet{friedel12}; (8) This work; (9) \citet{gio16}; (10) \citet{odell17}}
\end{deluxetable}

\section{Observations}
The data were taken with the $\lambda$ = 1 cm receivers on the Combined Array for Research in Millimeter-wave Astronomy 
(CARMA) 10 and 6 m telescopes, which operated between 26.938 and 34.938 GHz in single sideband mode.
CARMA was used in its C (24.3 m - 300.0 m baselines) and D (10.2 m - 119.0 m baselines) configurations to conduct observations in 2014 March - 2014 April and 2014 December - 
2015 April, respectively. The phase center for all observations was $\alpha$(J2000) = $05^h35^m14^s.35$ and 
$\delta$(J2000) = $-05{\degr}22{\arcmin}35{\arcsec}.0$. The synthesized beam sizes are $\sim7\arcsec\times5\arcsec$ for 
the C configuration and $\sim15\arcsec\times12\arcsec$ for the D configuration. One arcsecond is $\sim$414 AU at the 
distance of Orion-KL \citep{menten07}.
There are notable differences in the beam sizes for observations in the same array configuration 
due to issues with the receivers, no observations had the entire 15 element array, and on average had 13 dishes.

The correlator setup for the observations was to put eight 31 MHz wide windows end to end, overlapping a few channels 
and integrating for 2.5 hours. The IF frequency of the windows was then shifted by 240 MHz and the process was 
repeated. Each of the windows had 393 spectral channels, giving a velocity resolution of 0.842 - 1.07 \kms, depending 
on frequency. Numerous strong spectral birdies, due to emission from hardware components, were found in the lowest end 
of the spectrum (below 27.13 GHz) and the corresponding data were discarded. The rest of the data were closely 
inspected for additional birdies; none were found.
Due to hardware limitations, there was a small unobservable 
gap in our frequency coverage (30.885-30.945 GHz).
The flux density was 
calibrated using observations of Mars,  
and the antenna based gains were calibrated with observations of 0423-013 and 
0607-085. The flux density calibration is good to 10\%; all stated uncertainties in this paper are statistical in nature. A continuum map was constructed from line free channels. This map was subsequently used to self-calibrate the 
data to reduce overall noise. All calibration, continuum subtraction, and imaging were performed using the MIRIAD 
software package \citep{sault95}. Unless otherwise stated, all presented maps were produced using natural weighting.

\begin{figure}[!ht]
\includegraphics[scale=0.9]{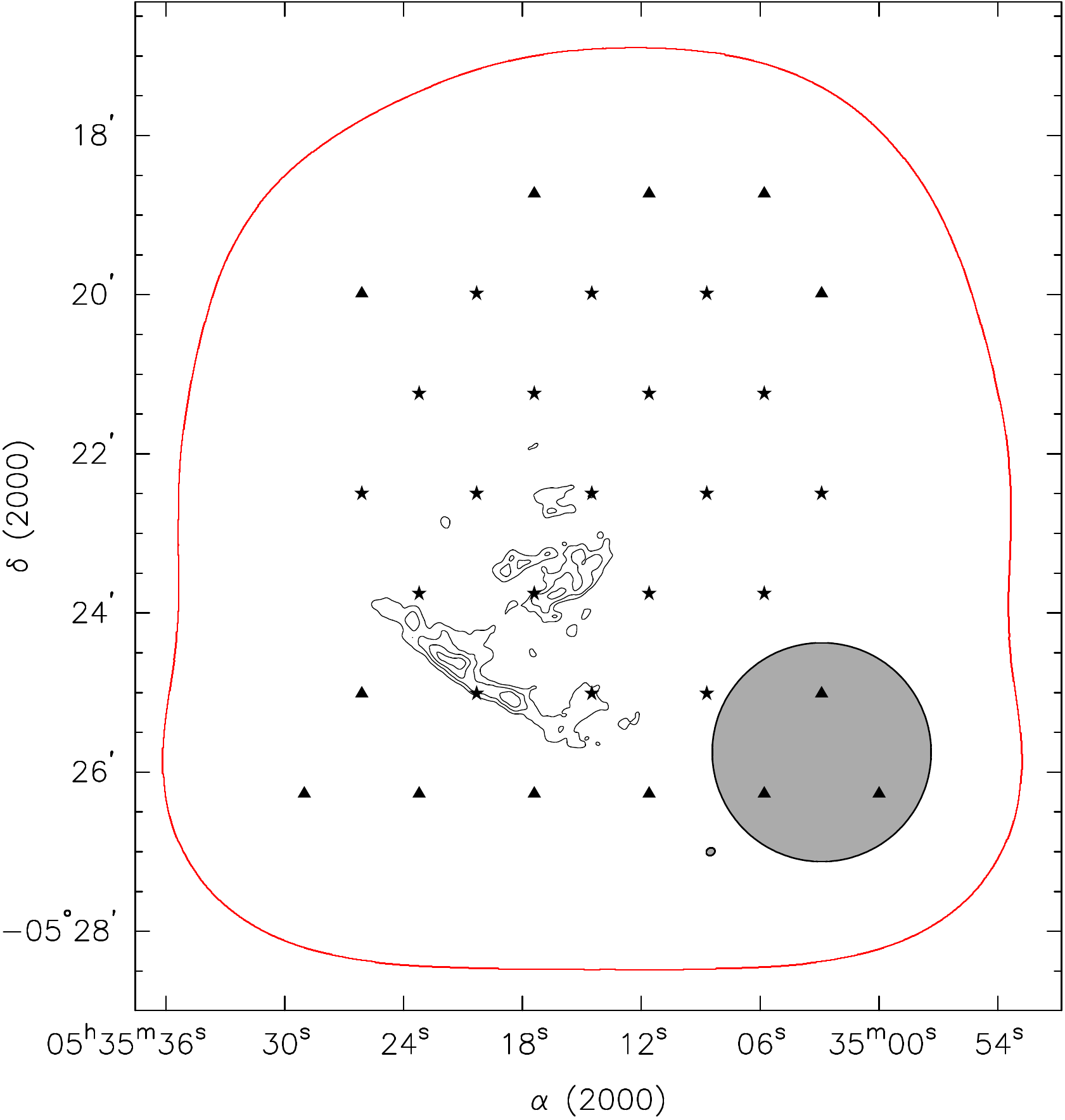}
\caption{Mosaic pattern and gain pattern of the single dish observations. The contours show the continuum from the 
array observations for reference. The stars indicate the original 19 point mosaic pattern, and the triangles indicate 
the positions added for the extended 32 point pattern. The red contour indicates the region of unity gain and the beams from
the array and single dish data are in the lower right corner.
\label{fig:cover}}
\end{figure}

Once the data were fully reduced some of the spectral lines appeared to have larger spatial scale components. We 
initiated further observations with CARMA in auto-correlation mode, effectively acting as an array of single dishes. 
Fourteen transitions, from 11 species were selected for this study (see notes in Table~\ref{tab:individual}). The 
observations were done by position switching between the on positions and an off position 1 degree east and 0.5 degrees 
south of the phase center position. The initial observations used a 19 point hexagonal mosaic centered on the phase 
center and had a primary beam FWHM of $165\arcsec$ at each pointing. In subsequent observations, this was expanded to a 
32 point mosaic to cover some potentially extended regions. Figure~\ref{fig:cover} shows the mosaic pattern and 
sensitivity coverage of the single dish observations. Only data from the 10.1 meter dishes were used, due to their 
higher sensitivity. 
For the single dish observations the correlator was set up with eight 31 MHz windows covering all of the 
selected transitions simultaneously. Data were calibrated and imaged using the MIRIAD software package. Due to the minimal overlap
between the single dish and array observations in {\it u-v} space, the two types of data could not be combined.

\section{Results and Discussion}
\begin{figure}
\includegraphics[scale=0.85]{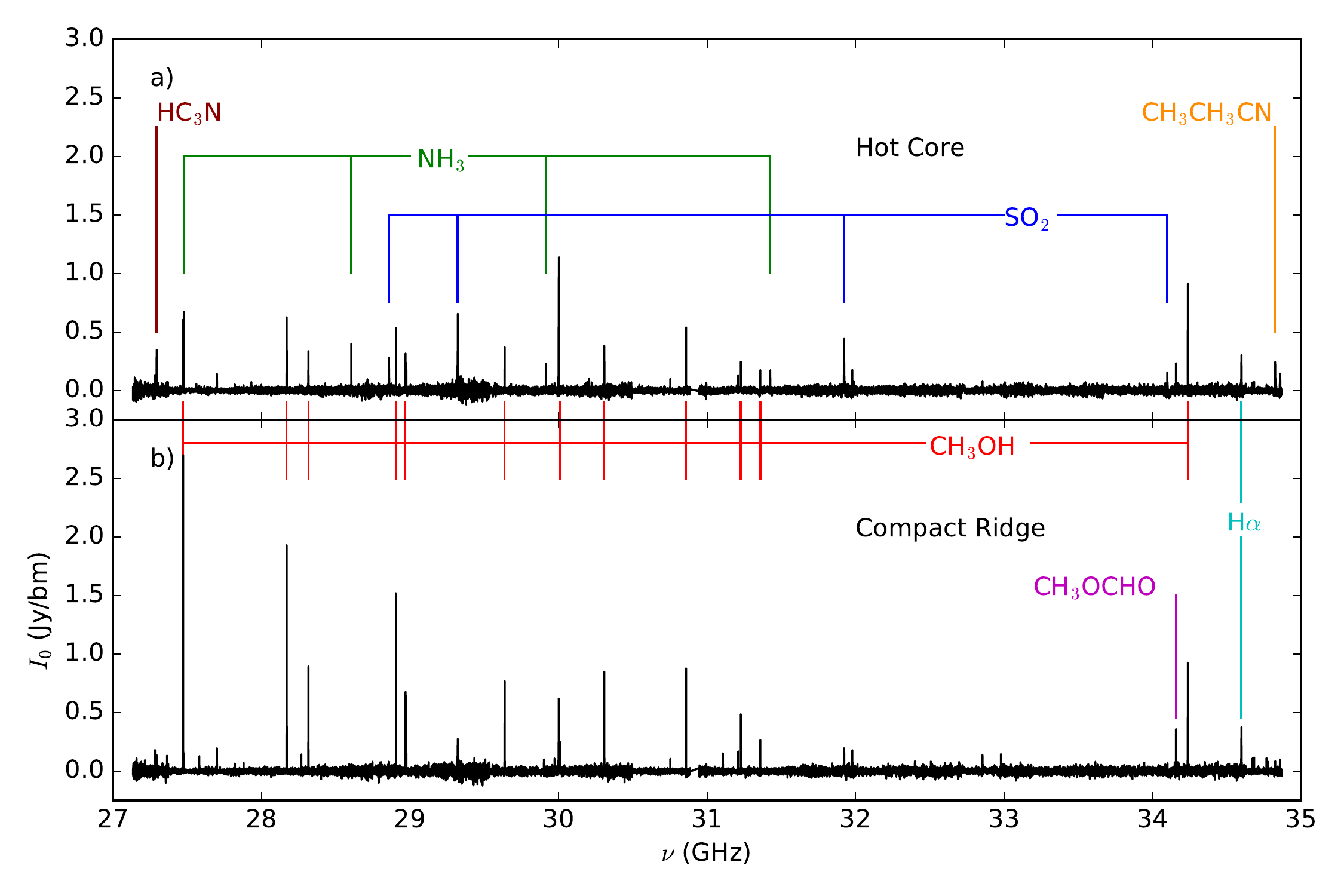}
\caption{The full spectra toward a) the Hot Core region and b) the Compact Ridge region. The 
ordinate is intensity in \jbm, and the abscissa is frequency in GHz. Prominent lines are labelled in the panels, 
specifically strong transitions of methanol [\mtoh] (red), SO$_2$ (blue), ammonia [NH$_3$] (green), H$_3$N (dark red), ethyl cyanide [\etcn] (orange), methyl formate [\mef] (magenta), and H$\alpha$ (light blue). 
\label{fig:fullspec}}
\end{figure}

The full spectra from the survey are shown in Figure~\ref{fig:fullspec}, where panel (a) displays the spectrum 
toward the Hot Core region ($\alpha$(J2000) = $05^h35^m14^s.53$, $\delta$(J2000) = $-05{\degr}22{\arcmin}30{\arcsec}$) and 
panel (b) shows the spectrum toward the Compact Ridge region ($\alpha$(J2000) = $05^h35^m14^s.2$, $\delta$(J2000) = $-05{\degr}22{\arcmin}40{\arcsec}$). Spectral line of select species are noted: methanol [\mtoh] (red), SO$_2$ (blue), ammonia [NH$_3$] (green), H$_3$N (dark red), ethyl cyanide [\etcn] (orange), methyl formate [\mef] (magenta), and H$\alpha$ (light blue).

\subsection{Continuum}
\begin{figure}[!ht]
\includegraphics[scale=0.45]{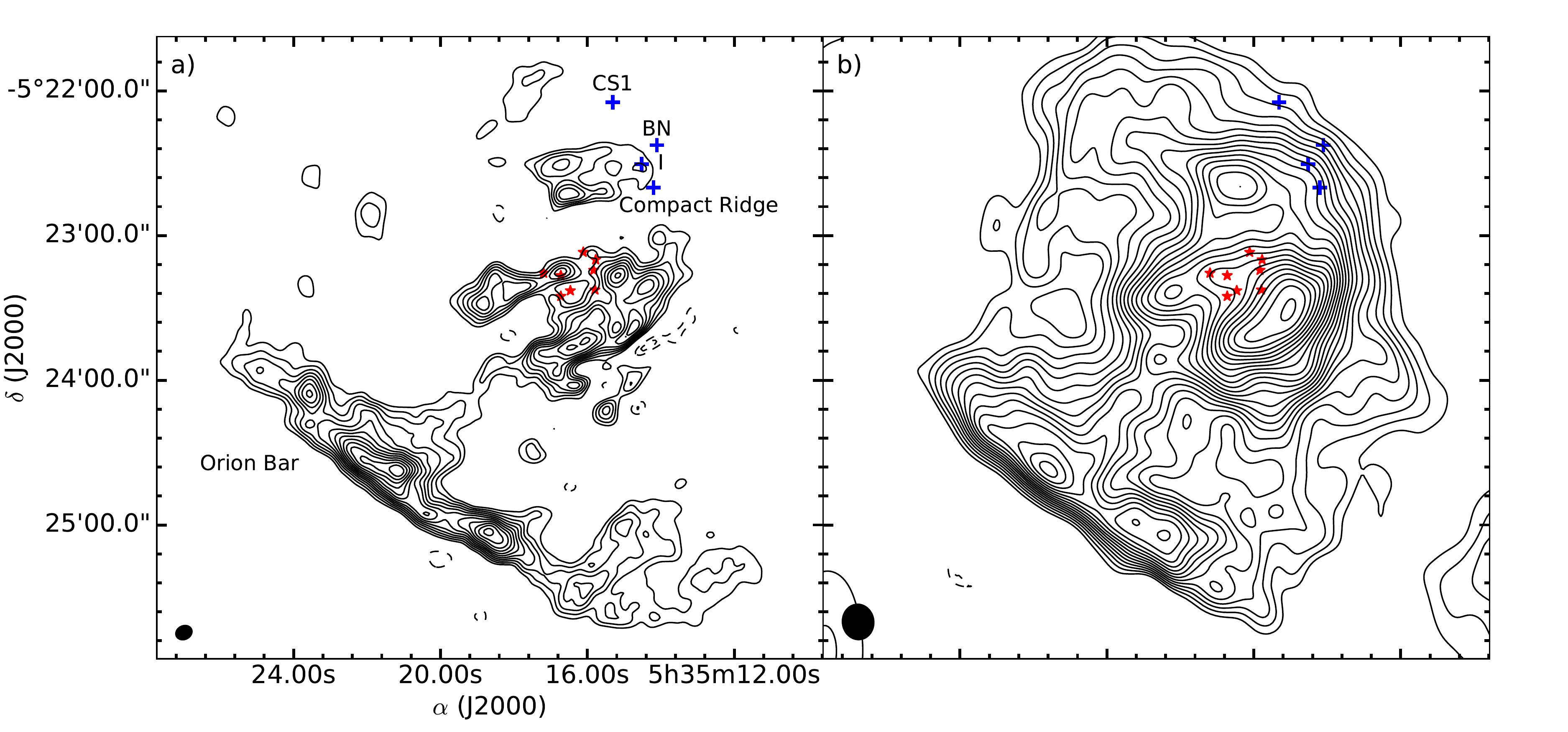}
\caption{Maps of the $\lambda$ = 1 cm continuum. a) C configuration ($6.8\arcsec\times5.6\arcsec$ resolution) data. 
b) D configuration ($14.7\arcsec\times12.7\arcsec$ resolution) data. The contours are $\pm12\sigma, 
\pm18\sigma, \pm24\sigma$, ... where $\sigma$=2.9 m\jbm\ for C configuration and $\sigma$=13.2 m\jbm\ for D 
configuration. 
The $\star$ icons in all maps indicate the position of Trapezium stars while the + 
icons denote the position of CS1, BN, Source I, and the Compact Ridge.\label{fig:contin}}
\end{figure}

\begin{figure}[!ht]
\includegraphics[scale=0.7]{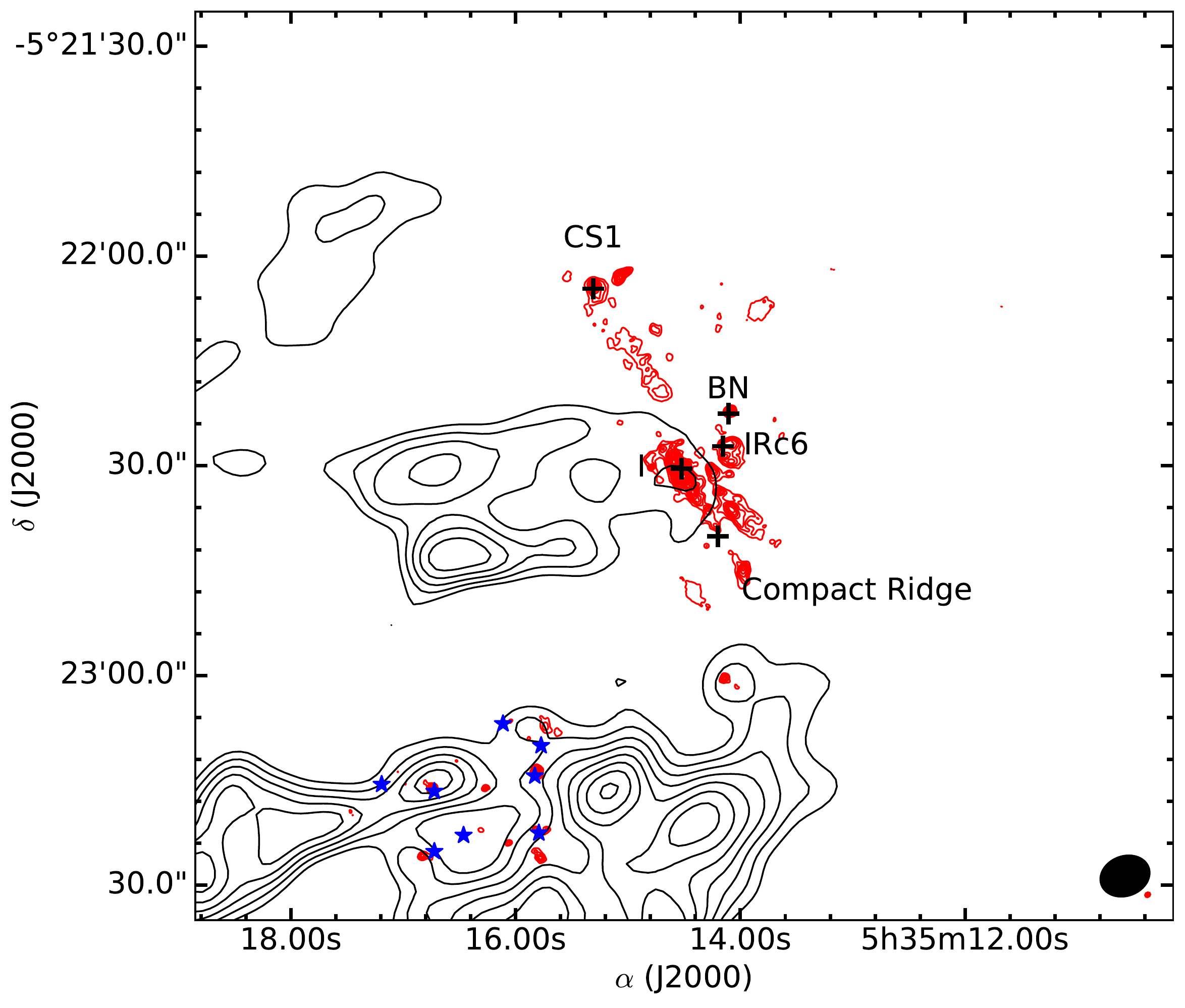}
\caption{Maps of the $\lambda$ = 1 cm continuum (black contours) and the $\lambda$ = 3mm continuum (red contours)from 
\citet{friedel11}. The contours are $12\sigma, 18\sigma, 24\sigma$, ... where $\sigma$=2.9 
m\jbm\ for 1 cm data and $\sigma$=312.3 $\mu$Jy for the 3 mm data. 
Note that the negative contours from spatial filtering
of large scale emission are not plotted to make the comparison easier. 
The $\star$ icons indicate the position of Trapezium stars and the + icons denote sources of note.\label{fig:ccomp}}
\end{figure}

The $\lambda$ = 1 cm continuum was mapped in both the C and D configurations, giving synthesized beams of 
$6.8\arcsec\times5.6\arcsec$ and $14.7\arcsec\times12.7\arcsec$ respectively. The maps were created by removing all 
channels with line emission from the spectrum and Fourier Transforming the remainder together, and they contain no zero spacing data 
from the single dish observations. Figure~\ref{fig:contin} shows the 
resulting maps. Figure~\ref{fig:contin}(a) is the C configuration data while (b) is the D configuration data. In this, and all other maps, the $\star$ icons denote the position of Trapezium stars and the + icons denote the positions of CS1, BN, Source I, and the Compact Ridge, for reference.
The two continuum maps show the same general shape 
to the continuum structure and peak at nearly the same place. The bulk of the continuum emission comes from the compact 
sources detected in the C configuration, but there is notable emission from extended structure.
Figure ~\ref{fig:ccomp} shows a comparison map of the 1 cm continuum (black contours, dominated by free-free emission) and 3 mm continuum (red contours, dominated by dust emission) 
from \citep{friedel11}. 
The map shows that different emission mechanisms are dominant in the different regions. There is some overlap near Source I, but all other 3 mm emission has no corresponding 1 cm emission. Several of 
the point-like sources to the southeast of the 3 mm map are coincident with the known optical/IR sources and local 1 cm 
peaks. Sources of note are labelled.

\clearpage

\begin{deluxetable}{lr}
\tablecolumns{2}
\tabletypesize{\scriptsize}
\tablewidth{0pt}
\tablecaption{Detected Molecular Species\label{t:summary}}

\tablehead{\colhead{Name} & \colhead{Detected Transitions}}
\startdata
\mtoh & 21\\
$^{13}$\mtoh & 8\\
H Recom. & 9\\
He Recom. & 2\\
\fmal & 2\\
\tfmal & 1\\
SO & 1\\
$^{34}$SO & 2\\
SO$_2$ & 6\\
$^{34}$SO$_2$ & 1\\
CH$_3$CCH & 2\\
NH$_3$ & 5\\
\mef & 8\\
\dme & 17\\
\etcn & 3\\
HC$_3$N & 1\\
\enddata

\end{deluxetable}

\begin{deluxetable}{lrrrcrrrrrc}
\tablecolumns{11}
\tabletypesize{\scriptsize}
\tablewidth{0pt}
\tablecaption{Molecular Parameters of Observed Lines\label{tab:individual}}

\tablehead{\colhead{Quantum} & \colhead{Frequency} &
	\colhead{$E_u$} & \colhead{S$\mu^2$} &
	\colhead{$\theta_a\times\theta_b$} & \colhead{$I_0$} & \colhead{\vlsr} & \colhead{FWHM} & \colhead{} & \colhead{RMS Noise} &
	\colhead{Array} \\
    \colhead{Numbers} & \colhead{(MHz)} & \colhead{(K)}
	&\colhead{($D^2$)}& \colhead{($\arcsec\times\arcsec$)} & \colhead{Jy/bm} & \colhead{km/s} & \colhead{km/s} & \colhead{Pos.\tablenotemark{a}}&
	\colhead{(mJy/bm)} & \colhead{Config.}}
\startdata
\cutinhead{\mtoh}
14$_{ -5, 9}$- 15$_{ -4,12}E$ & 27,283.14 (0.05) & 367.6 & 3.6 & $ 10.4 \times 9.6 $ &  0.02 (0) &  4.5 (0) & 5.2 (0) & & 18.6 & C \\
 & & & & &   0.21 (0) &  7.6 (0) & 1.9 (0) & & & \\
 & & & & &   0.04 (0) &  8.0 (0) & 5.3 (0) & & & \\
13$_{ 2,11}$- 13$_{ 1,12}E$\tablenotemark{b} & 27,472.53 (0.05) & 233.6 & 14.2 & $ 7.6 \times 6.5 $ &  0.13 (0) &  4.5 (0) & 7.2 (0) & & 12.7 & C \\
 & & & & &  1.77 (0) &  7.6 (0) & 1.9 (0)   & & &\\
 & & & & &  0.63 (0) &  8.0 (0) & 5.3 (0)   & & & \\
12$_{ 2,10}$- 11$_{ 1,11}A+$ $\nu=1$& 27,700.18 (0.05) & 479.2 & 7.1 & $ 12.7 \times 5.3 $ &  0.02 (0) &  4.5 (0) & 5.2 (0) & & 6.8 & C \\
 & & & & &   0.17 (0) &  7.6 (0) & 1.9 (0)   & & & \\
 & & & & &  0.06 (0) &  8.0 (0) & 5.3 (0)   & & & \\
19$_{ 5,15}$- 18$_{ 6,13}E$ & 27,820.84 (0.05) & 576.7 & 4.8 & $ 12.7 \times 5.3 $ &  0.07 (0) &  7.6 (0) & 1.9 (0)  & & 9.3 & C \\
 & & & & &  0.01 (0) &  8.0 (0) & 5.3 (0)   & &  &\\
14$_{ 2,12}$- 14$_{ 1,13}E$ & 28,169.47 (0.05) & 266.1 & 15.2 & $ 9.3 \times 5.5 $ &  0.15 (0) &  4.5 (0) & 5.2 (0)  & & 11.8 & C \\
 & & & & &  1.06 (0) &  7.6 (0) & 1.9 (0)   & &  &\\
 & & & & &  0.68 (0) &  8.0 (0) & 5.3 (0)   & &  &\\
 4$_{0,4}$ -   3$_{1,2}E$   & 28,316.07 (0.05) & 35.0  & 1.4  & $ 9.3 \times 5.5 $ &  0.03 (0) &  4.5 (0) & 5.2 (0) & & 10.7 & C \\
 & & & & &    0.29 (0) &  7.6 (0) & 1.9 (0)   & &  &\\
 & & & & &   0.45 (0) &  8.0 (0) & 5.3 (0)  & &  &\\
24$_{ 2,22}$- 24$_{ 1,23}E$ & 28,874.13 (0.17) & 718.4 & 12.8 & $ 7.7 \times 5.3 $ &  0.06 (0) & 4.4 (1) & 5.2 (3)  & & 11.5 & C \\
15$_{ 2,13}$- 15$_{ 1,14}E$ & 28,905.81 (0.05) & 301.0 & 16.1 & $ 7.7 \times 5.3 $ & 0.07 (0) &  4.5 (0) & 5.2 (0)   & & 13.2 & C \\
 & & & & &   0.99 (0) &  7.6 (0) & 1.9 (0)  & & & \\
 & & & & &   0.53 (0) &  8.0 (0) & 5.3 (0)  & & & \\
8$_{ 2, 7}$- 9$_{ 1, 8}A-$ \tablenotemark{b} & 28,969.96 (0.05) & 121.3 & 3.0 & $ 7.7 \times 5.3 $ &  0.04 (0) &  4.5 (0) & 5.2 (0) &  & 14.9 & C \\
 & & & & &   0.23 (0) &  7.6 (0) & 1.9 (0)  & & & \\
 & & & & &   0.38 (0) &  8.0 (0) & 5.3 (0)  & & & \\
16$_{ 2,14}$- 16$_{ 1,15}E$ & 29,636.94 (0.05) & 338.1 & 16.8 & $ 6.6 \times 5.4 $ &  0.10 (0) &  4.5 (0) & 5.2 (0) &  & 13.8 & C \\
 & & & & &   0.40 (0) &  7.6 (0) & 1.9 (0)  & & & \\
 & & & & &   0.37 (0) &  8.0 (0) & 5.3 (0)  & & & \\
23$_{ 2,21}$- 23$_{ 1,22}E$ & 29,972.84 (0.14) & 662.8 & 14.0 & $ 6.8 \times 5.5 $ &  0.06 (1) & 3.9 (4) & 4.8 (10) &  & 10.4 & C \\
8$_{ 1, 7}$- 8$_{ 1, 8}A-+$ & 30,010.55 (0.02) & 98.8 & 0.2 & $ 6.8 \times 5.5 $ &  0.11 (0) &  7.6 (0) & 1.9 (0) &  & 11.8 & C \\
 & & & & &   0.07 (0) &  8.0 (0) & 5.3 (0)  & & & \\
17$_{ 2,15}$- 17$_{ 1,16}E$ & 30,308.03 (0.05) & 377.6 & 17.2 & $ 7.0 \times 5.2 $ &  0.03 (0) &  4.5 (0) & 5.2 (0)  &  & 12.0 & C \\
 & & & & &  0.64 (0) &  7.6 (0) & 1.9 (0)  & & & \\
 & & & & &  0.29 (0) &  8.0 (0) & 5.3 (0)   & & & \\
22$_{ 2,20}$- 22$_{ 1,21}E$ & 30,752.08 (0.11) & 609.6 & 15.1 & $ 6.2 \times 5.3 $ &  0.09 (0) &  4.5 (0) & 5.2 (0) & & 6.1 & C \\
 & & & & &   0.01 (0) &  7.6 (0) & 1.9 (0)   & & & \\
 & & & & &   0.05 (0) &  8.0 (0) & 5.3 (0)  & & & \\
18$_{ 2,16}$- 18$_{ 1,17}E$ & 30,858.30 (0.05) & 419.4 & 17.3 & $ 6.2 \times 5.3 $ &   0.10 (0) &  4.5 (0) & 5.2 (0)  & & 10.3 & C \\
 & & & & &   0.55 (0) &  7.6 (0) & 1.9 (0) & & & \\
 & & & & &   0.35 (0) &  8.0 (0) & 5.3 (0)  & & & \\
21$_{ 2,19}$- 21$_{ 1,20}E$ & 31,209.68 (0.09) & 558.6 & 16.0 & $ 6.9 \times 5.0 $ &  0.07 (0) &  4.5 (0) & 5.2 (0)  & & 8.0 & C \\
 & & & & &  0.09 (0) &  7.6 (0) & 1.9 (0)   & & & \\
 & & & & &  0.04 (0) &  8.0 (0) & 5.3 (0)   & & & \\
19$_{ 2,17}$- 19$_{ 1,18}E$ & 31,226.75 (0.05) & 463.5 & 17.2 & $ 6.9 \times 5.0 $ &  0.02 (0) &  4.5 (0) & 5.2 (0) &  & 8.4 & C \\
 & & & & &  0.26 (0) &  7.6 (0) & 1.9 (0)   & & & \\
 & & & & &  0.20 (0) &  8.0 (0) & 5.3 (0)   & & & \\
20$_{ 2,18}$- 20$_{ 1,19}E$ & 31,358.42 (0.05) & 509.9 & 16.7 & $ 7.7 \times 4.9 $ &  0.26 (0) &  7.6 (0) & 1.9 (0)  & & 8.4 & C \\
 & & & & &   0.08 (0) &  8.0 (0) & 5.3 (0)  & & & \\
19$_{ 4,15}$- 20$_{ 3,17}E$ & 31,977.79 (0.05) & 536.7 & 6.3 & $ 11.0 \times 8.3 $ &  0.03 (0) &  4.5 (0) & 5.2 (0) & & 19.8 & C/D \\
 & & & & &   0.20 (0) &  7.6 (0) & 1.9 (0) & & & \\
 & & & & &   0.04 (0) &  8.0 (0) & 5.3 (0)  & & & \\
23$_{5, 18}- 24_{4, 21}A-$ & 34,003.14 (0.13) & 777.9 & 7.4 & $14.1 \times 12.2$ & 0.06 (2) & 4.5 (9) & 1.8 (18) &  & 9.0 & C/D\\
14$_{ -3,12}$- 15$_{ -2,14}E$ & 34,236.95 (0.05) & 306.4 & 4.7 & $ 16.4 \times 11.2 $ &  0.23 (0) &  4.5 (0) & 5.2 (0) & & 21.2 & D \\
 & & & & &  0.49 (0) &  7.6 (0) & 1.9 (0)  & & & \\
 & & & & &  0.58 (0) &  8.0 (0) & 5.3 (0)   & & & \\
\cutinhead{$^{13}$CH$_{3}$OH}
7$_{ 2, 5}$- 7$_{ 1, 6}$ & 27,215.57 (0.05) & 85.8 & 7.2 & $ 10.4 \times 9.6 $ &  0.13 (1) & 7.7 (2) & 3.4 (5)  & & 18.6 & C \\
8$_{2 , 6}$- 8$_{1 , 7}$ & 27,364.09 (0.05) & 103.9 & 8.4 & $10.4 \times 9.6 $ & \tablenotemark{c} & & & & 28.1 & C\\
9$_{ 2, 7}$- 9$_{ 1, 8}$ & 27,581.63 (0.05) & 124.3 & 9.5 & $ 7.6 \times 6.5 $ &  0.15 (1) & 7.6 (0) & 2.1 (2)  & & 6.0 & C \\
10$_{ 2, 8}$- 10$_{ 1, 9}$ & 27,880.03 (0.05) & 147.0 & 10.7 & $ 10.6 \times 5.1 $ &  0.10 (0) & 7.4 (0) & 2.9 (1) &  & 7.8 & C \\
11$_{ 2, 9}$- 11$_{ 1,10}$ & 28,267.77 (0.05) & 171.9 & 11.9 & $ 9.3 \times 5.5 $ &  0.16 (3) & 8.2 (2) & 2.3 (5)  & & 11.2 & C \\
12$_{ 2,10}$- 12$_{ 1,11}$ & 28,747.75 (0.05) & 199.1 & 13.0 & $ 8.0 \times 5.4 $ &  0.10 (0) & 8.3 (1) & 3.0 (4)  & & 13.9 & C \\
15$_{ 2,13}$- 15$_{ 1,14}$ & 30,643.72 (0.05) & 294.3 & 16.1 & $ 6.5 \times 5.1 $ &  0.06 (14) & 7.3 (11) & 1.4 (33) &  & 7.2 & C \\
4$_{ -1, 4}$- 3$_{ 0, 3}$ & 32,398.48 (0.05) & 28.3 & 2.5 & $ 5.7 \times 5.3 $ &  0.09 (2) & 7.2 (2) & 2.0 (5) &  & 12.7 & C \\
\cutinhead{Hydrogen Recombination Lines}
H(77)$\beta$ & 27,724.10 (0.00) & \nodata & \nodata & $ 9.3 \times 5.5 $ &  0.02 (0) & 0.3 (10) & 35.7 (24) & 1 & 6.4 & C \\
H(61)$\alpha$ & 28,274.87 (0.00) & \nodata & \nodata & $ 9.3 \times 5.5 $ & 0.04 (0) & -6.2 (6) & 14.3 (20) & 2 & 6.4 & C \\
 & & & & &  -0.11 (0) & -5.0 (2) & 15.6 (6)  & 3 & & \\
 & & & & &  0.04 (0) & 5.6 (4) & 13.4 (12) & 4 & & \\
 & & & & &   -0.03 (0) & 28.0 (6) & 2.8 (14)  & 4 & & \\
H(60)$\alpha$ & 29,700.36 (0.00) & \nodata & \nodata & $ 6.6 \times 5.4 $ &  0.03 (0) & 3.3 (10) & 20.2 (29) & 5 & 10.5 & C \\
 & & & & &  -0.04 (0) & -4.8 (5) & 10.8 (13)  & 6 & & \\
H(59)$\alpha$ & 31,223.31 (0.00) & \nodata & \nodata & $ 6.9 \times 5.0 $ &  0.05 (0) & 1.1 (5) & 22.5 (16) & 5 & 7.3 & C \\
 & & & & &   0.04 (0) & 0.4 (8) & 22.7 (23) & 7 & & \\
 & & & & &   0.06 (0) & 5.3 (15) & 41.5 (39) & 8 & & \\
H(83)$\gamma$ & 32,718.50 (0.00) & \nodata & \nodata & $ 15.9 \times 11.2 $ &  0.06 (1) & -17.9 (6) & 4.6 (17) & 11 & 11.6 & D \\
 & & & & &  0.05 (0) & -7.0 (7) & 7.4 (21)  & 11 & & \\
 & & & & & 0.07 (0) & -7.2 (7) & 34.0 (18)  & 10 & & \\
H(91)$\delta$ & 32,732.67 (0.00) & \nodata & \nodata & $ 15.9 \times 11.2 $ & -0.02 (1) & -20.6 (13) & 4.5 (28) & 4 & 10.6 & C/D \\
 & & & & & 0.04 (2) & -3.7 (5) & 1.6 (8) & 4 & & \\
 & & & & & 0.03 (0) & 9.7 (16) & 14.7 (44) & 4 & & \\
H(58)$\alpha$\tablenotemark{b} & 32,852.20 (0.00) & \nodata & \nodata & $ 15.9 \times 11.2 $ &  0.16 (0) & -2.5 (1) & 20.6 (3) & 4 & 23.5 & C/D \\
 & & & & &  0.23 (3) & 5.7 (3) & 15.2 (10) & 10 & & \\
 & & & & &  0.05 (1) & 19.8 (71) & 26.5 (79)  & 10 & & \\
 & & & & &  0.34 (0) & -4.5 (1) & 22.7 (2)  & 11 & & \\
 & & & & &  0.03 (0) & 15.8 (10) & 7.8 (28) & 11 & & \\
H(72)$\beta$ & 33,821.51 (0.00) & \nodata & \nodata & $ 8.1 \times 7.4 $ &  0.04 (1) & -6.3 (16) & 14.4 (40) & 2 & 14.4 & C/D \\
 & & & & &  -0.02 (0) & 12.1 (64) & 24.8 (119)  & 2 & & \\
H(57)$\alpha$ & 34,596.38 (0.00) & \nodata & \nodata & $ 17.2 \times 11.2 $ &  0.76 (0) & -7.0 (1) & 18.5 (2) & 2 & 45.4 & D\\
 & & & & &  0.13 (1) & -12.6 (3) & 5.6 (9)  & 4 & & \\
 & & & & &  0.25 (3) & -5.0 (2) & 7.1 (8) & 4 & & \\
 & & & & &  0.19 (1) & 3.0 (9) & 12.9 (13) & 4 & & \\
 & & & & &  0.38 (0) & 5.3 (2) & 18.1 (4) & 10 & & \\
 & & & & &  -0.10 (0) & -18.5 (8) & 19.0 (22) & 10 & & \\
\cutinhead{Helium Recombination Lines}
He(58)$\alpha$\tablenotemark{b} & 32,865.58 (0.00) & \nodata & \nodata & $ 15.9 \times 11.2 $ &\tablenotemark{d} && & 1,2 & 9.5 & C/D \\
He(57)$\alpha$ & 34,610.48 (0.00) & \nodata & \nodata & $ 17.2 \times 11.1 $ &  0.07 (1) & 7.4 (5) & 4.8 (10)  & 3 & 15.8 & D \\
 & & & & &  0.07 (0) & 1.8 (8) & 9.7 (22)  & 4 & & \\
\cutinhead{H$_{2}$CO}
3$_{1,2}$-3$_{1,3}$\tablenotemark{b} & 28,974.80 (0.01) & 33.4 & 0.6 & $ 7.7 \times 5.3 $ &  0.12 (0) & 4.1 (7) & 11.1 (9) & I & 11.2 & C \\
 & & & & &  0.14 (2) & 8.5 (1) & 4.2 (6)  & I & & \\
 & & & & &  0.10 (2) & 10.7 (1) & 1.6 (3) & HC-NE & \\
 & & & & &  0.46 (3) & 7.5 (0) & 1.7 (0) & CR-SE & & \\
10$_{ 2, 8}$-10$_{ 2, 9}$ & 34,100.05 (0.01) & 240.7 & 4.0 & $ 14.1 \times 12.2 $ & 0.07 (3) & 7.8 (14) & 6.8 (38)  & & 12.0 & C/D \\
\cutinhead{H$_{2}$CS}
1$_{0,1}$-0$_{0,0}$ & 34,351.43 (0.02) & 1.6 & 1.0 & $16.4 \times 11.2  $ &  0.08 (1) & 8.4 (2) & 2.3 (6) & CR & 17.8 & D \\
 & & & & &  0.09 (5) & 10.6 (4) & 1.4 (10) & CS1 & & \\
\cutinhead{SO}
1$_{0}$-0$_{1}$\tablenotemark{b} & 30,001.52 (0.10) & 1.4 & 1.0 & $ 6.8 \times 5.5 $ &  0.85 (0) & 5.3 (0) & 12.4 (1) & HC/CR & 8.6 & C \\
 & & & & & 0.16 (0) & 20.9 (3) & 11.5 (8)  & HC/CR & & C\\
 & & & &  &  0.18 (1) & 9.2 (81) & 7.2(42) & HC-NE & 8.6 & C\\
 & & & & &  0.40 (1) & 10.7 (0) & 4.0 (2) & HC-NE & & C\\
 & & & & &  0.10 (1) & 21.6 (5) & 8.1 (12) & HC-NE & & C\\
 & & & &  &  0.24 (0) & 1.6 (5) & 25.0 (9) &IRc6 & 8.6 & C\\
 & & & & &  0.33 (4) & 11.7 (2) & 6.8 (4) & IRc6 & & C\\
 & & & & &  0.22 (2) & 18.5 (87) & 4.3(51) & IRc6 & & C\\
\cutinhead{$^{34}$SO}
1$_{0}$-0$_{1}$ & 29,678.98 (0.10) & 1.4 & 1.0 & $ 6.6 \times 5.4 $ &  0.02 (0) & 8.0 (9) & 22.8 (27) &  & 11.7 & C \\
% uncertainties could not be calculated
2$_{3}$-2$_{2}$ & 34,857.16 (0.20) & 20.9 & 0.7 & $ 15.1 \times 11.5 $ & 0.12 \tablenotemark{e} & 2.4 & 15.5  & & 15.9 & D \\
 & & & & &   0.05  & 17.6 & 10.5  & & & \\
\cutinhead{SO$_{2}$}
25$_{ 2,24}$-24$_{ 3,21}$ & 27,932.41 (0.00) & 303.7 & 3.1 & $ 10.6 \times 5.1 $ & 0.02 (0) & 4.5 (9) & 18.3 (23) & I & 8.6 & C \\
17$_{2,16}$-16$_{3,13}$ & 28,858.04 (0.02) & 149.2 & 2.1 & $ 7.7 \times 5.3 $ & 0.02 (2) & -6.4 (19) & 2.8 (47) & I & 10.2 & C \\
& & & & &  0.17 (1) & 5.0 (2)  & 5.4 (5)  & I & & \\
& & & & &  0.08 (1) & 5.6 (7) & 15.7 (17)  & I & & \\
& & & & & 0.05 (0) & -4.2 (4) & 9.9 (11) & IRc6 &  &  \\
& & & & &  0.10 (0) & 13.2 (1) & 10.2 (5)   & IRc6 & & \\
4$_{0,4}$-3$_{1,3}$ & 29,321.33 (0.02) & 9.2 & 1.7 & $ 7.4 \times 5.6 $ & 0.08 (0)\tablenotemark{f} & 2.9 (9) & 40.4 (33)  & I & 28.0 & C \\
& & & & &  0.27 (2)\tablenotemark{f} & 4.5 (1) & 9.7 (6)  & I & & \\
& & & & &  0.29 (2)\tablenotemark{f} & 4.9 (1)  & 2.7 (3)  & I & & \\
& & & & &  0.10 (1)\tablenotemark{f} & 7.5 (9) & 17.8 (21) & HC-NE & & \\
& & & & &   0.15 (9)\tablenotemark{f} & 10.5 (5) & 1.9 (13)  & HC-NE & & \\
& & & & &   0.11 (0) & 6.2 (15)  & 26.8 (22) & IRc6 &  &  \\
& & & & & 0.17 (2) & 12.0 (8) & 5.6 (17)   & IRc6 & & \\
& & & & &  0.11 (3) & 18.6 (16) & 6.1 (30)  & IRc6 & & \\
16$_{2,14}$-17$_{1,17}$ & 30,205.52 (0.10) & 137.5 & 0.5 & $ 6.7 \times 5.3 $ &  0.07 (0) & 3.8 (4) & 4.0 (8)  & I & 11.4 & C \\
16$_{ 4,12}$- 17$_{ 3,15}$ & 31,922.21 (0.03) & 164.5 & 6.7 & $ 11.0 \times 8.3 $ &  0.09 (1)\tablenotemark{f} & -1.5 (15) & 7.5 (27) & I & 14.4 & C/D \\
& & & & &  0.36 (2)\tablenotemark{f} & 4.9 (2)  & 5.9 (5)  & I & & \\
& & & & &  0.09 (0)\tablenotemark{f} & 11.4 (13) & 32.9 (22)  & I & & \\
& & & & &  0.07 (1) & 4.4 (4) & 3.7 (10)  & HC-NE & & \\
& & & & &  0.08 (0) & 8.4 (8) & 19.8 (22)  & HC-NE & & \\
& & & & & 0.09 (1) & 8.3 (18) & 26.1 (26) & IRc6 &  &\\
& & & & &  0.04 (1) & 15.3 (15) & 36.0 (47)  & IRc6 & & \\
26$_{ 6,20}$- 27$_{ 5,23}$\tablenotemark{b} & 34,097.72 (0.03) & 411.4 & 10.9 & $ 14.1 \times 12.2 $ &  0.15 (1) & 5.5 (2) & 6.2 (7)  & I & 9.0 & C/D \\
& & & & &  0.08 (1) & 4.9 (3) & 5.4 (8) & HC-NE & & \\
\cutinhead{$^{34}$SO$_{2}$}
4$_{0,4}$-3$_{1,3}$ & 31,011.18 (0.05) & 9.1 & 1.7 & $ 6.3 \times 5.3 $ &  0.07 (15) & 8.5 (9) & 1.2 (22) & & 9.9 & C \\
& & & & &  0.02 (0) & 15.9 (3) & 2.0 (8)  & & & \\
\cutinhead{CH$_{3}$CCH}
2$_{1}$-1$_{1}$\tablenotemark{b} & 34,182.76 (0.04) & 9.7 & 1.5 & $ 16.4 \times 11.2 $ &  0.16 (2) & 10.6 (0) & 1.6 (1)  & CS1 & 25.4 & D \\
 & & & & &  0.02 (1) & 8.8 (0) & 1.1 (3)   & CR-SW & & \\
2$_{0}$-1$_{0}$\tablenotemark{b} & 34,183.42 (0.04) & 2.5 & 2.0 & $ 16.4 \times 11.2 $ &  0.29 (3) & 10.6 (0) & 1.6 (1)  & CS1 & 14.3 & D \\
 & & & & &  0.15 (4) & 8.8 (0) & 1.1 (3)   & CR-SW & & \\
\cutinhead{NH$_{3}$}
9$_{ 9}$0a- 9$_{-9}$0s\tablenotemark{b} & 27,477.94 (0.01) & 852.8 & 145.8 & $ 7.6 \times 6.5 $ &  0.21 (0) &  1.0 (0) & 6.3 (0)  & & 8.9 & C \\
 & & & & &  0.62 (0) &  6.0 (0) & 5.7 (0)   & & & \\
10$_{ 10}$0a - 10$_{ 10}$0s & 28,604.74 (0.01) & 1035.6 & 90.0 & $ 8.1 \times 5.4 $ &  0.04 (0) &  1.0 (0) & 6.3 (0)  & & 14.4 & C \\
 & & & & &  0.36 (0) &  6.0 (0) & 5.7 (0)   & & & \\
11$_{ 11}$0a - 11$_{ 11}$0s & 29,914.49 (0.01) & 1234.8 & 99.4 & $ 6.8 \times 5.5 $ &  0.20 (0) &  6.0 (0) & 5.7 (0)  & & 9.8 & C \\
12$_{ 12}$0a - 12$_{ 12}$0s & 31,424.94 (0.01) & 1450.0 & 108.8 & $ 7.7 \times 4.9 $ &  0.17 (0) &  6.0 (0) & 5.7 (0)  & & 7.6 & C \\
13$_{ 13}$0a - 13$_{ 13}$0s & 33,156.85 (0.01) & 1691.1 & 104.8 & $6.7 \times 4.4$ & 0.03 (0) &  6.0 (0) & 5.7 (0)  & & 9.9 & C\\
\cutinhead{CH$_{3}$OCHO}
9$_{2,7}$-9$_{2,8}E$ & 30,670.58 (0.05) & 30.2 & 1.8 & $ 6.5 \times 5.1 $ &  0.06\tablenotemark{e} & 7.4 & 0.5 & & 8.4 & C \\
9$_{2,7}$-9$_{2,8}A$ & 30,695.15 (0.05) & 30.2 & 1.8 & $ 6.5 \times 5.1 $ &  0.02 (0) & 8.9 (11) & 5.5 (31)  & & 10.2 & C \\
3$_{1,3}$-2$_{1,2}E$\tablenotemark{b} & 34,156.88 (0.05) & 4.0 & 7.2 & $ 16.4 \times 11.2 $ &  0.36 (0) & 7.9 (0)\tablenotemark{g} & 2.4 (0)  & & 17.1 & D \\
3$_{1,3}$-2$_{1,2}A$\tablenotemark{b} & 34,158.12 (0.05) & 3.9 & 7.2 & $ 16.4 \times 11.2 $ &  0.30 (0) & 7.9 (0) & 2.4 (0) &  & 24.8 & D \\
6$_{2,4}$-6$_{1,5}E$ & 34,671.76 (0.05) & 15.4 & 2.4 & $ 15.1 \times 11.5 $ &  0.11 (1) & 7.6 (1) & 2.0 (2)  &  & 10.4 & D \\
6$_{2,4}$-6$_{1,5}A$ & 34,682.81 (0.05) & 15.3 & 2.4 & $ 15.1 \times 11.5 $ &  0.11 (0) & 8.0 (0) & 2.6 (1)  &  & 8.0 & D \\
5$_{2,3}$-5$_{1,4}E$ & 34,766.00 (0.05) & 11.7 & 1.9 & $ 15.1 \times 11.5 $ &  0.13 (2) & 7.8 (1) & 1.2 (2)  & & 10.9 & D \\
5$_{2,3}$-5$_{1,4}A$ & 34,775.65 (0.05) & 11.7 & 1.9 & $ 15.1 \times 11.5 $ &   0.08 (6) & 7.6 (5) & 1.6 (9) &  & 7.1 & D \\
\cutinhead{CH$_{3}$OCH$_{3}$}
1$_{1,0}$-1$_{0,1}AE$ & 29,900.48 (0.03) & 2.3 & 12.1 & $ 6.8 \times 5.4$ &  0.03 (1) & 7.8 (0) & 1.5 (1) &  & 10.8 & C \\
1$_{1,0}$-1$_{0,1}EA$ & 29,900.48 (0.03) & 2.3 & 8.1 & $ 6.8 \times 5.5 $ &  0.03 (1) & 7.8 (0) & 1.5 (1) &  & 10.8 & C \\
1$_{1,0}$-1$_{0,1}EE$ & 29,901.40 (0.03) & 2.3 & 32.2 & $ 6.8 \times 5.5 $ &  0.11 (1) & 7.8 (0) & 1.5 (1) &  & 10.8 & C \\
1$_{1,0}$-1$_{0,1}AA$ & 29,902.29 (0.03) & 2.3 & 20.1 & $ 6.8 \times 5.5 $ &  0.05 (3) & 7.8 (0) & 1.5 (1) &  & 10.8 & C \\
2$_{1,1}$-2$_{0,2}AE$ & 31,105.26 (0.10) & 4.2 & 6.6 & $ 6.9 \times 5.0 $ &  0.03 (0) & 8.1 (0) & 1.4 (0) &  & 8.0 & C \\
2$_{1,1}$-2$_{0,2}EA$ & 31,105.26 (0.10) & 4.2 & 13.2 & $ 6.9 \times 5.0 $ &  0.03 (0) & 8.1 (0) & 1.4 (0) &  & 8.0 & C \\
2$_{1,1}$-2$_{0,2}EE$ & 31,106.20 (0.05) & 4.2 & 52.6 & $ 6.9 \times 5.0 $ &  0.12 (1) & 8.1 (0) & 1.4 (0) &  & 8.0 & C \\
2$_{1,1}$-2$_{0,2}AA$ & 31,107.12 (0.10) & 4.2 & 19.7 & $ 6.9 \times 5.0 $ &  0.05 (0) & 8.1 (0) & 1.4 (0) &  & 8.0 & C \\
9$_{2,7}$-8$_{3,6}AA$ & 31,996.10 (0.03) & 47.0 & 15.8 & $ 8.2 \times 4.2 $ &  0.01 (0) & 8.0 (2) & 1.7 (6) &  & 7.9 & C \\
9$_{2,7}$-8$_{3,6}EE$ & 31,999.33 (0.03) & 47.0 & 25.2 & $ 8.2 \times 4.2 $ & 0.03(1) & 8.0 (2) & 1.7 (6) &  & 7.9 & C \\
3$_{1,2}$-3$_{0,3}AE$ & 32,977.34 (0.14) & 7.0 & 26.8 & $ 6.7 \times 4.3 $ &  0.03 (1) & 8.2 (2) & 1.6 (5) &  & 14.6 & C \\
3$_{1,2}$-3$_{0,3}EA$ & 32,977.34 (0.14) & 7.0 & 17.8 & $ 6.7 \times 4.3 $ &  0.03 (1) & 8.2 (2) & 1.6 (5) &  & 14.6 & C \\
3$_{1,2}$-3$_{0,3}EE$ & 32,978.29 (0.05) & 7.0 & 71.4 & $ 6.7 \times 4.3 $ &  0.09 (5) & 8.2 (2) & 1.6 (5) &  & 14.6 & C \\
3$_{1,2}$-3$_{0,3}AA$ & 32,979.24 (0.10) & 7.0 & 44.6 & $ 6.7 \times 4.3 $ &  0.07 (6) & 8.2 (2) & 1.6 (5) &  & 14.6 & C \\
8$_{4,4}$-9$_{3,7}EE$ & 33,943.17 (0.01) & 55.3 & 13.1 & $ 14.1 \times 12.2 $ & 0.09 (0)\tablenotemark{h} & 7.9 (0) & 1.6 & & 17.8 & C/D \\
8$_{4,4}$-9$_{3,7}AA$ & 33,943.96 (0.01) & 55.3 & 6.8 & $ 14.1 \times 12.2 $ & 0.05 (0)\tablenotemark{h} & 7.9 (0) & 1.6 & & 17.8 & C/D \\
8$_{4,4}$-9$_{3,7}EA$ & 33,944.05 (0.02) & 55.3 & 2.8 & $ 14.1 \times 12.2 $ & 0.02 (0)\tablenotemark{h} & 7.9 (0) & 1.6 & & 17.8 & C/D \\
\cutinhead{CH$_{3}$CH$_{2}$CN}
9$_{3,6}$-10$_{2,9}$ & 27,363.78 (0.05) & 29.4 & 1.4 & $ 10.4 \times 9.6 $ &  \tablenotemark{c} &  &  & & 15.1 & C \\
3$_{1,2}$-2$_{1,1}$\tablenotemark{b} & 27,561.61 (0.05) & 3.8 & 39.5 & $7.6 \times 6.5$ & 0.04 (0) & 4.8 (2) & 8.5 (6) & & 7.6 & C \\
 & & & & &  0.02 (0) & -8.1 (4) & 19.1 (9)   & & \\
4$_{1,4}$-3$_{1,3}$ & 34,824.07 (0.05) & 5.3 & 55.6 & $ 15.1 \times 11.5 $ &  0.21 (1) & 5.3 (3) & 8.4 (8) &  & 11.5 & D \\
& & & & &  0.06 (1) & -5.0 (12) & 11.7 (26)   & & & \\
\cutinhead{HC$_{3}$N}
3-2,F=3-3\tablenotemark{b} & 27,292.90 (0.01) & 2.6 & 1.5 & $ 10.4 \times 9.6 $ &  0.02 (0) & 10.0 (0) & 1.8 (0) & CS1 & 17.6 & C \\
 & & & & &   0.006 (0)\tablenotemark{i} & 2.2 (15) & 16.8 (16)  & I & & \\
 & & & & &   0.008 (1) & 5.8 (2) & 5.5 (9)  & I & & \\
 & & & & &   0.002 (0) & 18.2 (8) & 4.7 (21)  & I & & \\
3-2,F=2-1\tablenotemark{b} & 27,294.06 (0.03) & 2.6 & 8.3 & $ 10.4 \times 9.6 $ &  0.11 (0) & 10.0 (0) & 1.8 (0) & CS1 & 17.6 & C \\
 & & & & &   0.033 (4) & 2.2 (15) & 16.8 (16)  & I & & \\
 & & & & &   0.045 (7) & 5.8 (2) & 5.5 (9)  & I & & \\
 & & & & &   0.012 (3) & 18.2 (8) & 4.7 (21)  & I & & \\
3-2,F=3-2\tablenotemark{b} & 27,294.31 (0.03) & 2.6 & 12.3 & $ 10.4 \times 9.6 $ &  0.16 (1) & 10.0 (0) & 1.8 (0) & CS1 & 17.6 & C \\
 & & & & &   0.049 (7) & 2.2 (15) & 16.8 (16)  & I & & \\
 & & & & &   0.067 (10) & 5.8 (2) & 5.5 (9)  & I & & \\
 & & & & &   0.018 (4) & 18.2 (8) & 4.7 (21)  & I & & \\
3-2,F=4-3\tablenotemark{b} & 27,294.31 (0.03) & 2.6 & 17.8 & $ 10.4 \times 9.6 $ &  0.24 (1) & 10.0 (0) & 1.8 (0) & CS1 & 17.6 & C \\
 & & & & &   0.070 (9) & 2.2 (15) & 16.8 (16)  & I & & \\
 & & & & &   0.097 (15) & 5.8 (2) & 5.5 (9)  & I & & \\
 & & & & &   0.025 (7) & 18.2 (8) & 4.7 (21)  & I & & \\
3-2,F=2-2\tablenotemark{b} & 27,296.23 (0.01) & 2.6 & 1.5 & $ 10.4 \times 9.6 $ &  0.02 (0) & 10.0 (0) & 1.8 (0) & CS1 & 17.6 & C \\
 & & & & &   0.006 (0) & 2.2 (15) & 16.8 (16)  & I & & \\
 & & & & &   0.008 (1) & 5.8 (2) & 5.5 (9)  & I & & \\
 & & & & &   0.002 (0) & 18.2 (8) & 4.7 (21)  & I & & \\
\enddata
\tablenotetext{a}{Position where the spectra were taken, see text in each section for specific meanings, if it is blank, then spectra are taken from the emission peak.}
\tablenotetext{b}{This transition was part of the mosaic and single dish observations.}
\tablenotetext{c}{These transitions of $^{13}$\mtoh\ and \etcn\ are blended.}
\tablenotetext{d}{Spectral profile is too complex for a reliable fit}
\tablenotetext{e}{Uncertainties could not be calculated.}
\tablenotetext{f}{There could be more components, and likely are, but due to high noise no reliable fits could be made.}
\tablenotetext{g}{Intensities are free while \vlsr and FWHM were forced to be the same for each pair.}
\tablenotetext{h}{In order to obtain line fits, the ratio of the lines was fixed as was the FWHM.}
\tablenotetext{i}{Intensity ratios were fixed for low opacity.}

\end{deluxetable}

\clearpage
	
\subsection{Individual Molecular Results}
Table~\ref{t:summary} lists a count of all lines detected, by molecule. Table~\ref{tab:individual} lists the individual 
transitions that were detected\footnote{All rest frequencies and transition information were obtained from the \citet{jpl} and \citet{cdms} databases presented through the Splatalogue Database for Astronomical Spectroscopy (http://www.cv.nrao.edu/php/splat/).}. The columns give the quantum numbers, rest frequency in MHz, upper state energy in K, 
the line strength times the appropriate dipole moment in Debye$^2$, the synthesized beam in arcseconds, fitted intensity 
in \jbm, fitted \vlsr\ in \kms, fitted FWHM in \kms, position the spectrum was taken from, rms noise in m\jbm, and the 
array configuration the data were taken in, respectively. There were no unidentified lines in the entire survey. The 
following sections detail the results for each molecular species. For those where rotation temperatures and column 
densities were calculated the following equations were used. Assuming local thermodynamic equilibrium the upper state 
column density ($N_u$) for a single transition can be calculated from \citet{friedelphd}:

\begin{equation}
\frac{N_u}{g_u}=\frac{2.04W}{B\theta_a\theta_bS\mu^2\nu^3}\times10^{20}~{\rm cm}^{-2},
\label{eqn:nu}
\end{equation}

\noindent where $g_u$ is the upper state degeneracy, $W$ is the integrated intensity of the line in Jy beam$^{-1}$ 
\kms, $B$ is the beam filling factor, $\theta_a$ and $\theta_b$ are the major and minor axes of the synthesized beam, 
$S\mu^2$ is the product of the line strength and the square of the relevant dipole moment in Debye$^2$, and $\nu$ is 
the transition frequency in GHz. The beam filling factor is defined as:

\begin{equation}
B=\frac{\Theta^2_S}{\Theta^2_S+\Theta^2_B},
\label{eqn:B}
\end{equation}

\noindent where $\Theta^2_S$ is the source size in square arcseconds and $\Theta^2_B$ is the area of the synthesized 
beam. The source size can be determined by either fitting the data in the $u-v$ plane, fitting a Gaussian in the image 
plane with the MIRIAD task {\it imfit}, or by iterating over source sizes with the rotation temperature diagram method 
(see below) to minimize the $\chi^2$ of the fit. In the case of multiple velocity components that are unresolved, the 
only reliable option is with the rotation temperature diagram method.

From a single transition the total beam averaged column density can be calculated as:
\begin{equation}
\langle N_T\rangle=\frac{N_u}{g_u}Q_{rv}e^{E_u/T_{r}}~{\rm cm}^{-2},
\label{eqn:nt}
\end{equation}

\noindent where $Q_{rv}$ is the rotational-vibrational partition function, $E_u$ is the upper state energy of the 
transition, and $T_r$ is the rotation temperature. The rotation temperature diagram is an extension of the calculations 
for a single transition in that it is a plot of the natural log of equation~(\ref{eqn:nu}) versus $E_u$. A weighted fit 
to the plot gives the rotation temperature (the negative inverse of the slope) and $N_u/g_u$ at $E_u=0$ (the y-axis 
intercept). The total column density can be calculated from equation~(\ref{eqn:nt}).

Traditionally, when calculating column densities and relative abundances, only the rotational part of the partition 
function has been used \citep[e.g.,][]{remijan03,remijan04c}. The vibrational part of the partition function is usually 
only employed when vibrationally excited transitions of a species have been detected \citep[see e.g.,][]
{numm99,mehringer04,remijan05}. The rotational transitions of vibrationally excited states have been catalogued for 
only a few states of a few species. Thus it is highly likely that transitions of vibrationally excited states are in 
spectra, but their identity may not be known. In general the vibrational part of the partition function should be 
included in the analysis, {\it even if} vibrationally excited transitions have not been detected. While the lowest 
lying vibrationally excited state of many simple species are over 1000 K above ground, for many larger species the 
lowest lying vibrational states can be very low. For example l-C$_3$H has its lowest vibrationally excited state only 
39 K above ground, HC$_3$N has 3 vibrationally excited states below 720 K, and the asymmetric top acetone [\acetone] 
has 2 vibrationally excited states below 200~K \citep{cdms,groner05}. Thus, at temperatures commonly seen in 
Hot Cores, these low lying states will contain a significant population, which needs to be included in the column 
density.

The addition of the vibrational part of the partition function is rather straightforward. From \citet{ww05} we have:
\begin{equation}
Q_{rv}\approx\sum^n_{i=0}e^{-E_i/T_r}Q_r,
\label{eqn:qrv}
\end{equation}
where $Q_{rv}$ is the rotational-vibrational partition function, $n$ is the number of vibrationally excited states in 
the calculation, and $E_i$ is the energy above ground of the vibrationally excited state. Note that for $n=0$ this 
reduces to just $Q_{rv}=Q_r$. Table~\ref{t:qrv} gives the known vibrational energies (below 1000 K) for each molecular 
species detected in this survey.  The columns give the name of the species, rotational partition function, vibrational energies  
in K, the value of the rotational partition function at a temperature of 150 K, the value of the ro-vibrational partition function at 
150 K, the percent difference between the two partition function values, and the references for the molecular 
parameters, respectively. All of the species listed have vibrational states above 1000 K; however at Hot Core 
temperatures, these states will contribute a negligible amount to the column density. In addition to the energies 
listed in column~(3), there are combination bands (whose energies are the sum of any two or more of the energies listed 
in column~(3)) and overtones (integer multiples of the energies listed in column~(3)), which need to be taken into 
account. Both of these types have been detected in several species, such as \vycn\ (overtones) \citep{numm99, 
friedelphd} and HC$_3$N (overtones and combinations) \citep{remijan05, friedelphd}. To include overtones and combination 
bands, additional terms need to be added to equation~(\ref{eqn:qrv}) with the energy of the overtone or combination.

One must be careful when calculating partition functions for molecules with one or more internal rotors where the 
ground state energy between the torsional states is noticeable (e.g. \mtoh\ has a 14 K difference between the ground 
state energies of the A and E torsional states). Under typical hot core conditions the energy difference is 
insignificant and the partition function is just two times that for the A torsional state \citep{turner91}. However, in 
the extremes (i.e. cold dark clouds and very hot cores) the partition function must be calculated for each torsional state 
separately. For cold dark clouds the effect is obvious, the higher energy torsional state may not be populated. But for 
hot cores the effect is not as obvious, until you look at the energies of the vibrational states. The 14 K energy 
difference between the $A$ and $E$ ground torsional states grows to a $\sim$120~K difference between the $A$ and $E$ 
torsional states of the first vibrationally excited state.
With such a difference in energy the partition function and column density for each torsional state must be calculated 
separately and then combined.

\begin{deluxetable}{lcllrrr}
\tablecolumns{76}
\tablewidth{0pt}
\tabletypesize{\scriptsize}
\tablecaption{Partition Functions and Vibrationally Excited Energies\tablenotemark{a}\label{t:qrv}}
\tablehead{\colhead{ } & \colhead{ Rotational Partition} & \colhead{ Vibrational} & \colhead{ }& \colhead{ } &\colhead{\%}& \colhead{ }\\
\colhead{ Species} & \colhead{ Function ($Q_r$)} & \colhead{ Energies (K)} & \colhead{ $Q_r$(150 K)}& \colhead{ $Q_{rv}$(150 K)}& \colhead{Diff.} & \colhead{ Refs.}}
\startdata
 \mtoh  &   \phn$1.20T^{3/2}$           & 300.5, 423.6\tablenotemark{b,c} &  2263.5   &  2776.0   & 23 & 1\\
 H$_2$CO   & \phn0.55$T^{3/2}$                    & \tablenotemark{d}        &           &           &    & 2, 3\\
 H$_2$CS   &   \phn$1.15T^{3/2}$      & \tablenotemark{d}        &           &           &    & 2, 3\\
 SO        &   \phn$0.98T$\phantom{$^{3/2}$}      &   \tablenotemark{d}      &           &           &    & 1\\
 SO$_2$\tablenotemark{c}  &   \phn$2.30T^{3/2}$   &   744.8                  &  4164.7   &  4193.8   & 1  & 1\\
 CH$_3$CCH & \phn1.04$T^{3/2}$                    & 471.8, 910.6             & 1910.6    & 2000.8    & 5  & 1, 3\\
 NH$_3$    &  \phn$0.70T^{3/2}$                   &   \tablenotemark{d}      &           &           &    & 2, 3\\
 \mef      &   $12.45T^{3/2}$                     & 187, 457, 478            &  2.3\e{4} &  3.5\e{4} & 52 & 2, 4\\
 \dme      &   \phn$2.86T^{3/2}$\tablenotemark{e} & 292, 348, 601            &  1.7\e{5} &  2.2\e{5} & 29 & 2, 5\\
 \etcn     &   \phn$7.17T^{3/2}$                  & 297, 306, 544, 784       &  1.3\e{4} &  1.8\e{4} & 38 & 2, 6\\
 HC$_3$N   &   \phn$4.57T$\phantom{$^{3/2}$}      & 320.8, 717.8, 954.0      &  685.5    &  783.8    & 14 & 1\\
\enddata
\tablenotetext{\footnotesize a}{\footnotesize Only vibrationally excited states below 1000 K are listed, because even at hot core temperatures the population of vibrationally excited states above 1000 K will be less than 10\% of the total.}
\tablenotetext{\footnotesize b}{\footnotesize Isotopomers are assumed to have the same energies for their vibrationally excited states, unless otherwise noted.}
\tablenotetext{\footnotesize c}{\footnotesize The energies refer to the first vibrational state of the A and E torsional states.}
\tablenotetext{\footnotesize d}{\footnotesize Lowest vibrationally excited state lies above 1000 K}
\tablenotetext{\footnotesize e}{\footnotesize In order to calculate the column density one needs to multiply by 64/2 and weight the integrated intensity by the spin weight of the transition (see \citet{friedelphd} (Appendix I))}
\tablerefs{\footnotesize (1) \citet{cdms}; (2) \citet{nist}; (3) \citet{jpl}; (4) \citet{oest99}; (5) \citet{gron02}; (6) \citet{mehringer04} }

\end{deluxetable}

%--------------------------------------------------------------------------------------------------

\subsubsection{Methanol [\mtoh\ and $^{13}$\mtoh]}
\begin{figure}[!ht]
\includegraphics[scale=0.45]{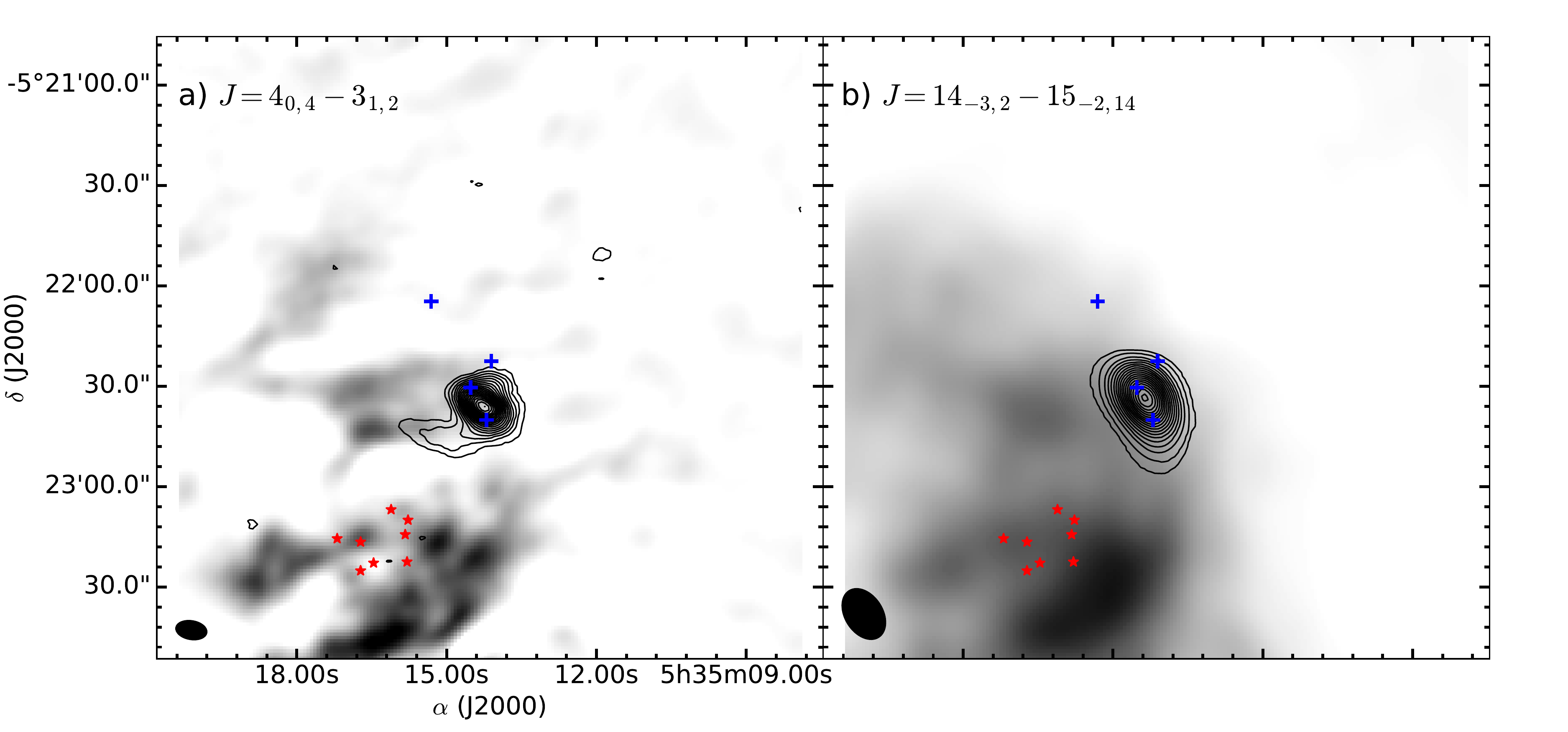}
\caption{Maps of \mtoh\ at two different resolutions, overlaid on the appropriate grayscale continuum. a) shows the $4_{0,4}-3_{1,2}$ transition and has a resolution of $9.3\arcsec\times5.5\arcsec$. b) shows the 
$14_{-3,1}-15_{-2,4}$ transition and has a resolution of $16.4\arcsec\times11.2\arcsec$. Contours are $\pm3\sigma, 
\pm7\sigma, \pm11\sigma, ...$ where $\sigma$=3.0 and 5.6 m\jbm\ respectively.
\label{fig:mtoh}}

\end{figure}

\begin{figure}[!ht]
\includegraphics[scale=0.5]{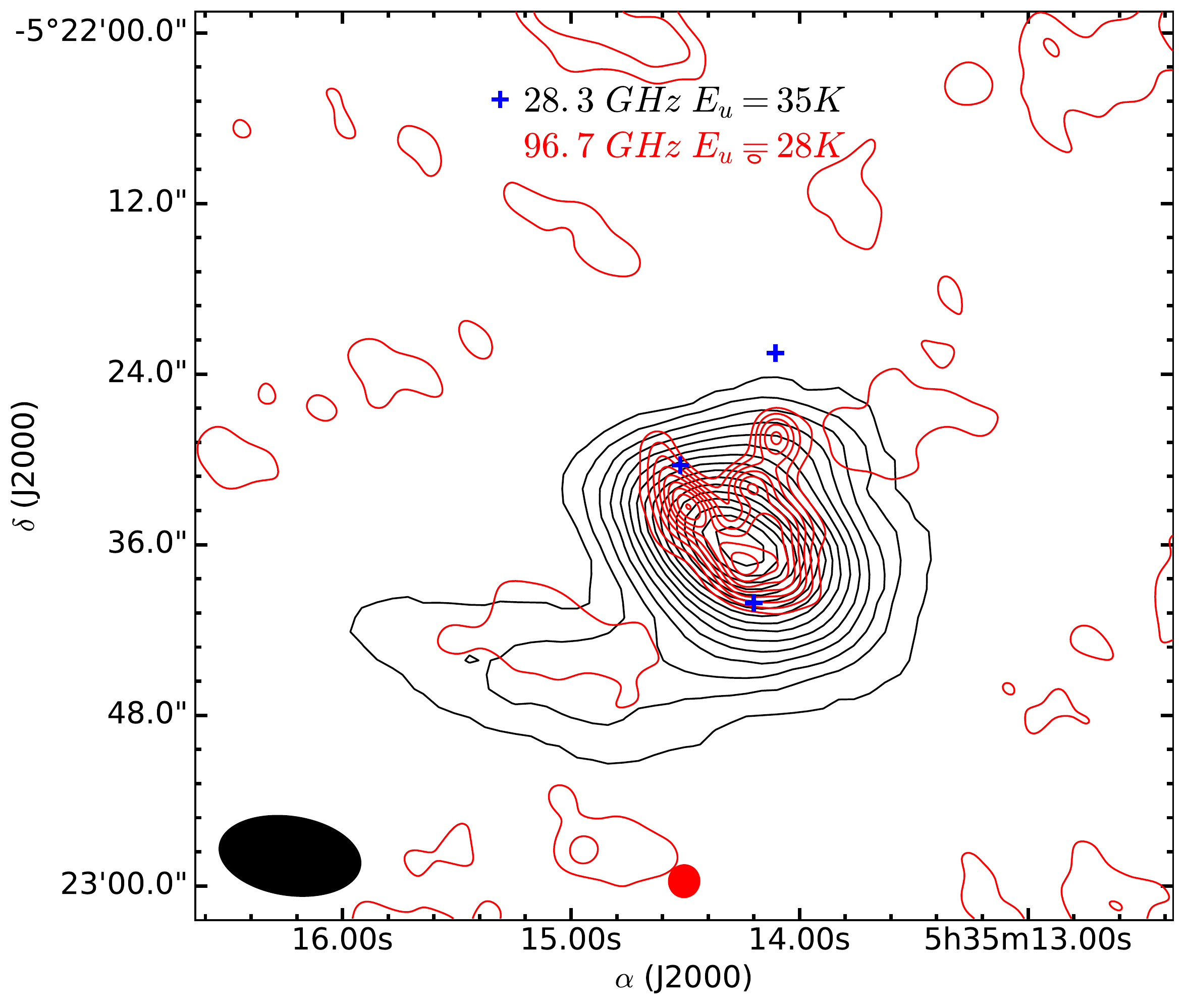}
\caption{Comparison of \mtoh\ emission at 96.7 GHz (red) and 28.3 GHz (black). The 96.7 GHz data are from \citet{friedel12}
and have a beam of 
$2.3\arcsec\times2.0\arcsec$.  The 28.3 GHz emission has a beam of $9.3\arcsec\times5.5\arcsec$.\label{fig:mtoh-compare}}
\end{figure}

Methanol had the largest number of detected transitions (21 for the main isotopologue and 8 for $^{13}$\mtoh) of all 
detected species. The analysis presented here focuses on the main isotopologue, but applies to $^{13}$\mtoh\ as well. 
The average intensity maps (over the width of the line) of the $4_{0,4}-3_{1,2}$ and $14_{-3,1}-15_{-2,4}$ transitions
in Figure~\ref{fig:mtoh} show that the 
emission is compact around the Hot Core/Compact Ridge region, with the exception of a weakly emitting region
extending to the southeast.
The \mtoh\ emission at 28.3 GHz  and 96.7 GHz are compared in Figure~\ref{fig:mtoh-compare}.
Both transitions have very similar 
upper state energy 35 K and 28 K, respectively. The 96.7 GHz data are at a considerably higher resolution,
but the vast majority of the 
emission appears to come from the region designated by the 96.7 GHz contours.

\begin{figure}[!ht]
\includegraphics[scale=0.75]{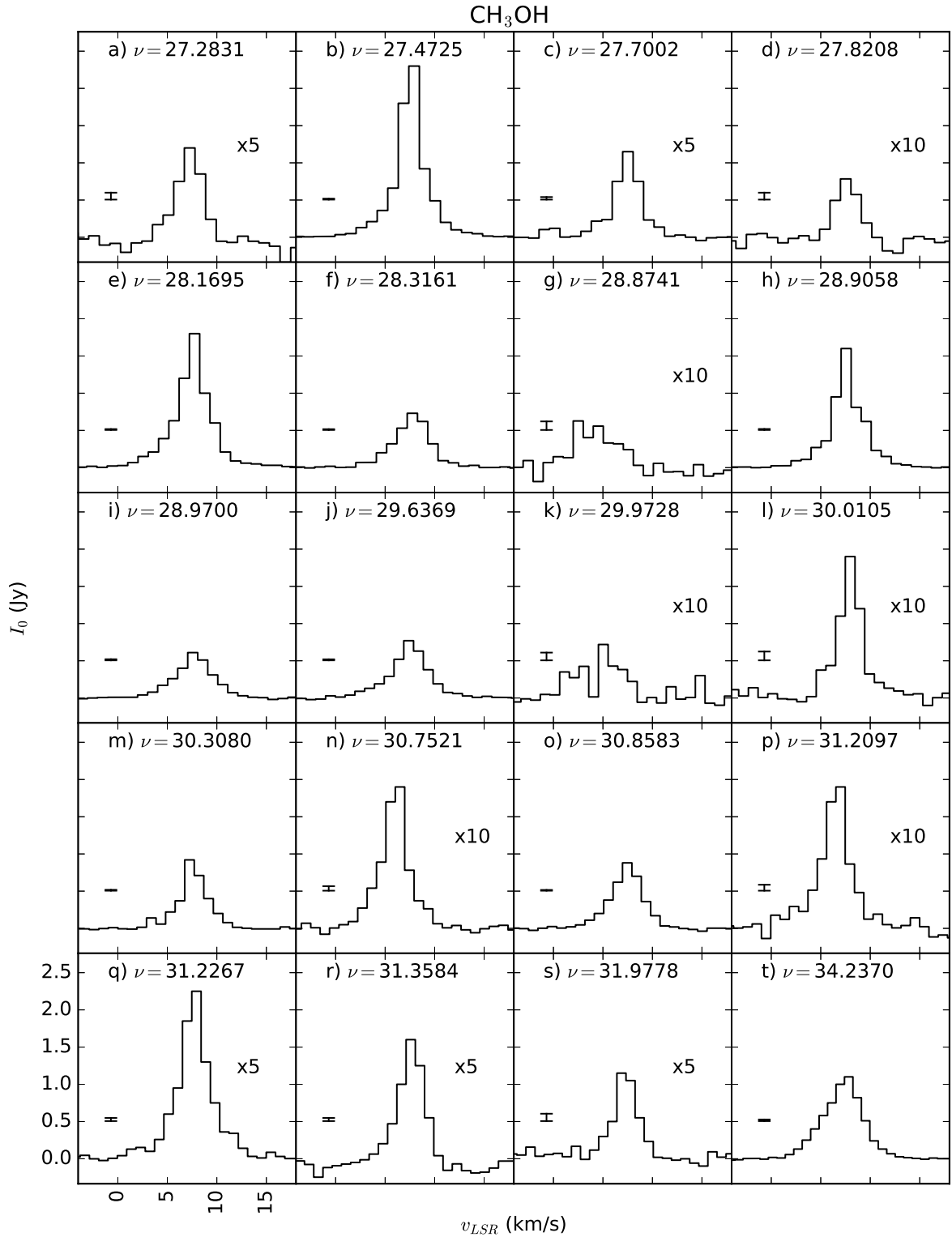}
\caption{Spectra of each detected \mtoh\ transition taken at the peak of emission. The abscissa is velocity in \kms\ and 
the ordinate is intensity in \jbm. Some of the spectra are scaled in intensity, denoted by ``$\times5$'' or ``$\times10$'' 
in the plot. The frequency of each transition is given at the top of the sub-plot and the 'I' bar denotes the 1$\sigma$ 
rms noise for each spectra.\label{fig:mtoh-spec}}
\end{figure}

\begin{figure}[!ht]
\includegraphics[scale=.3]{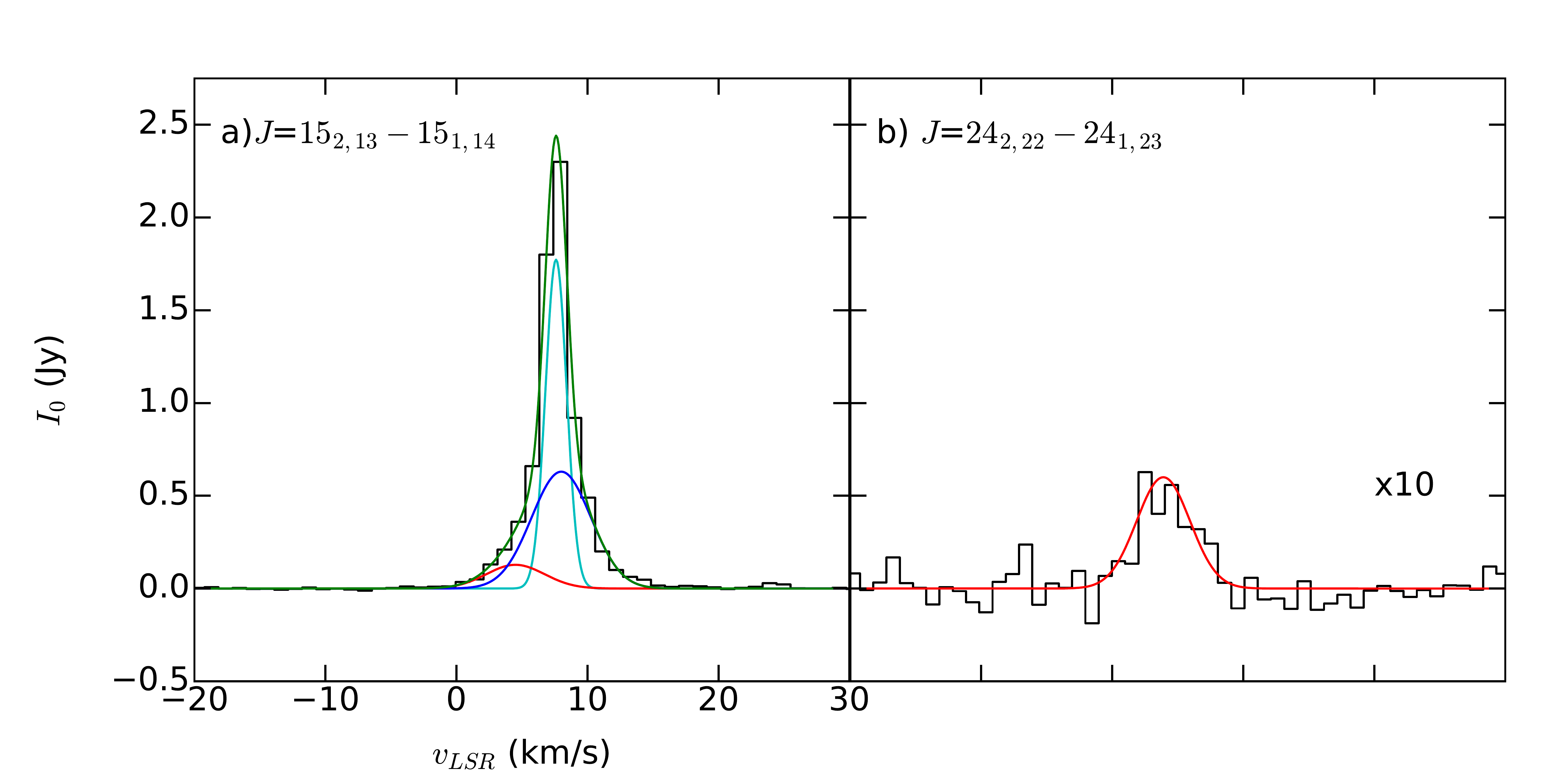}
\caption{Example fits to a lower energy, three component, line and a high energy, single component, line. The abscissa 
is velocity in \kms\ and the ordinate is intensity in \jbm. a) shows the three component fit with the 
4.5 \kms\ component shown in red, the 7.6 \kms\ component in cyan, the 8.0 \kms\ component in blue, and the sum total 
fit in green. b) shows a single component fit, shown in red, to a high energy line, centered at 4.4 
\kms. \label{fig:mtoh-fits}}
\end{figure}

In order to  create a rotation-temperature diagram to estimate the temperature
of the \mtoh\ gas, we must resolve the individual gas components along the line of sight.
Figure~\ref{fig:mtoh-spec} shows the spectra from the 20 lowest energy \mtoh\ lines taken at the emission peak of each 
map (all peaks are within half a beam of each other). 
Fits to each profile are described in 
Table~\ref{tab:individual}. The three highest energy transitions (662, 718, and 778 K) are best fit with a single 
Gaussian with a \vlsr\ of 4.4 (0) \kms\ and FWHM of 5.1 (1) \kms. The remaining 18 transitions were best fit with three 
Gaussians with \vlsr 's of 4.5 (0), 7.6 (0), and 8.0 (0) \kms\ and FWHMs of 5.2 (0), 1.9 (0), and 5.3 (0) \kms, respectively. The 
fitting was done simultaneously on all 18 transitions by constraining the \vlsr\ and FWHM of the 4.5 \kms\ component to 
be close to the values found for the three highest energy transitions and constraining all intensities to be positive valued. For a few of the transitions the intensity of the 4.5 \kms\ component is so small as to be 
undetectable in our observations, and are not reported. Figure~\ref{fig:mtoh-fits} shows an example of the three 
component fit in the panel (a). The fit for each Gaussian is shown individually (red for the 4.5 \kms\ component, 
green for the 7.6 \kms\ component, and blue for the 8.0 \kms\ component) and the sum of the Gaussians is shown by the 
green fit. The right hand plot shows the fit (in red) for a single component in the high energy lines.

\begin{figure}[!ht]
\includegraphics[scale=.2]{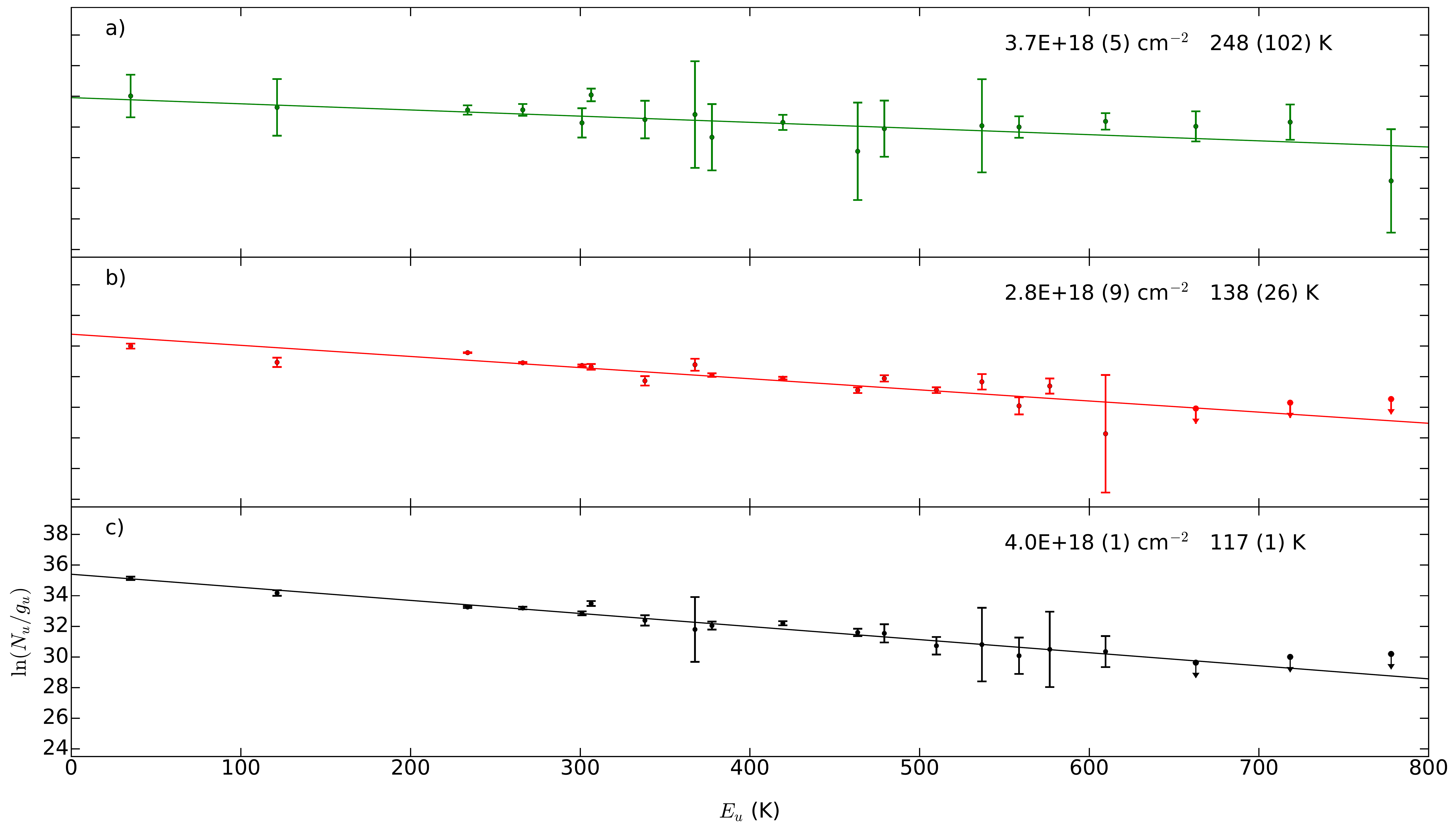}
\caption{Rotation temperature diagrams of the three \mtoh\ velocity components, plotting the natural log of the upper 
state column density ($N_u/g_u$) versus the upper state energy ($E_u$). a) is from the 4.5 \kms\ velocity 
component, b) is from the 7.6 \kms\ velocity component, and c) is from the 8.0 \kms\ 
component. The error bars are 3$\sigma$ and are dominated by the rms noise of each spectrum. The upper limits (middle 
and lower plots, three highest energy components) are based on our $3\sigma$ detection limit.
\label{fig:mtoh-rtd}}
\end{figure}

With the velocity components fit, we constructed a rotation-temperature diagram for each of the three components
(Figure~\ref{fig:mtoh-rtd}).  Panel (a) shows the fit for 
the 4.5 \kms\ component, giving a beam average total column density of \mtohNa\ and a rotation temperature of \mtohTa. 
These data were also best fit by a source size of $\sim 2\arcsec^2$. The source size was determined by allowing 
$\Theta_S$ to vary in equation~(\ref{eqn:B}) and minimizing the $\chi^2$. The uncertainties are dominated by the low 
intensity of the components compared to the rms noise. Panel (b) shows the fit for the 7.6 \kms\ component, 
giving a beam average total column density of \mtohNb\ and a rotation temperature of \mtohTb. These data were best fit 
by a source size of $\sim7\arcsec^2$. Panel (c) shows the best fit for the 8.0 \kms\ component, giving a beam 
averaged total column density of \mtohNc\ and a rotation temperature of \mtohTc. These data were best fit with a source 
size of $\sim10\arcsec^2$. The uncertainties for each point are 3$\sigma$, and the upper limits for the three highest 
energy transitions are based on a 3$\sigma$ detection threshold. The transition at 30.01055 GHz was excluded from these 
calculations because it had an abnormally high intensity; it also has the lowest line strength of all transitions by 
nearly an order of magnitude, and is the only parity changing transition. 
Based on the calculated total column 
densities and rotation temperatures there are no missing transitions from \mtoh\ in this survey. A missing transition 
is one that should have been detected above our 3$\sigma$ cutoff, but was not.

\begin{figure}[!ht]
\includegraphics[scale=0.75]{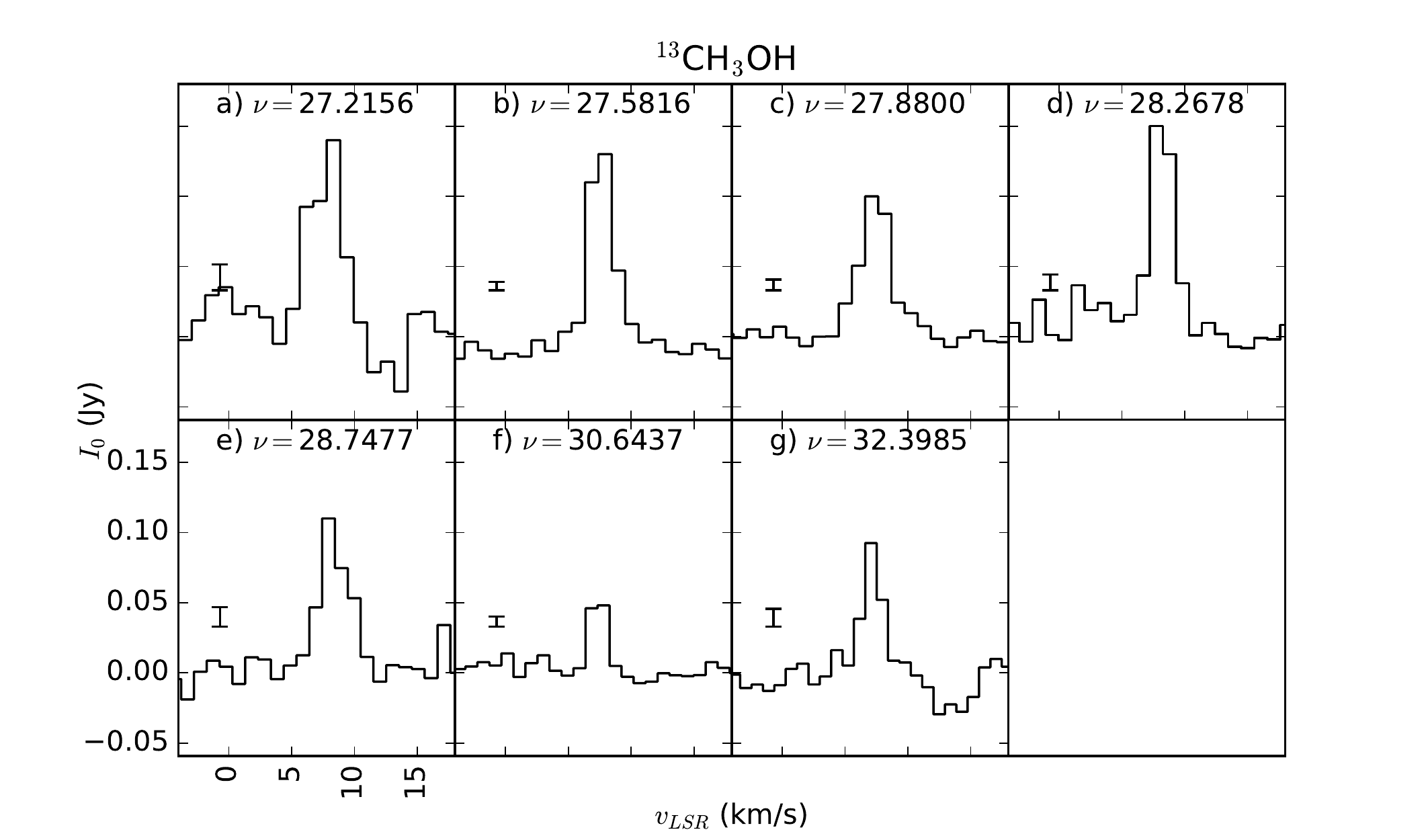}
\caption{Spectra of each detected $^{13}$\mtoh\ transition taken at the peak of emission. The abcissa is velocity in 
\kms\ and the ordinate is intensity in \jbm. The frequency of each transition is given at the top of the sub-plot and 
the 'I' bar denotes the 1$\sigma$ rms noise for each spectra.\label{fig:13mtoh-spec}}
\end{figure}

The same procedures were followed for the $^{13}$\mtoh\ transition (Figure~\ref{fig:13mtoh-spec}).
Fits to each profile are described in Table~\ref{tab:individual}. Unlike the main isotopomer, $^{13}$\mtoh\ 
was best fit with only a single component with a \vlsr\ of 7.6 (0) \kms\ and a FWHM of 2.7 (3) \kms. The \vlsr\ is 
consistent with the strongest component from the main isotopomer, but the FWHM is 50\% larger, indicating that it too 
may be blended by other unresolved velocity components. Given that the $^{13}$\mtoh\ line intensities are much weaker 
than those of \mtoh\ the presence of other velocity components similar to those of \mtoh\ would be undetectable in our 
observations. Based on the rotation temperature of the 7.6 \kms\ \mtoh\ component, the total beam averaged column 
density of $^{13}$\mtoh\ is \thmtohN.
Based on this total column 
density and rotation temperature there are no missing $^{13}$\mtoh\ lines in this survey.

\begin{figure}
\includegraphics[scale=0.5]{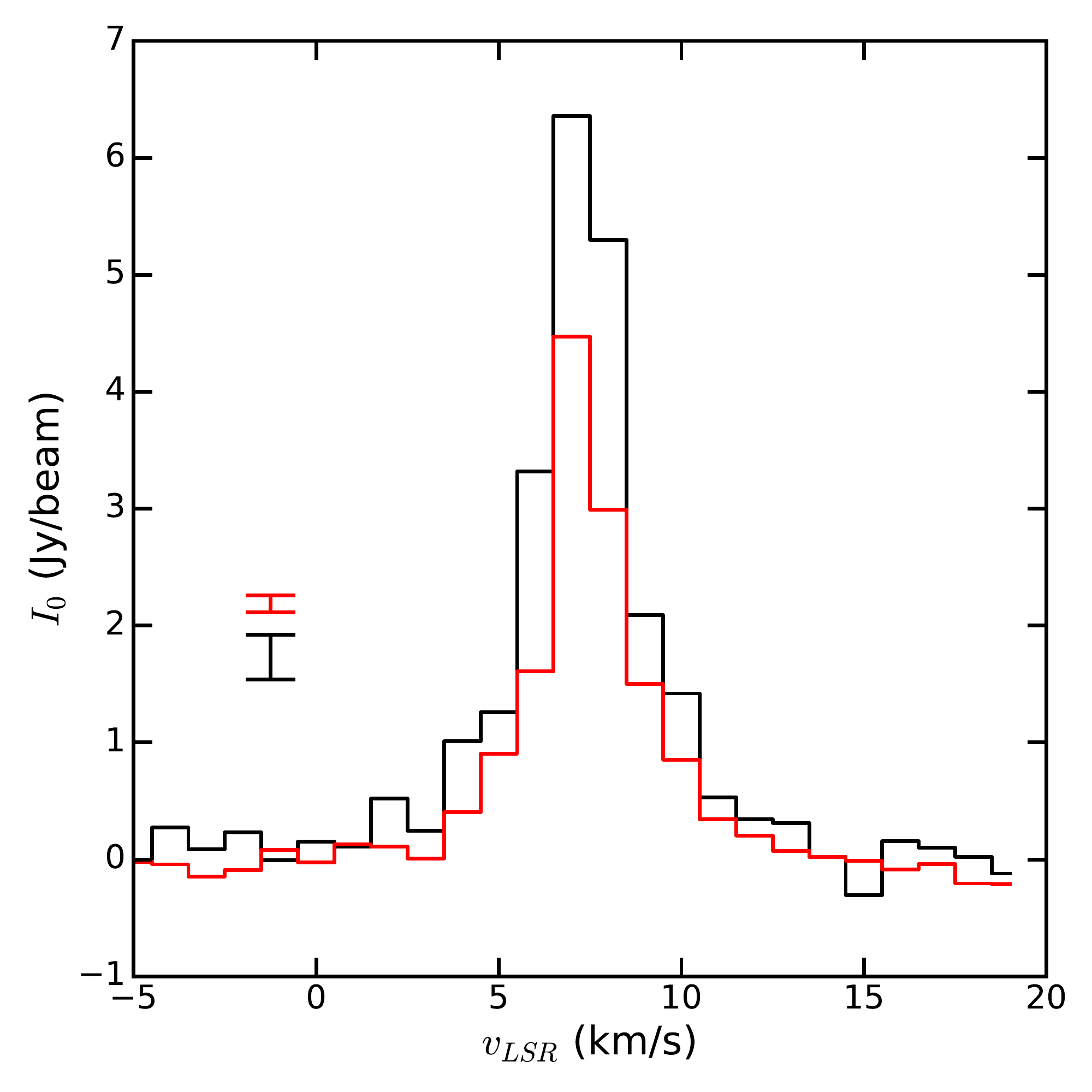}
\caption{Spectra from single dish observations (black) and array observations (red) of the $13_{2,11}-13_{1,12} E$ 
transition of \mtoh. The abscissa is velocity in \kms\ and the ordinate is intensity in \jbm. The 'I' bars denote the 
1$\sigma$ rms noise for the respective spectrum. \label{fig:mtoh-sd-spec}}
\end{figure}

\begin{figure}
\includegraphics[scale=.5]{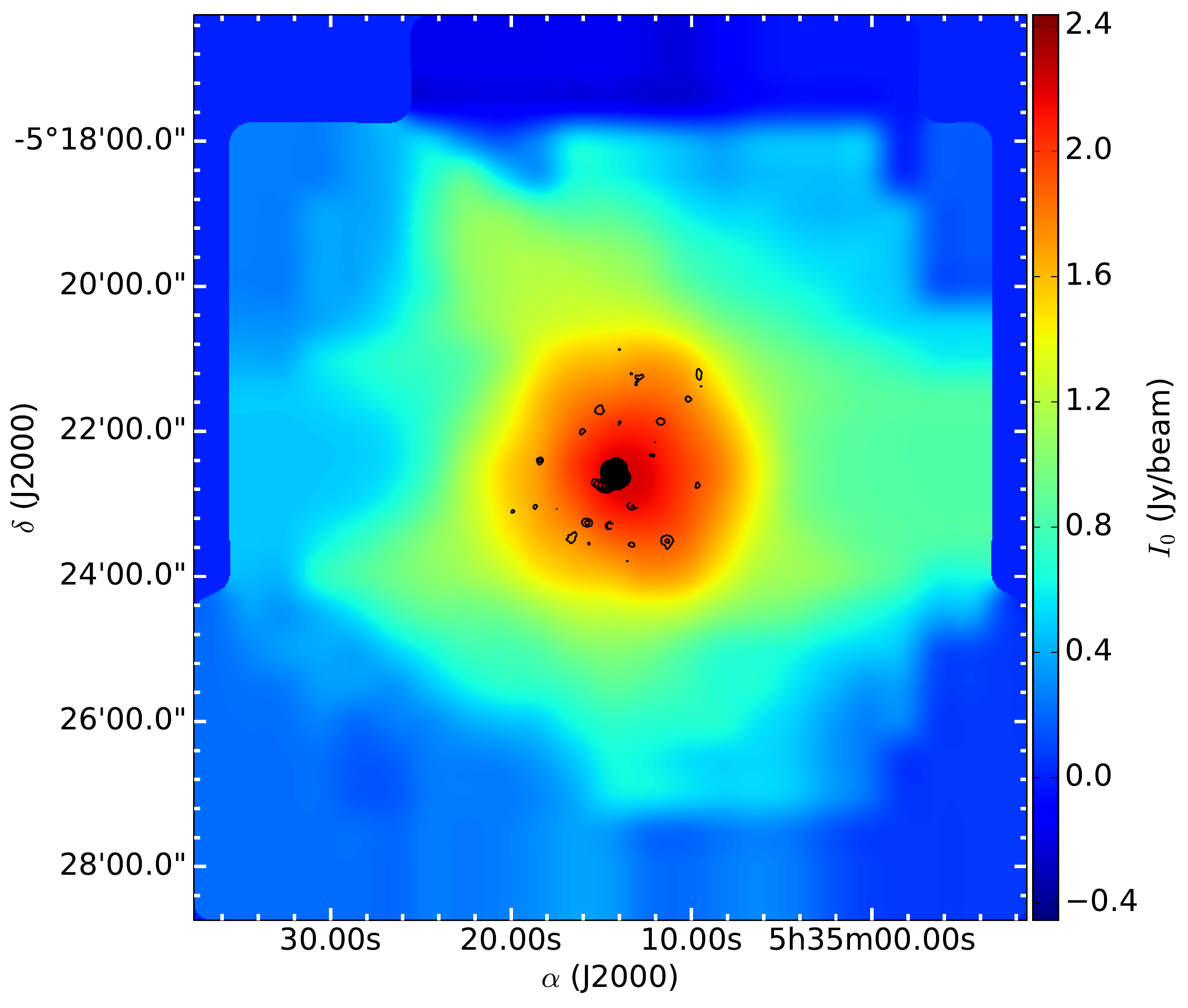}
\caption{Array average intensity map of the $13_{2,11}-13_{1,12} E$ transition of \mtoh\ (contours) overlaid on the color scale map 
of the same transition from single dish observations. Contours are $\pm3\sigma, \pm6\sigma, \pm9\sigma, ...$ where 
$\sigma$=7.2 m\jbm.
\label{fig:mtoh-sd-map}}
\end{figure}

From Figure~\ref{fig:mtoh}(a) there is an extension to the southeast that indicates that there may 
be some extended structure to the \mtoh\ emission. To investigate the extent of the large scale structure the 
$13_{2,11}-13_{1,12}$ ($E_u=233$ K) transition was observed in single dish mode\footnote{The lowest energy \mtoh\ 
transitions could not be observed due to a limitation on the number of spectral windows available.}. 
Figure~\ref{fig:mtoh-sd-spec} shows the single dish (black) and interferometric (red) spectra of the 
$13_{2,11}-13_{1,12}$ transition of \mtoh. The interferometric spectra were generated by convolving the array map with 
a $165\arcsec$ 2-D Gaussian. By comparing the spectra it can be seen that the interferometric observations detect 
nearly all of the flux seen by the single dish, indicating that there is little extended emission from this transition. 
The average intensity maps of this transition also show that the emission is from a 
compact source (Figure~\ref{fig:mtoh-sd-map}). The interferometric average intensity map is the contours overlaid on the single dish average intensity map. Note that the single dish data are clipped at $2\sigma$ before calculating the average intensity maps.

\clearpage

%--------------------------------------------------------------------------------------------------

\subsubsection{Hydrogen Recombination}
\begin{figure}[!ht]
\includegraphics[scale=0.33]{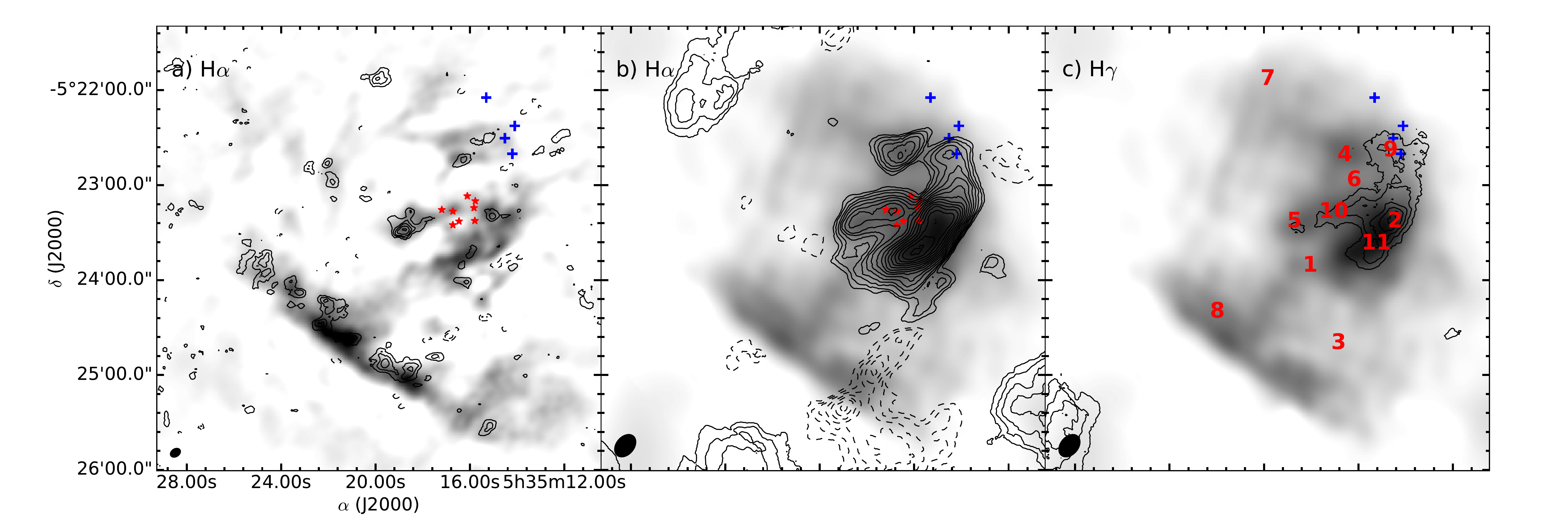}
\caption{Maps of three representative H recombination lines: a) H(59)$\alpha$, b) H(58)$\alpha$, and c) H(83)$\gamma$. Contours 
are $\pm3\sigma$, $\pm6\sigma$, $\pm9\sigma$, ..., where $\sigma$ = 1.1, 3.4, and 1.0 m\jbm, respectively.
\label{fig:h}}
\end{figure}

A total of nine H recombination lines (5$\alpha$, 2$\beta$, 1$\gamma$, and 1$\delta$) were detected in this survey. The 
detection of the emission was highly dependent on the size of the synthesized beam. The emission appears to emanate 
from a large region, as the C configuration only picked up the tightest knots of emission, while the D configuration 
data show significantly more structure. Figure~\ref{fig:h} shows maps from the H(59)$\alpha$, H(58)$\alpha$, and 
H(83)$\gamma$ transitions, from the C, D, and D configurations, respectively. 
The compact H recombination emission (panel (a)) is concentrated along the Bar with a few knots associated with the more prominent continuum peaks. 
Contrasting this, the more extended emission is concentrated to the southwest of the Trapezium, tracing the bulk of the 
continuum. The emission peaks toward the edge of the map and the large bowl are due to sidelobes and the array's 
response to more extended structure. Emission from the Bar is notably absent from the panel (b). This is likely due 
to emission that is both weaker than our detection threshold (due to higher noise in this region of the spectrum) and 
the sidelobes. The emission from the H(83)$\gamma$ transition (panel (c)) also traces the main continuum peak, 
although it is offset to the northwest from the H$\alpha$ peak. The red numbers denote the positions of the spectra.

\begin{figure}[!ht]
\includegraphics[scale=0.7]{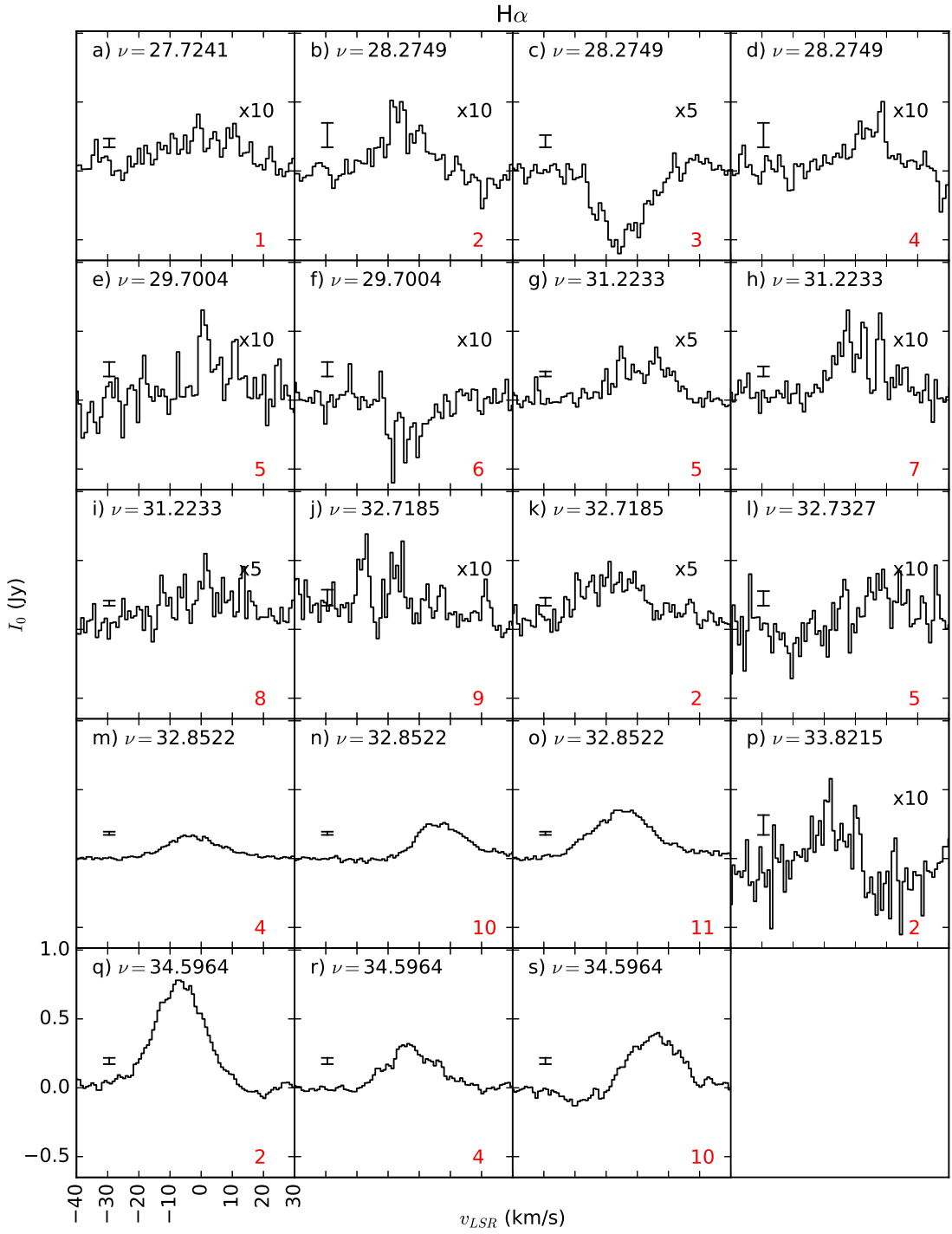}
\caption{Spectra from the H recombination transitions. The rest frequency of each is given at the top of the panel, the 
abscissa is \vlsr\ in \kms and the ordinate is intensity in \jbm. The rest frequency for each transition is given in 
each panel and the `I' bars denote the 1$\sigma$ rms noise.\label{fig:h_spec}}
\end{figure}

Figure~\ref{fig:h_spec} shows the spectra of the detected H recombination lines from different regions. 
The position at which the spectra were taken is denoted by 
the red number in the lower right corner of each panel. These numbers correspond to those in Figure~\ref{fig:h}(c).
There are numerous missing H 
recombination lines, but this is to be expected since the detection is highly correlated with the size of the 
synthesized beam.

\begin{figure}
\includegraphics[scale=0.5]{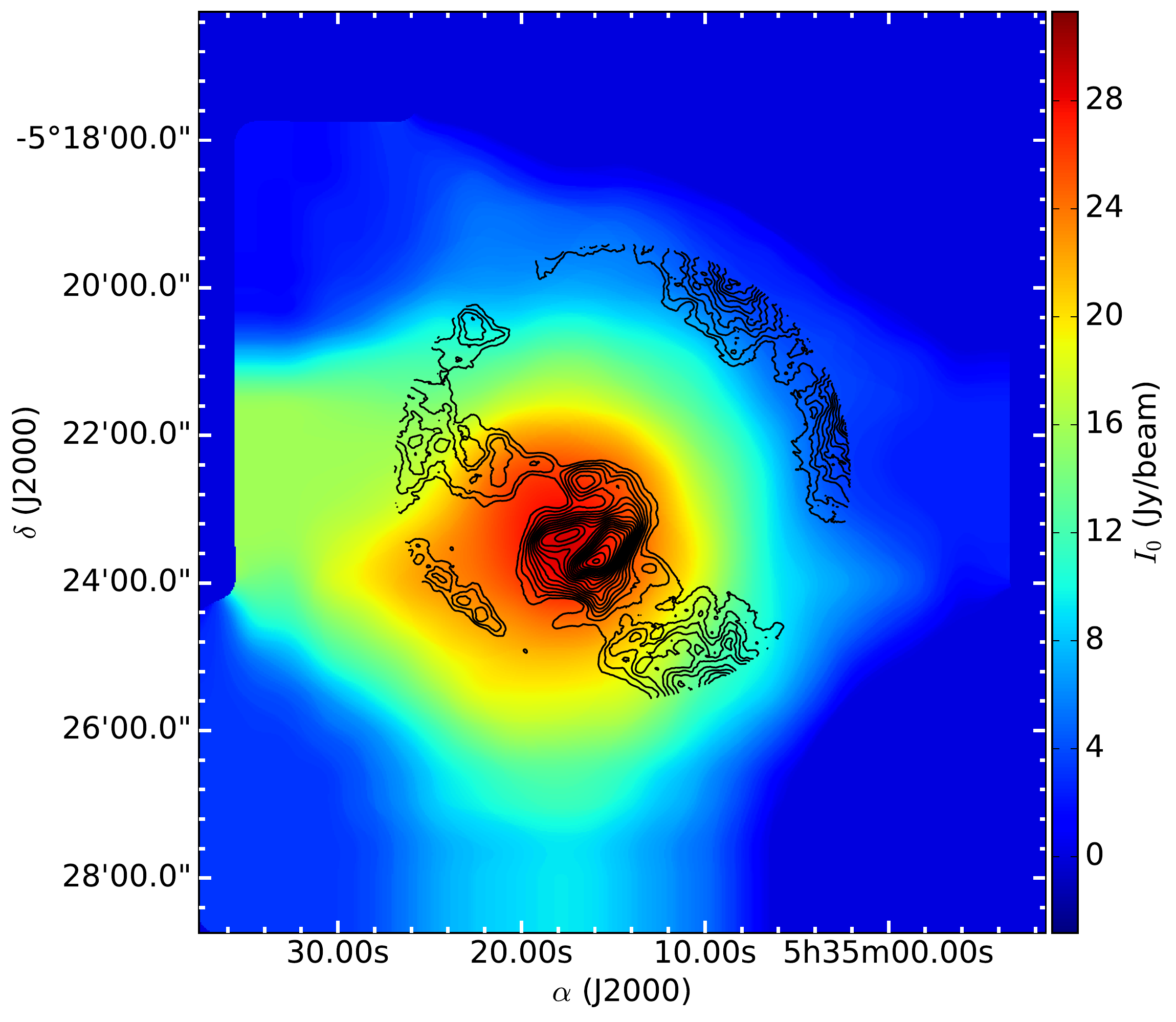}
\caption{Average intensity map of the single dish (color) map of H(58)$\alpha$ transition, overlaid with the interferrometric average intensity map. Contours are $\pm3\sigma, \pm6\sigma, \pm9\sigma, ...$, where $\sigma$=5 m\jbm.
\label{fig:hsd}}
\end{figure}

\begin{figure}
\includegraphics[scale=0.5]{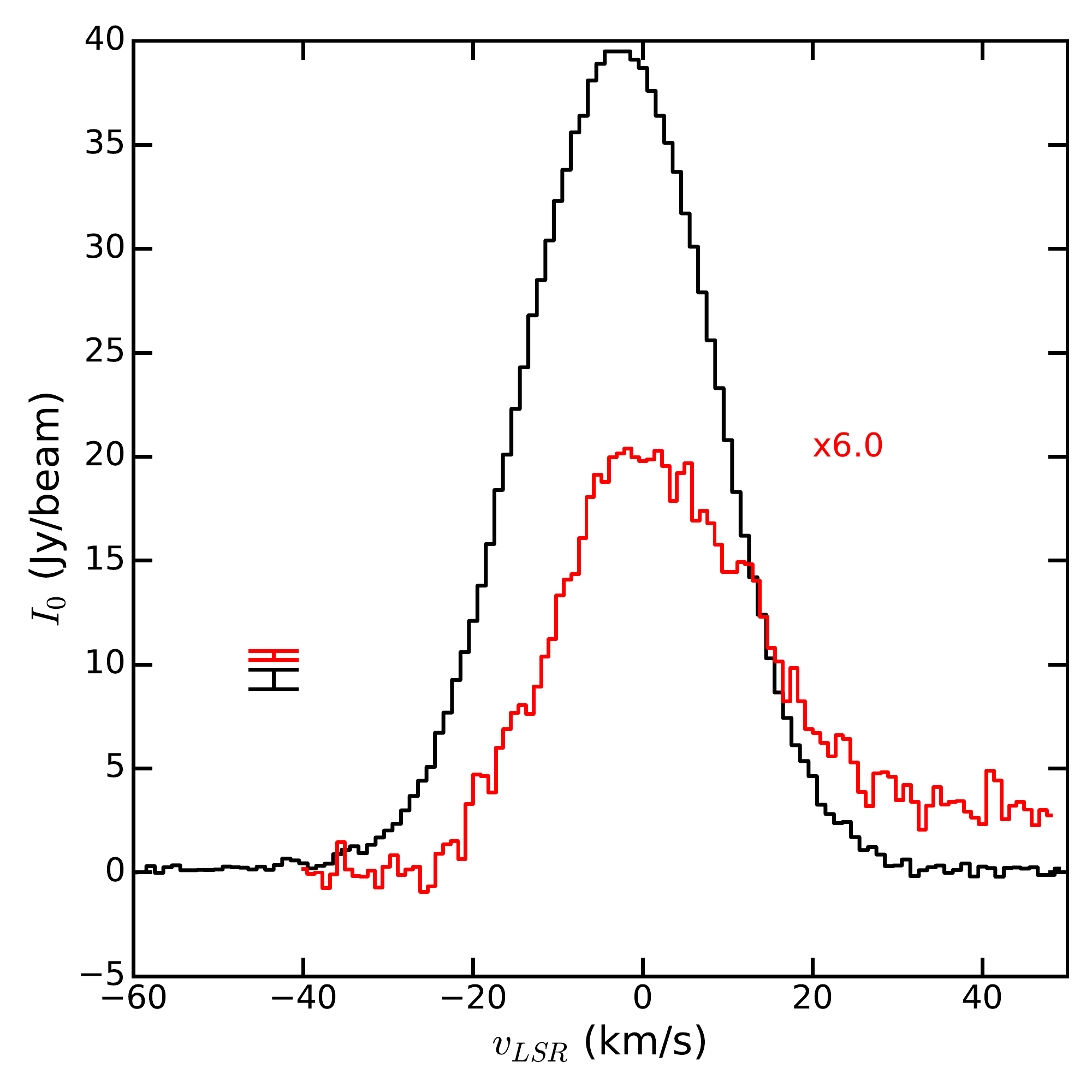}
\caption{Spectra from the single dish H(58)$\alpha$ observations (black) and interferrometric observations (red). The abscissa is 
\vlsr\ in \kms and the ordinate is intensity in \jbm\ and the `I' bars denote the 1$\sigma$ rms noise of the respective 
spectrum. \label{fig:hsdspec}}
\end{figure}

In order to determine how much flux was being resolved out by the array, the H(58)$\alpha$ transition was also observed 
in single dish mode with a 32 point mosaic pattern. Figure~\ref{fig:hsd} shows an average intensity map of the line in color, 
overlaid by the array map of the same transition. The single dish map indicates that the vast majority of the 
recombination emission comes from the region of Orion-KL, with an extension toward the Bar in the southeast, and an 
additional extension to the east, which is outside of the field-of-view of the interferometer. Figure~\ref{fig:hsdspec} 
shows the spectra from the single dish (black) and interferometric (red) observations. The black spectrum was taken at 
the peak of the single dish emission and the red spectrum is from the peak of the interferometric emission. The 
interferometric data are scaled by a factor of 6.0 to show them on the same scale and were best fit by three Gaussians 
with \vlsr's of -3.5(32), 10.5(49), and 39.8(456) \kms and FWHM of 20.4(22), 20.3(67) and 17.0(1580) \kms, 
respectively. Contrasting the the single dish data are best fit by a single large Gaussian with a \vlsr\ of -2.8(0) 
\kms and a FWHM of 26.1(0) \kms, indicating that there is a significant extended component that is resolved out by the interferometer.

\clearpage

%--------------------------------------------------------------------------------------------------

\subsubsection{He Recombination}
\begin{figure}[!ht]
\includegraphics[scale=0.7]{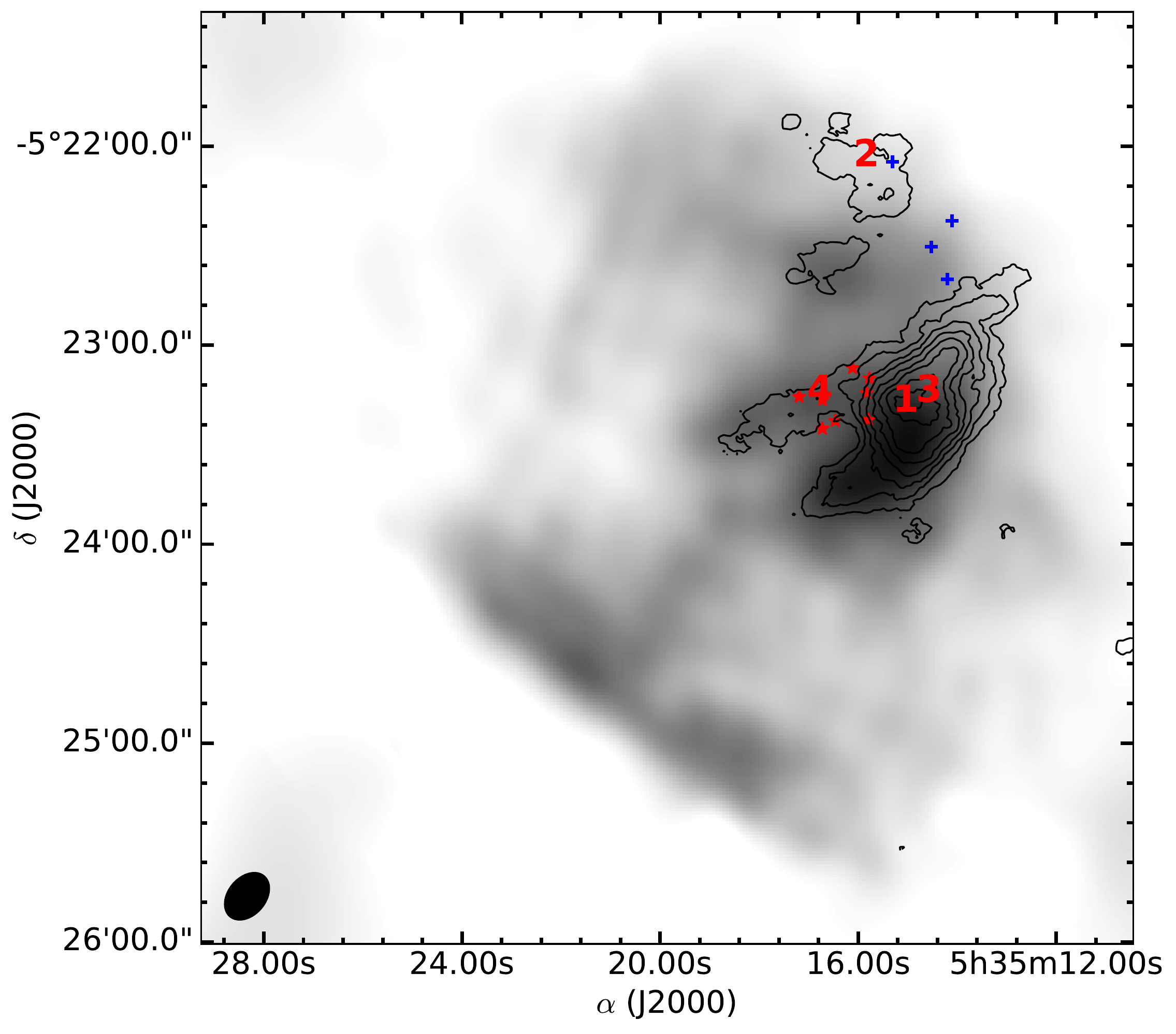}
\caption{Map of the He(58)$\alpha$ line, overlaid on the continuum. Contours are $\pm3\sigma$, $\pm6\sigma$, 
$\pm9\sigma$,... where $\sigma$=1.3 m\jbm. The red numbers indicate the position of each spectrum shown in 
Figure~\ref{fig:he_spec}.\label{fig:he}}
\end{figure}

\begin{figure}[!ht]
\includegraphics[scale=0.8]{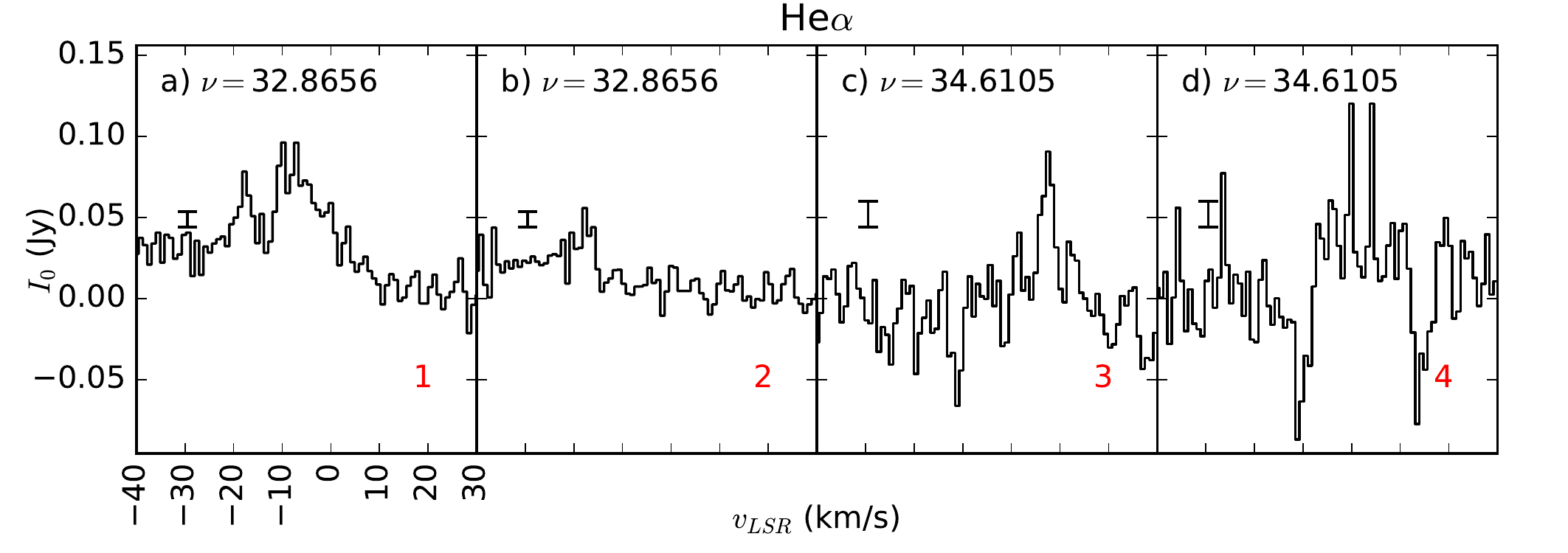}
\caption{Spectra of the detected He recombination transitions. The abscissa is \vlsr\ in \kms and the ordinate is 
intensity in \jbm. The red numbers indicate the point where each spectrum was taken.  The rest frequency for each transition is given in each panel and the 'I' bars denote the 1$\sigma$ 
rms noise. \label{fig:he_spec}}
\end{figure}

A total of three He recombination lines were detected in our survey. Due to its extended nature and weaker emission 
than H, it was only detected with the larger D configuration beam. Figure~\ref{fig:he} shows the emission map of the 
He(58)$\alpha$ transition overlaid on the D configuration continuum,
and Figure~\ref{fig:he_spec} shows the spectra.
The He 
emission is very similar to that of the H$\gamma$ emission, as expected.

\begin{figure}
\includegraphics[scale=.5]{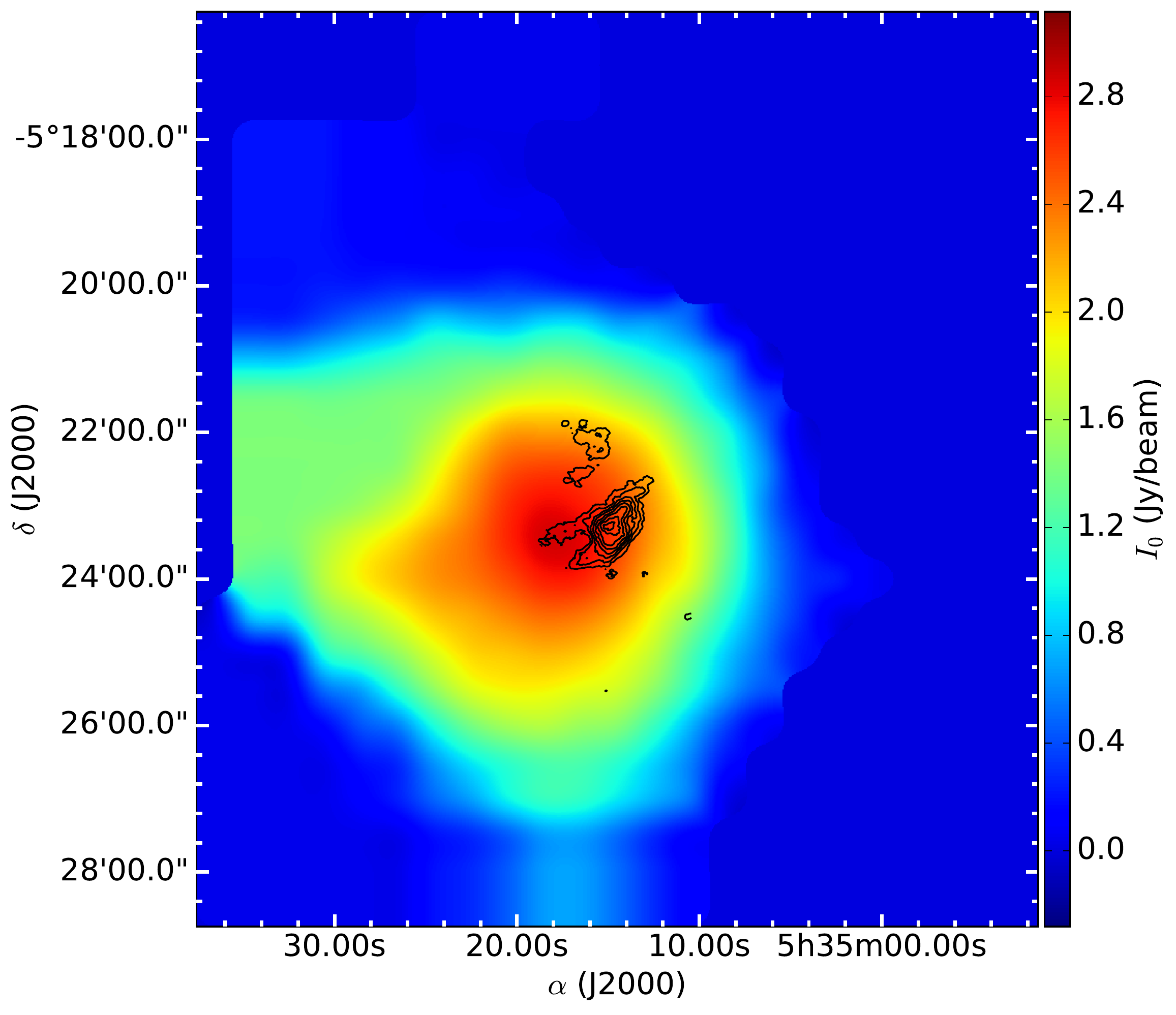}
\caption{Average intensity map of the He(58)$\alpha$ transition from single dish observations (colorscale) and interferometric 
observations (contours). The contours are $\pm3\sigma, \pm6\sigma, \pm9\sigma, ...$ where $\sigma$=1.3 m\jbm.
\label{fig:he-sd-map}}
\end{figure}

\begin{figure}
\includegraphics[scale=0.5]{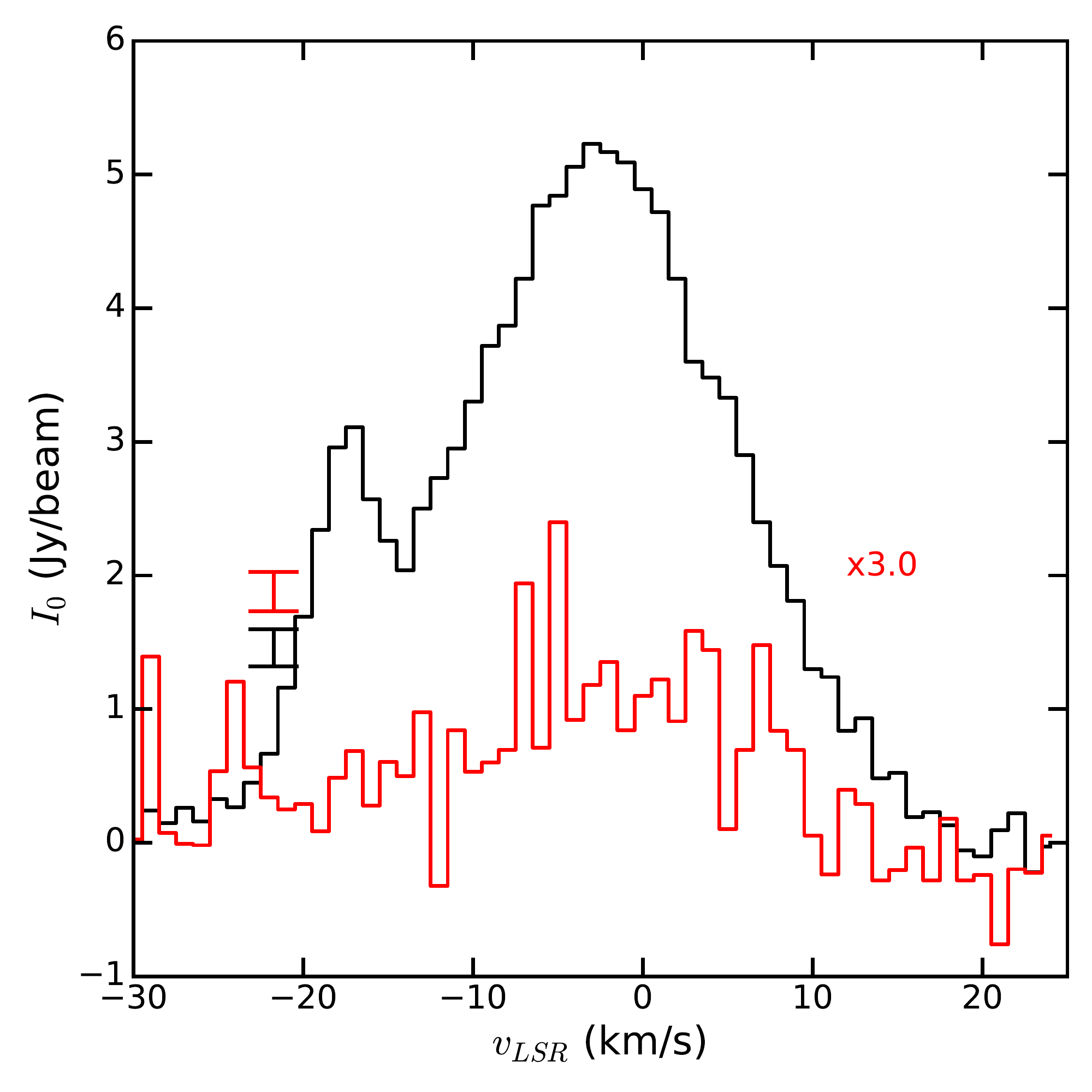}
\caption{Spectrum of the single dish observations of the He(58)$\alpha$ line (black) and interferometric 
observations(red). The interferometric data have been scales by a factor of 3.0. The abscissa is \vlsr\ in \kms and the 
ordinate is intensity in \jbm\ and the 'I' bars denote the 1$\sigma$ rms noise of the respective spectrum.
\label{fig:he-sd-spec}}
\end{figure}

Figure~\ref{fig:he-sd-map} shows the single dish average intensity map of the He(58)$\alpha$ transition 
overlaid by the interferometric average intensity map. 
The single dish map is very similar to the H(58)$\alpha$ map in structure. Figure~\ref{fig:he-sd-spec}
 shows the single dish spectrum (black) and interferometric spectrum (red) of the same transition. The 
interferometric data are scaled by a factor of 3.0 and were best fit by a single Gaussian with a \vlsr\ of -2.3(8) 
\kms\ and FWHM of 16.4(20) \kms. The single dish spectra were best fit with a pair of Gaussians with \vlsr's of 
-17.9(0) and -2.7(0) \kms with FWHM of 4.8(2) and 6.5(2) \kms, respectively. While the -17.9 \kms\ velocity component 
is not detected in the convolved interferometric map, it is detected in the un-convolved spectra of this transition in 
the first two panels of Figure~\ref{fig:he_spec}. It disappears from the convolved spectra due to its overall weak 
emission and strong sidelobes produced by the interferometer's response to the large scale structure. While the 
He(58)$\alpha$ single dish transition shows this -17.9 \kms\ feature, the corresponding H transition does not. H was 
detected in some of the interferometric observations. Its absence in the H spectra may be due to much stronger 
components at other velocities overpowering it.

\clearpage

%--------------------------------------------------------------------------------------------------

\subsubsection{Methyl Acetylene [CH$_3$CCH]}
\begin{figure}[!htb]
\includegraphics[scale=0.7]{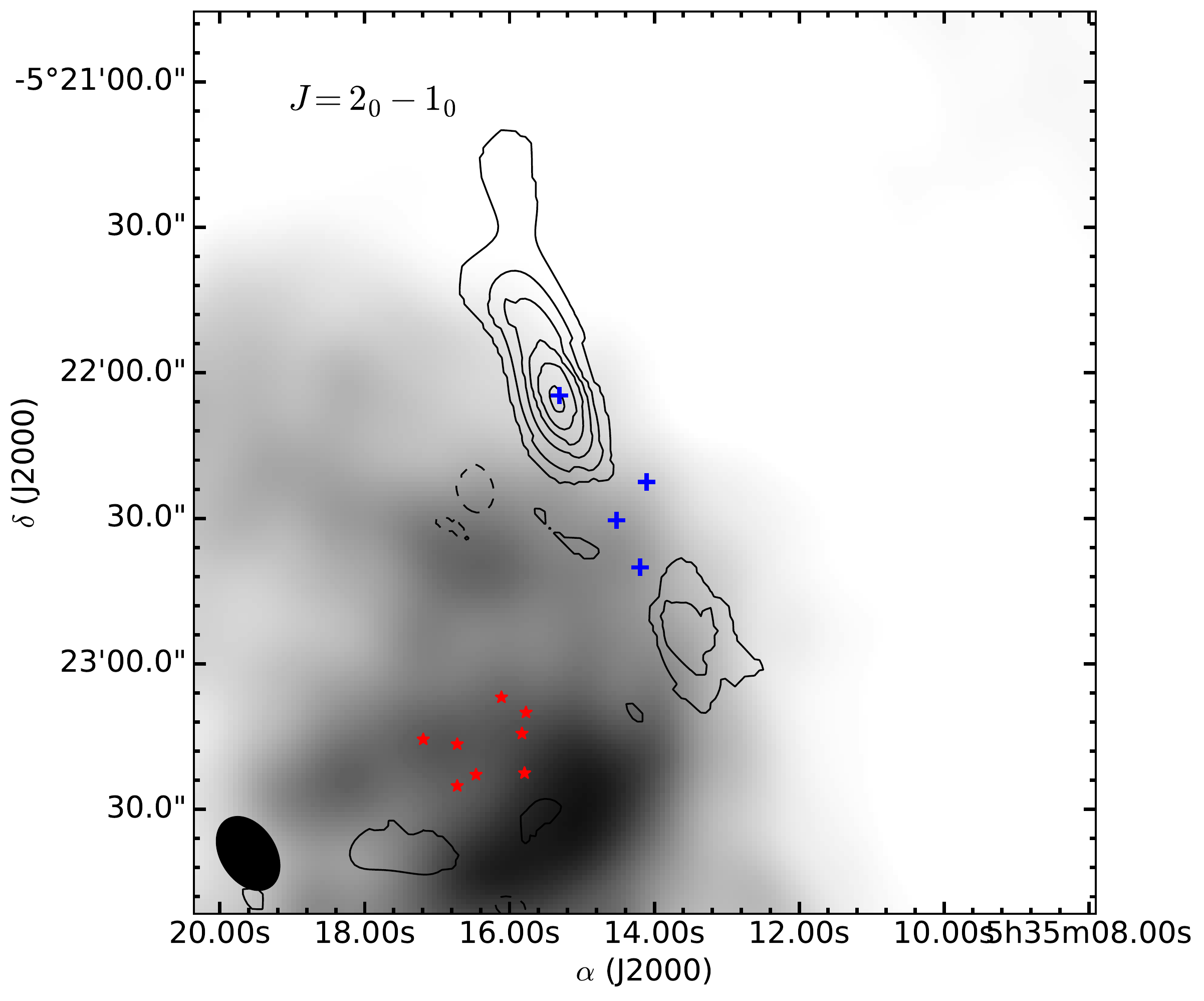}
\caption{Average intensity map of the lowest energy methyl acetylene transition as contours overalyed on the D configuration 
continuum.
The emission is primarily confined to two regions, one northeast of the Hot Core and one to the southwest of 
the Hot Core. The contours are $\pm2\sigma, \pm4\sigma, \pm6\sigma, ...$, where $\sigma$=0.316 \jbm\ and the 
synthesized beam is in the lower left corner. \label{fig:ch3cch}}
\end{figure}

\begin{figure}[!htb]
\includegraphics[scale=0.7]{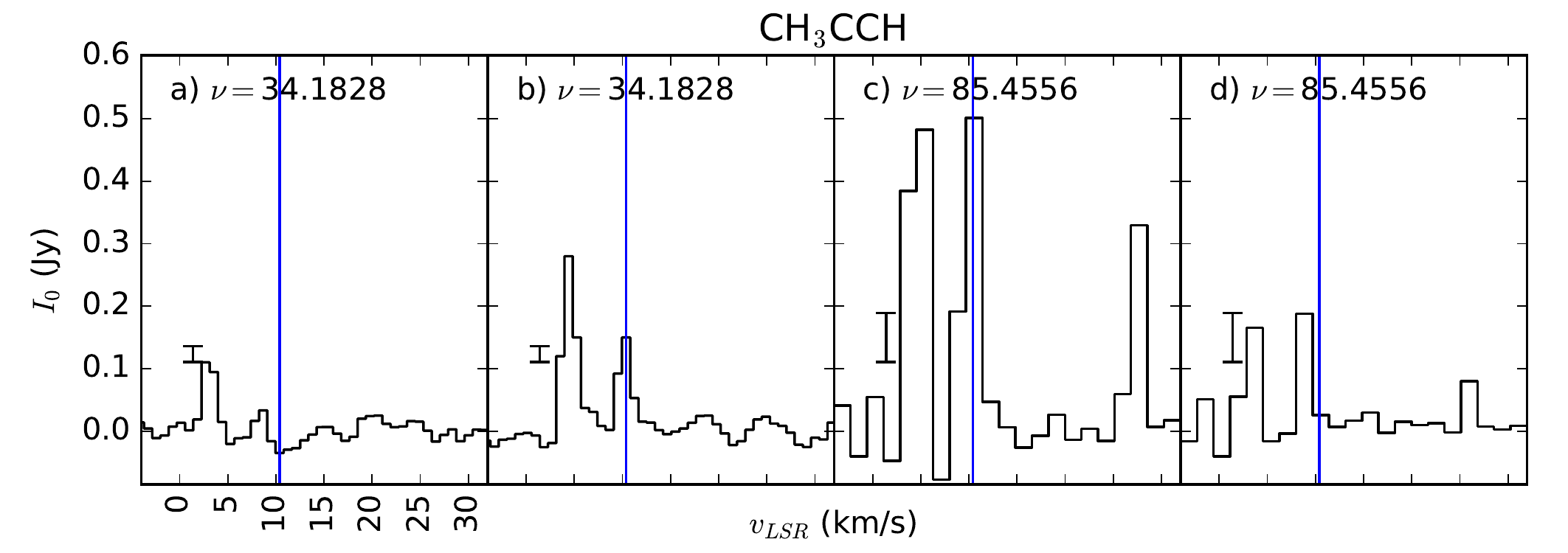}
\caption{Spectra of methyl acetylene taken from the two main peaks. a) shows the spectrum from the 
southwest peak, b) is from the northeast peak, and c) is also from the northeast peak,
but from transitions at 85.4556 GHz. All detected \mae\ lines are displayed in the 
spectra, the velocity scale is referenced to the higher energy transition (34.1828 GHz). The abscissa is \vlsr\ in \kms and the 
ordinate is intensity in \jbm. The rest frequency for each transition is given in each panel and the 'I' bars denote 
the 1$\sigma$ rms noise.\label{fig:ch3cch_spec}}
\end{figure}

All methyl acetylene [\mae] transitions in our frequency range were detected. Figure~\ref{fig:ch3cch} shows the average intensity map 
of the lower energy \mae\ transition as contours overlaid on the D configuration continuum. 
The emission is primarily in two regions, one to 
the southwest of the Compact Ridge (CS-SW), the other centered on CS1 and extending to the northeast. The 
spectrum from each peak is shown in Figure~\ref{fig:ch3cch_spec}. Panel (a) is from the peak to the 
southwest, panel (b) is from the peak near CS1, and panel (c) is from the $J = 5_* - 4_*$ transitions at 85.4556 GHz. All detected \mae\ transitions are shown in each spectra 
and the velocity axis is referenced to the $2_0-1_0$ transition. Each set of transitions was fit together and each is 
best fit by a single, narrow Gaussian component. The southwest peak has a \vlsr\ of 8.8(0) \kms\ and FWHM of 1.1(3) \kms\ 
and the CS1 peak has a \vlsr\ of 10.6(0) \kms\ and FWHM of 1.6(1) \kms.

\begin{figure}[!htb]
\includegraphics[scale=.7]{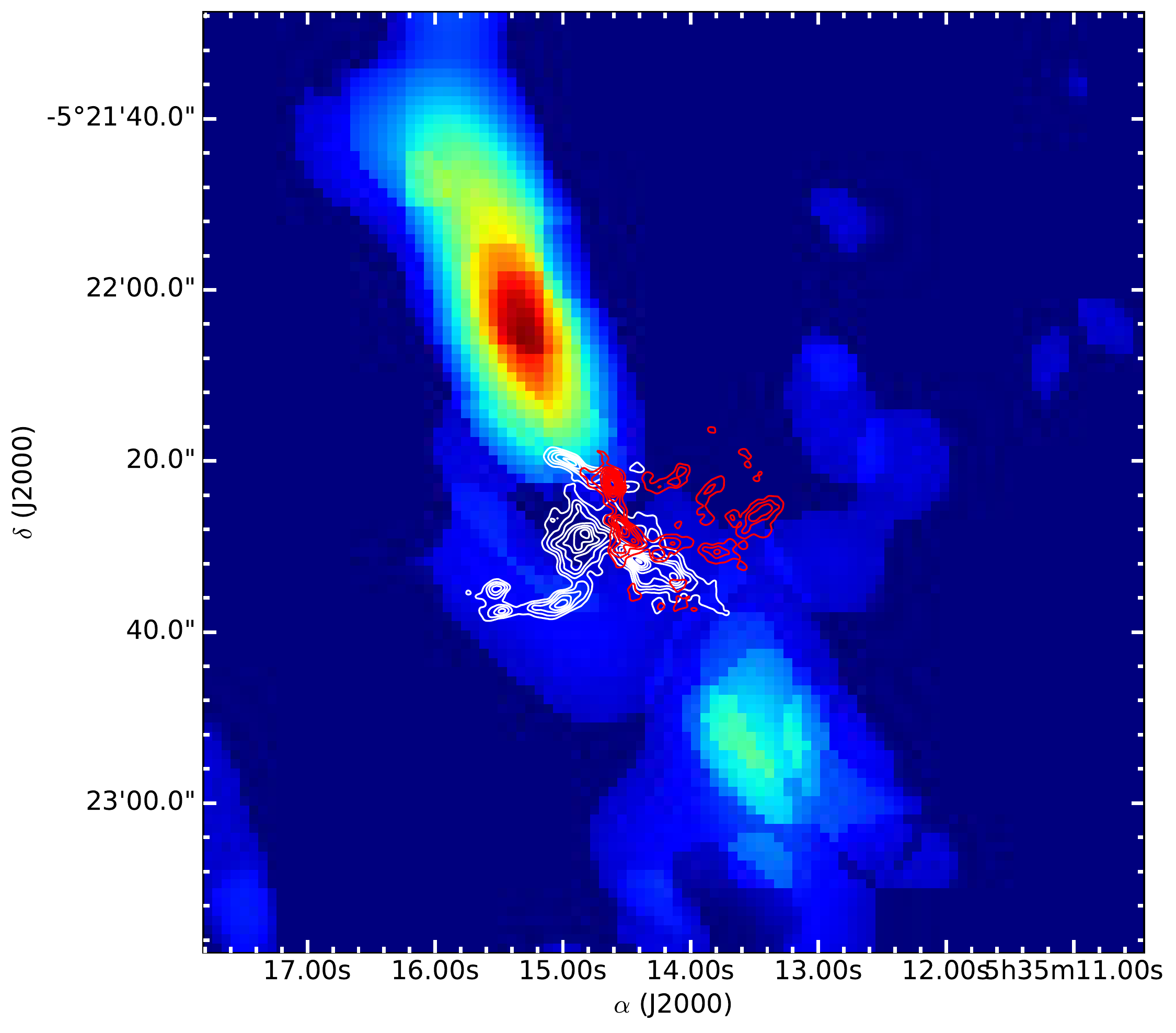}
\caption{Methyl acetylene color scale overlaid by SiO as contours. The white contours are the blueward (-27- -13\kms) 
SiO emission, while the red contours are the redward velocity (20-47\kms) SiO emission. All SiO data are from 
\citet{plambeck09}. \label{fig:ch3cch-sio}}
\end{figure}

\begin{figure}[!htb]
\includegraphics[scale=.7]{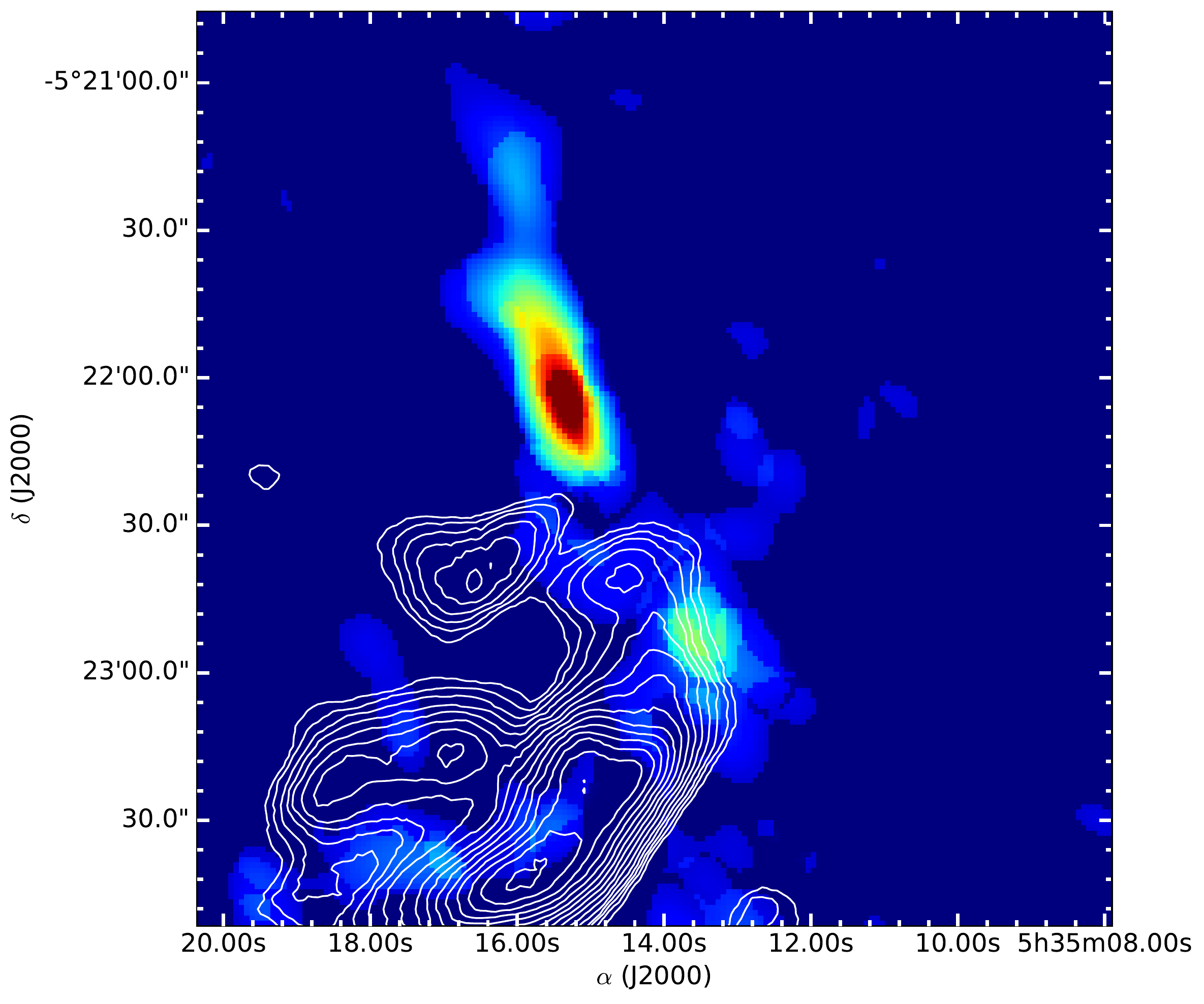}
\caption{Methyl acetate color scale overlaid by H recombination emission from the D configuration. \label{fig:ch3cch-h}}
\end{figure}

In order to understand the physics of the \mae\ emission regions, we compare the \mae\ emission to the emission of 
tracers of different physical regions: SiO and H recombination transitions. 
The SiO emission (a shock and outflow tracer)
from \citet{plambeck09} is 
shown overlaid on a colorscale of \mae\ emission in Figure~\ref{fig:ch3cch-sio}.
The peak to the southwest appears to be in line with the SiO outflow, but it is 
notably separated from the SiO emission. The peak near CS1 appears to be a direct extension of the SiO outflow and may 
be forming/liberated from grains where the outflow is impacting a dustier region. 
Since H recombination is a tracer of ionized
regions, we compared the H recombination line emission  with the \mae\ emission in 
Figure~\ref{fig:ch3cch-h}.
The southeast component boarders along the H emission and a faint ridge of \mae\ emission roughly follows the 
H emission ridge to the southwest, indicating that the ionizing radiation may be destroying it.

Although \mae\ is easily dissociated by UV radiation, it is thought to form in the gas phase and is a dense gas 
tracer \citep{charnley92,bisschop07,kuiper84}. However, the transitions detected in this work have critical densities 
$\lesssim10^4$, which is too low to consider these transitions dense gas tracers. 

To further understand the physical conditions of the \mae\ emission regions we obtained data of the $J=5_* - 4_*$ (85.4556 GHz) transitions of \mae\ toward Orion-KL from the CARMA data archive\footnote{http://carma-server.ncsa.uiuc.edu:8181/}. The data were taken in 2012 March, in C configuration, giving a beam of $3.6\arcsec\times2.5\arcsec$ and channel spacing of 1.7 \kms. The $J=5_K - 4_K$ ($K$ = 0, 1, 2) transitions were detected with upper state energies of 12.3, 19.5, and 41.1 K. The resulting spectrum from the northwestern peak is shown in Figure~\ref{fig:ch3cch_spec}(c) and from the southern peak is shown in Figure~\ref{fig:ch3cch_spec}(d). Fits to these lines are given in Table~\ref{tab:ch3cch}.  

\begin{deluxetable}{lrrrcrrrrrc}
\tablecolumns{11}
\tabletypesize{\scriptsize}
\tablewidth{0pt}
\tablecaption{Molecular Parameters of High frequency \mae\ Transitions\label{tab:ch3cch}}

\tablehead{\colhead{Quantum} & \colhead{Frequency} &
	\colhead{$E_u$} & \colhead{S$\mu^2$} &
	\colhead{$\theta_a\times\theta_b$} & \colhead{$I_0$\tablenotemark{a}} & \colhead{\vlsr} & \colhead{FWHM} & \colhead{} & \colhead{RMS Noise} &
	\colhead{Array} \\
    \colhead{Numbers} & \colhead{(MHz)} & \colhead{(K)}
	&\colhead{($D^2$)}& \colhead{($\arcsec\times\arcsec$)} & \colhead{Jy/bm} & \colhead{km/s} & \colhead{km/s} & \colhead{Position\tablenotemark{a}}&
	\colhead{(mJy/bm)} & \colhead{Config.}}
\startdata
$5_2 - 4_2$ & 85.45073 (6) & 41.1 & 2.4 & $3.6 \times 2.5$ &  0.69 (4) &  10.4 (0) & 2.1 (1) & N & 4.5 & C \\
 & & & & &   0.22 (12) &  9.0 (2) & 1.6 (4) & S & & \\
$5_1 - 4_1$ & 85.45562 (6) & 19.5 & 2.7 & $3.6 \times 2.5$ &  0.55 (2) &  10.4 (0) & 2.1 (1) & N & 4.5 & C \\
 & & & & &   0.20 (8) &  9.0 (2) & 1.6 (4) & S & & \\
$5_0 - 4_0$ & 85.45727 (6) & 12.3 & 2.8 & $3.6 \times 2.5$ &  0.32 (7) &  10.4 (0) & 2.1 (1) & N & 4.5 & C \\
 & & & & &   0.08 (29) &  9.0 (2) & 1.6 (4) & S & & \\
\enddata
\tablenotetext{a}{The uncertainties to the fits are dominated by the fact that the lines are just over a single channel wide.}
\end{deluxetable}

\begin{figure}[!htb]
\includegraphics[scale=0.2]{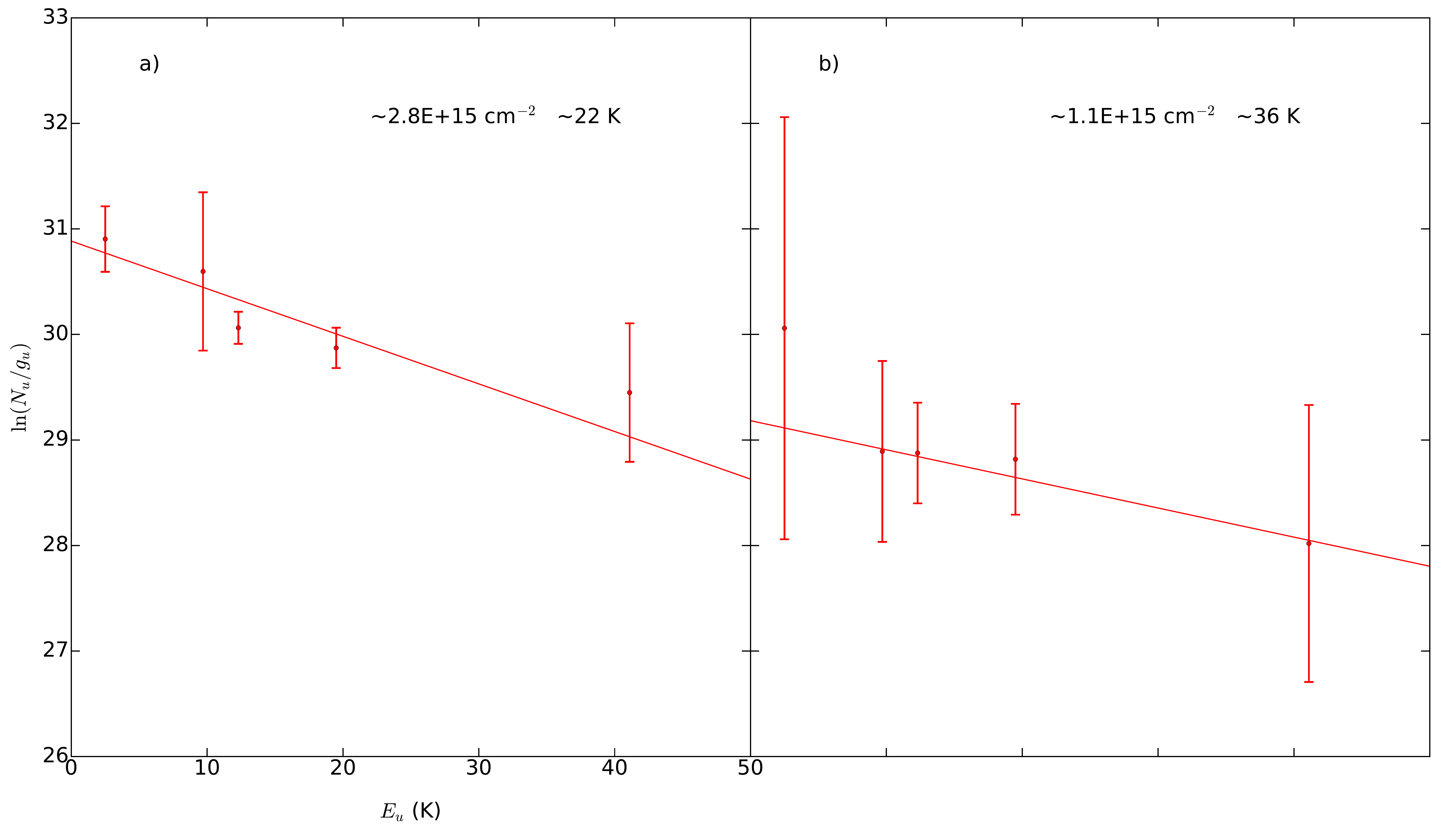}
\caption{Rotation temperature diagrams for the two detected \mae\ peaks. a) is from the northwest peak and b) is from the southeast peak.\label{fig:ch3cch_rot}}
\end{figure}

Using these fits we constructed rotational-temperature diagrams for the CS1 and CR-SW peaks. These fits are shown in Figure~\ref{fig:ch3cch_rot}. Panel (a) is from the northeastern core and yields a total column density of $\sim$2.8\e{15} \cms\ and a rotation temperature of $\sim$22 K. Panel (b) is from the CR-SW peak and yields a total column density of $\sim$1.1\e{15} \cms\ and rotation temperature of $\sim$36 K. Bolstering the results of these calculations are the lack of any high ($>$50 K) temperature line, from any molecular species, in either region. Additionally since no complex molecules, not even \mtoh, in this survey were detected toward these sources, this indicates that not only are these regions cold, but the gas is also dominated by gas-phase reaction products, as any grain surface reaction products are still locked on the grains.

\begin{figure}
\includegraphics[scale=.5]{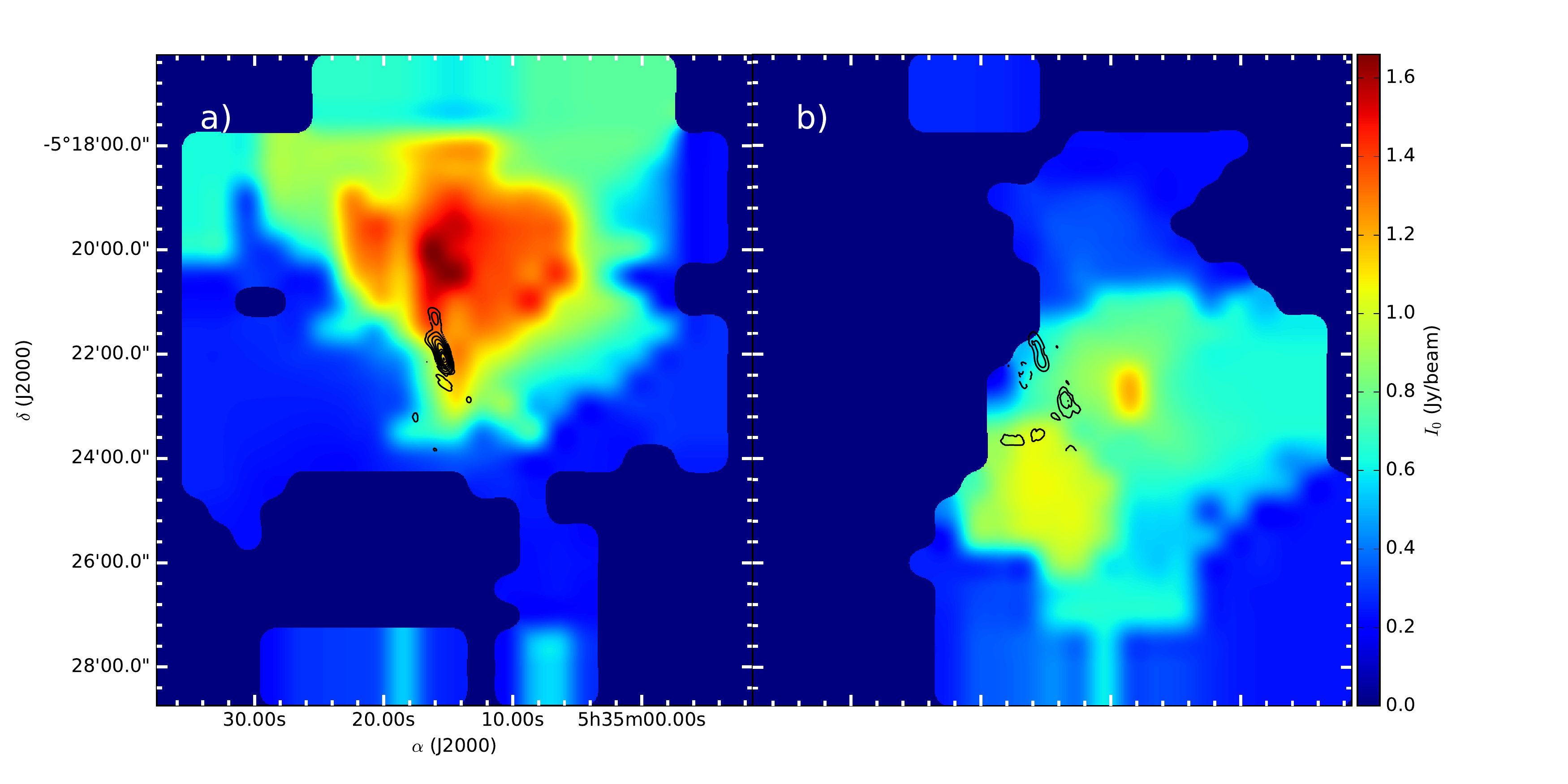}
\caption{Array average intensity map of the $2_1-1_1$ transition of \mae\ in black contours overlaid on the single dish average intensity map of the same 
transition in color scale. The contours are $\pm3\sigma, \pm6\sigma, \pm9\sigma, ...$ where $\sigma$=14.3 m\jbm. a) 
shows the single dish average intensity map from \vlsr=9-11 \kms\ and b) shows the single dish average intensity map of from 
\vlsr=6-8 \kms\, the array average intensity map spans the entire line width in both panels.\label{fig:ch3cch-sd-map}}
\end{figure}

\begin{figure}
\includegraphics[scale=0.4]{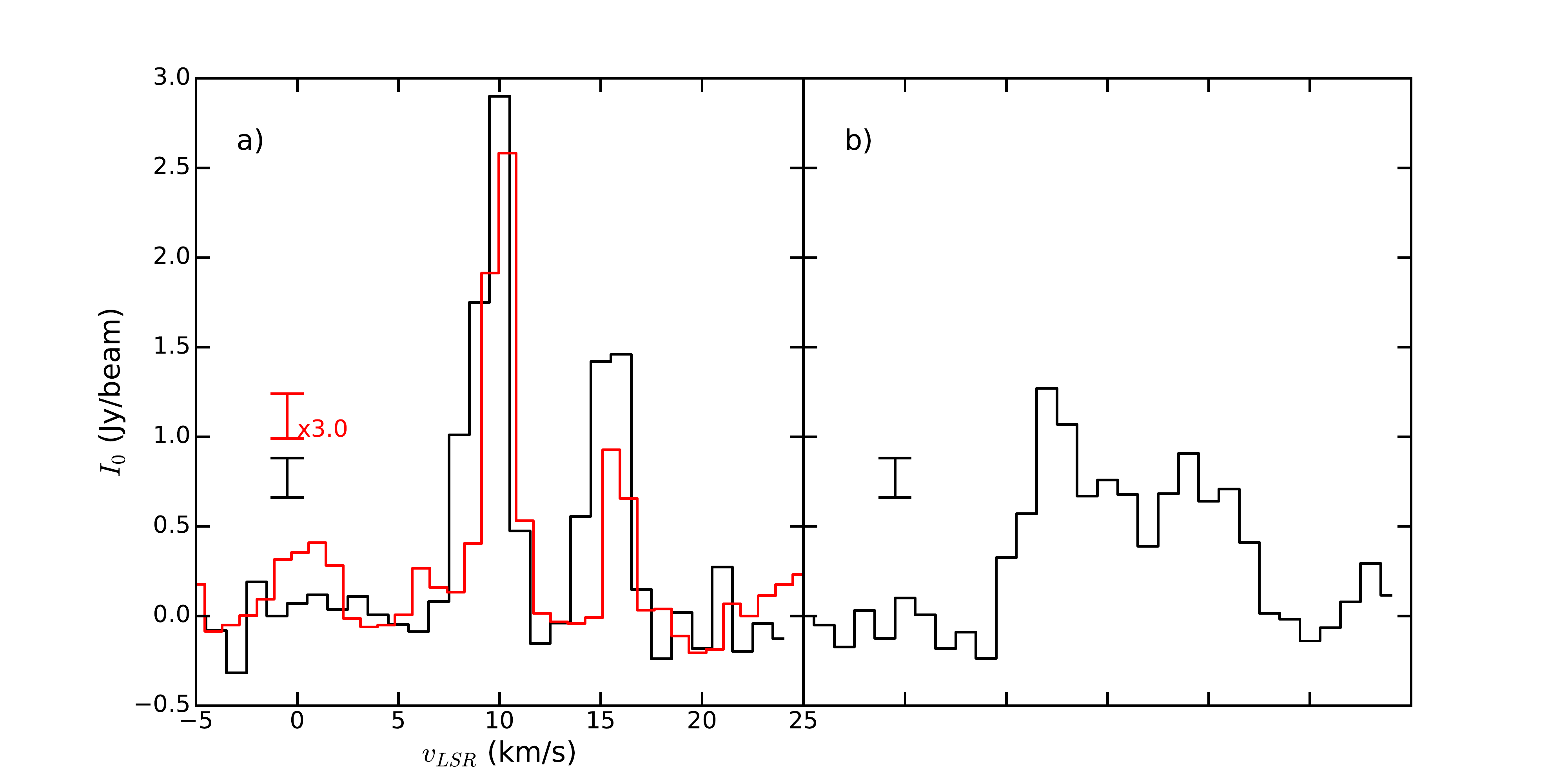}
\caption{Spectra of single dish observations (black) and array observations (red) of the $2_1-1_1$ transition of \mae. 
The array data are scaled by a factor of 3.0. a) shows the northern single dish peak is shown in the and b) shows the 
southern single dish peak. The abscissa is \vlsr\ in \kms and the ordinate is intensity 
in \jbm\ and and the 'I' bars denote the 1$\sigma$ rms noise of the respective spectrum.\label{fig:ch3cch-sd-spec}}
\end{figure}

Since it forms so easily in the cool gas phase and the interferometric maps show potential extended structure, we 
observed \mae\ in single dish mode too. Figure~\ref{fig:ch3cch-sd-map} shows the average intensity maps of the $2_1-1_1$ transition of 
\mae\ for both array and single dish. 
From these maps we see that there are at least three distinct large 
scale regions of emission of \mae. The \vlsr\ of both the northern interferometric and single dish peaks share a common 
velocity, as do the southern peaks. The large scale peaks may be extension of the outflows.
The spectra from both are in Figure~\ref{fig:ch3cch-sd-spec}.
Panel (a), from the northern peak of the single dish map, was best fit with a 
single Gaussian with a \vlsr\ of 9.6(0) \kms\ and FWHM of 2.1(0) \kms. The array data were also best fit with a single 
Gaussian with a \vlsr\ of 10.1(1) \kms\ and FWHM of 1.5(2) \kms. Panel (b), from the southern peak, was best 
fit with a pair of Gaussians with \vlsr's of 7.6(1) and 10.4(2) \kms\ and FWHM of 2.9(3) and 1.7(5) \kms. The 
interferometric data have been scaled by a factor of 3.0, indicating that there is significant emission that is  
resolved out by the array, which is also supported by the maps. The emission from the northern peak has a rather narrow 
line profile while the emission from the southern peak has a notably wider line profile.
\clearpage

%--------------------------------------------------------------------------------------------------

\subsubsection{Formaldehyde [\fmal]}
\begin{figure}[!ht]
\includegraphics[scale=0.45]{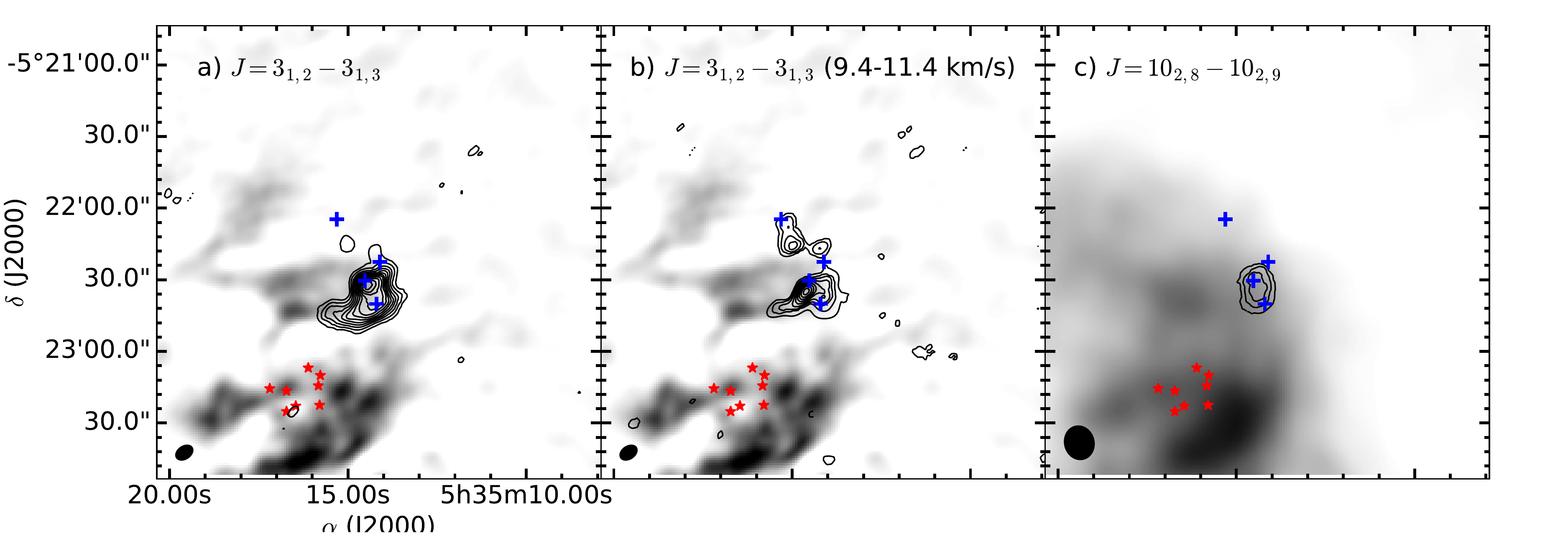}
\caption{Average intensity maps of the two detected formaldehyde transitions. a) is the lower energy transition and 
shows an extension of emission to the southeast from the Hot Core region, b) is the same transition as a) but only the 9.4-11.4 \kms\ velocity components showing the emission from near CS1, and  c) is from the higher energy transition 
which is confined to the Hot Core region. The contours are $\pm3\sigma, 6\sigma, 9\sigma, ...$, where $\sigma$=2.2 and 
3.3 m\jbm\ respectively. \label{fig:fmal}}
\end{figure}

\begin{figure}[!ht]
\includegraphics[scale=0.8]{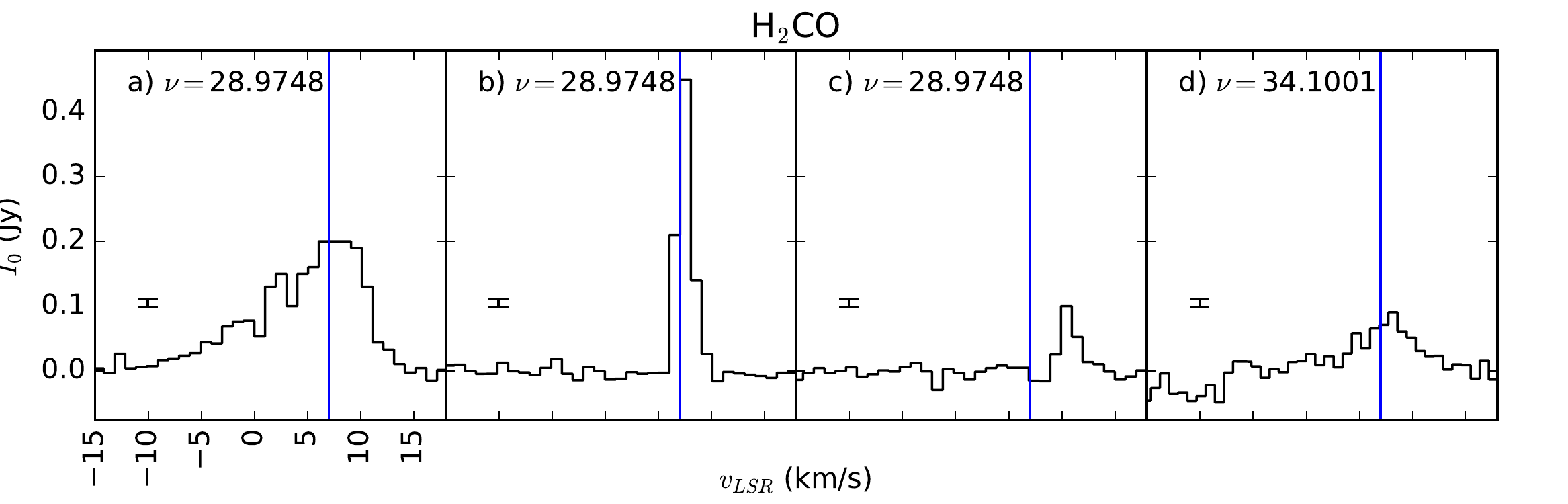}
\caption{Spectra from the \fmal\ data. a) is from the intensity peak of the lowest energy transition, b)
 is from the southeast wing of the lowest energy transition, c) is from the CS1 region of the lowest energy transition, and d) is from the intensity 
peak of the highest energy transition. The abscissa is \vlsr\ in \kms and the ordinate is intensity in \jbm. The rest 
frequency for each transition is given in each panel and the 'I' bars denote the 1$\sigma$ rms noise.
\label{fig:fmal_spec}}
\end{figure}

A total of two \fmal\ transitions were detected in this survey (Figure~\ref{fig:fmal}).
The 
lower energy transition (panel (a)) peaks near the Hot Core and IRc6 with an extension to the southeast, just as \mtoh\ does.
Panel (b) show the emission from the same transition, but over a very narrow velocity range (9.4-11.4 \kms) which highlights the weak emission coming from HC-NE. The 
higher energy transition (panel (c)) has a single peak near the Hot Core. Figure~\ref{fig:fmal_spec} shows that spectra from the 
transitions. Panel (a) is from the peak of the lower energy transition. The spectrum was 
best fit by a pair of Gaussians, with \vlsr\ of 4.1(7) and 8.5(1) \kms\ and FWHM of 11.1(9) and 4.2(6) \kms, respectively. There may 
be more components, but without more lines to use for the analysis the uncertainties grow well beyond the values of the 
fit. Panel (b) is a spectrum taken through the southeast wing of the low energy transition. It was best fit by a 
single narrow Gaussian with a \vlsr\ of 7.5(0) \kms\ and a FWHM of 1.7(0) \kms. Panel (c) shows the narrow line from the HC-NE region. It is best fit by a single Gaussian with a \vlsr\ of 10.7(1) \kms and FWHM of 1.6(3) \kms. Panel (d) shows the spectrum 
from the intensity peak of the higher energy transition and was best fit by a single Gaussian with a \vlsr\ of 7.8(14) 
\kms\ and FWHM of 6.8(38) \kms; however the spectral profile does indicate the presence of other components (similar to 
panel (a)), but below our detection threshold. Both the low and high energy intensity peaks are within 
3\arcsec\ (a fraction of the beam) of each other. There were not enough transitions to form a rotation temperature 
diagram for this molecule, so we assume a rotation temperature of 138 K from the similarly distributed \mtoh. 
This yields beam averaged column densities of 4.8(9)\e{17} and 1.1(0)\e{18} \icm\ for the two component low energy 
spectrum, 6.4(4)\e{17} \icm\ for the southeast wing, and 1.3(10)\e{17} \icm\ for the high energy transition. Using a rotation temperature of 30 K we find a beam averaged total column density of 3.2(0)\e{16} \icm\ toward HC-NE. The 
sources were assumed to fill the beam (giving a beam filling factor of 0.5). Based on these values there are no missing 
\fmal\ transitions.

\begin{figure}
\includegraphics[scale=0.5]{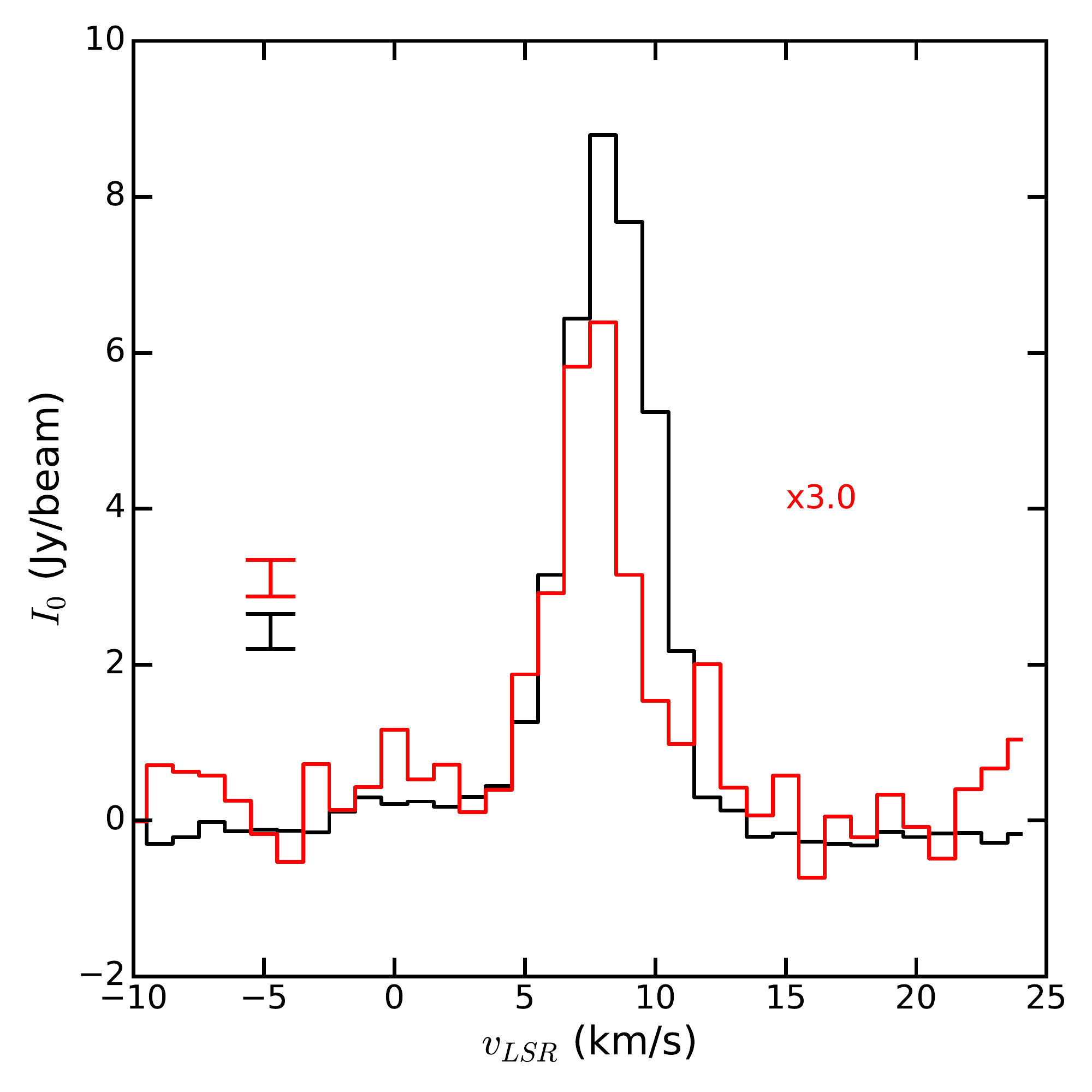}
\caption{Spectra from single dish observations (black) and array observations (red) of the $3_{1,2}-3_{1,3}$ transition 
of \fmal. The array data have been scaled by a factor of 3.0. The abscissa is \vlsr\ in \kms and the ordinate is 
intensity in \jbm\ and the 'I' bars denote the 1$\sigma$ rms noise of the associated spectrum. The blue line indicates a \vlsr\ of 7 \kms.\label{fig:fmal-sd-spec}}
\end{figure}

\begin{figure}
\includegraphics[scale=.5]{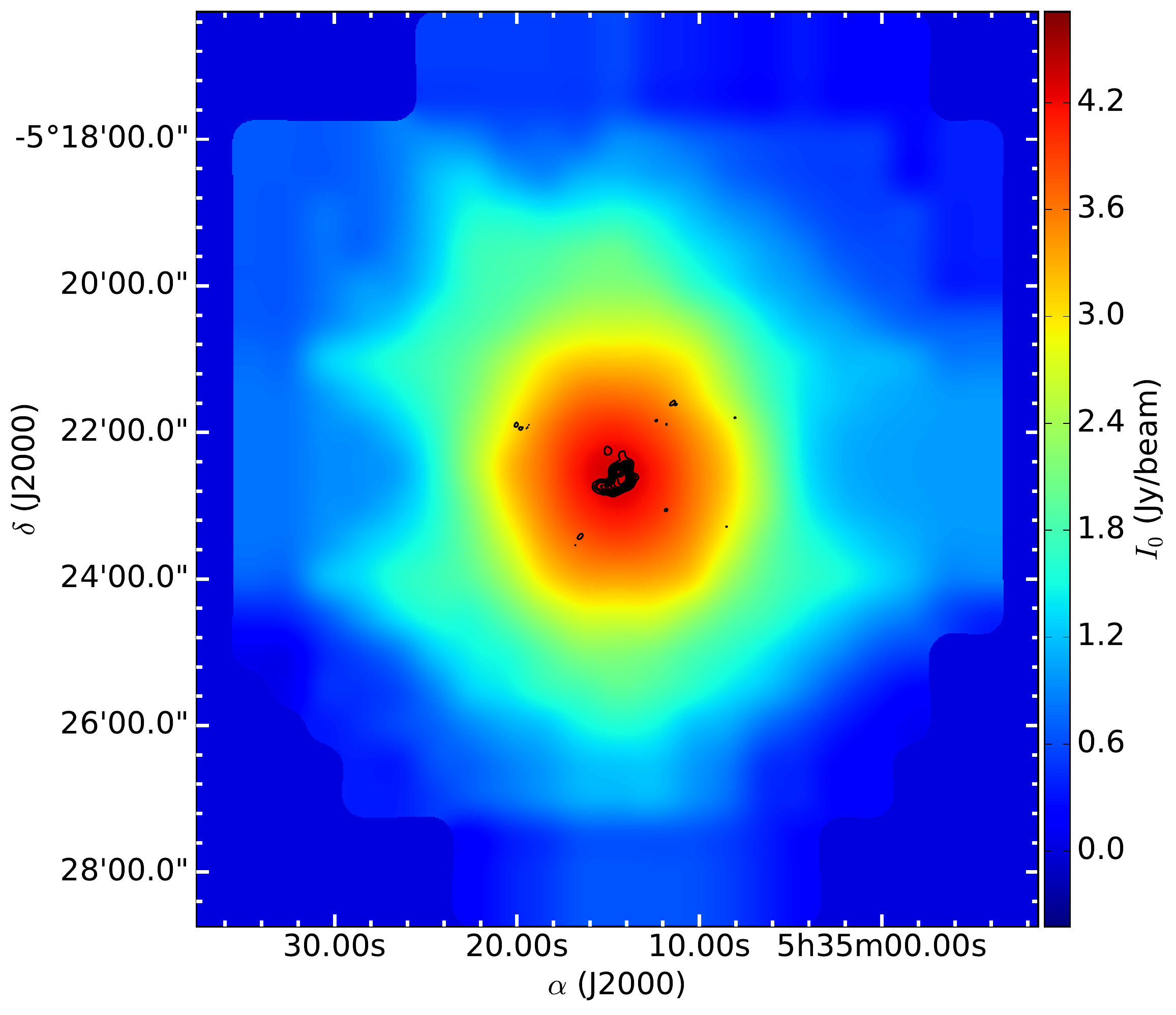}
\caption{Array average intensity map of the $3_{1,2}-3_{1,3}$ transition of \fmal\ (contours) overlaid on the color scale map from 
single dish observations of the same transition. Contours are $\pm3\sigma, \pm6\sigma, \pm9\sigma, ...$ where 
$\sigma$=11.2 m\jbm. \label{fig:fmal-sd-map}}
\end{figure}

Like \mtoh, \fmal\ shows an extension to the southeast and was observed in single dish mode. Figure~\ref{fig:fmal-sd-spec}
shows the single dish and interferometric spectra from the $3_{1,2}-3_{1,3}$ transition of \fmal. 
The spectra indicate that there is a significant extended 
component, possibly from the HC-NE/CS1 interface region which has a \vlsr\ of $\sim$10.6 \kms. that is resolved out by the interferometer. There is a slight velocity offset between the peaks ($\sim1$ 
\kms), indicating that the extended structure is at a slightly higher velocity.  Figure~\ref{fig:fmal-sd-map} shows the 
interferometric average intensity map overlaid on the single dish average intensity map of this transition. The map 
indicates that any extended emission comes from the the same region as the compact emission and has little extension as 
it is confined to a single beam.
\clearpage

%--------------------------------------------------------------------------------------------------

\subsubsection{Thioformaldehyde [H$_2$CS]}
\begin{figure}[!htb]
\includegraphics[scale=.5]{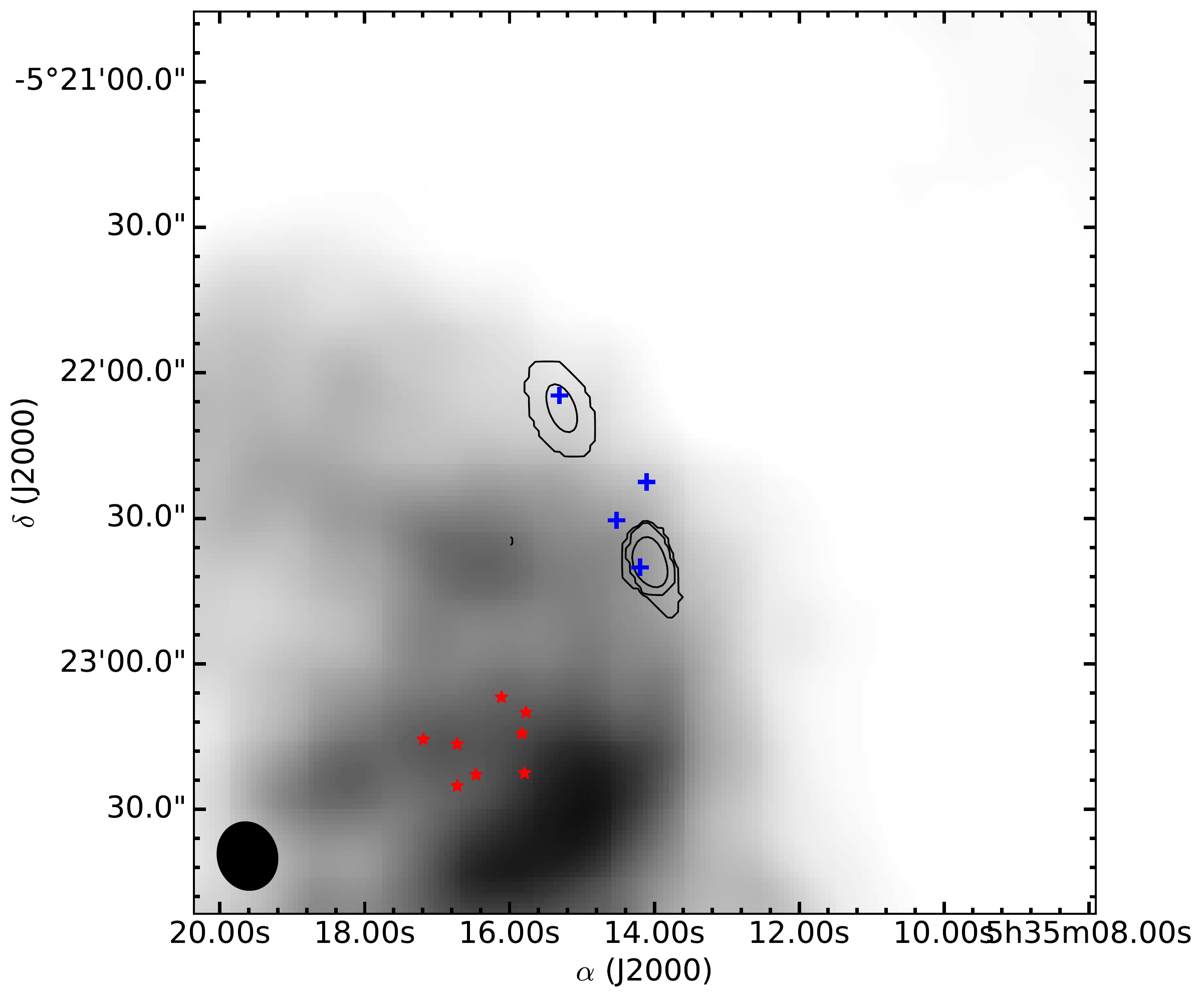}
\caption{Contour map of the H$_2$CS transition overlaid on the grayscale continuum. Contours are $\pm3\sigma$, 
$\pm6\sigma$, $\pm9\sigma$, ... where $\sigma$=3.7 m\jbm. The synthesized beam is given in the lower left corner of the 
map.\label{fig:h2cs-map}}
\end{figure}

\begin{figure}[!htb]
\includegraphics[scale=.5]{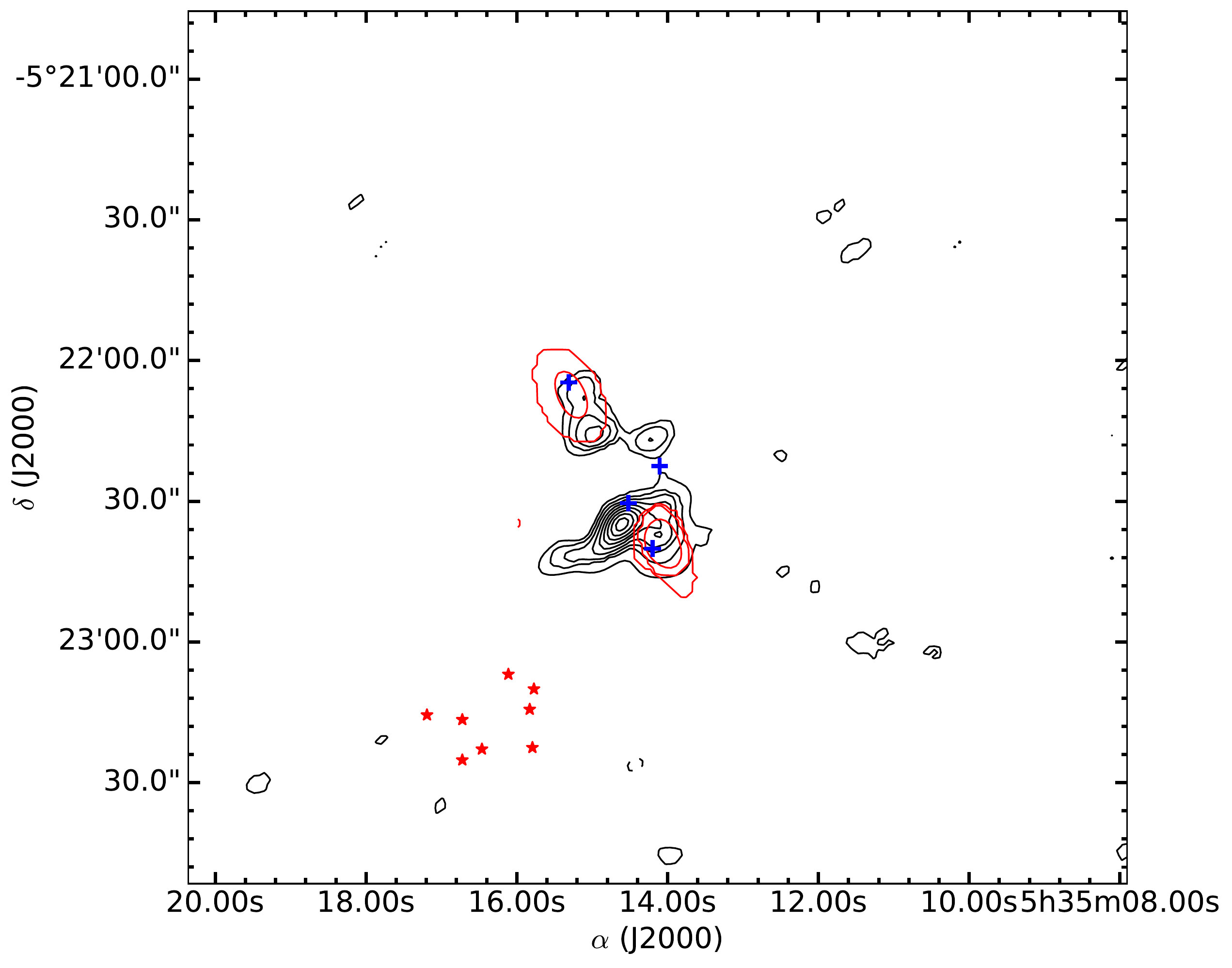}
\caption{Comparison map of \fmal\ (black contours) and H$_2$CS (red contours). Contours are $\pm3\sigma$, $\pm6\sigma$, 
$\pm9\sigma$, ... where $\sigma$= 2.2 and 3.7 m\jbm, respectively.\label{fig:h2cs-fmal}}
\end{figure}

The fundamental transition was the only \tfmal\ transition detected in our survey. Figure~\ref{fig:h2cs-map} shows the 
average intensity map of the transition. 
It has two notable components, the strongest near CS1 and a slightly weaker one near the HC/CR. This is in 
contrast to its cousin, \fmal, which has the vast bulk of its emission coming from the Hot Core and Compact Ridge 
regions and only a trace from CS1, but with similarly narrow lines and \vlsr. Figure~\ref{fig:h2cs-fmal} shows a comparison of the 
two tracers. 

While they are not equivalent transitions (e.g. 
the \fmal\ transition is not its fundamental transition) they are both low upper state energy transitions. The weak \fmal\ emission from the overlapping regions and lack of \tfmal\ emission from the HC
could be explained by a stark difference between the temperatures of the Hot Core and the overlap regions.  The 
differences in emission could also distinctly different formation mechanisms for these structurally similar molecules. 
Formaldehyde is thought to form in high abundance in icy grain mantles \citep[e.g.,][]{tielens87,cuppen09,madzunkov09} 
but may also have a notable gas formation pathway \citep{leurini10}. It is likely released from the grain surfaces by 
the shock and higher temperatures\footnote{\fmal\ evaporates at $\sim$40 K from grain surfaces \citep{Garrod08}.} of 
the SiO outflow as it interacts with the dust. Its emission is well correlated with the 3mm continuum of the region 
(see Figure~\ref{fig:ccomp}). Contrasting this, thioformaldehyde is thought to have a strong gas phase formation path 
\citep{ghosh79}. This is supported by the map as the emission is from the more extended, cooler gas. Additionally, the 
\tfmal\ may be destroyed in the denser, hotter region around the Hot Core. Thus if the regions of \tfmal\ emission are cold $\le40$ K (as the \mae\ data indicate) then the \fmal\ emission would be suppressed as most would still be in the solid state on grains. Then as the
outflow interacts with these regions the \fmal\ will be liberated from the grain surfaces into the gas phase while the \tfmal\ is destroyed.

\begin{figure}[!htb]
\includegraphics[scale=1.0]{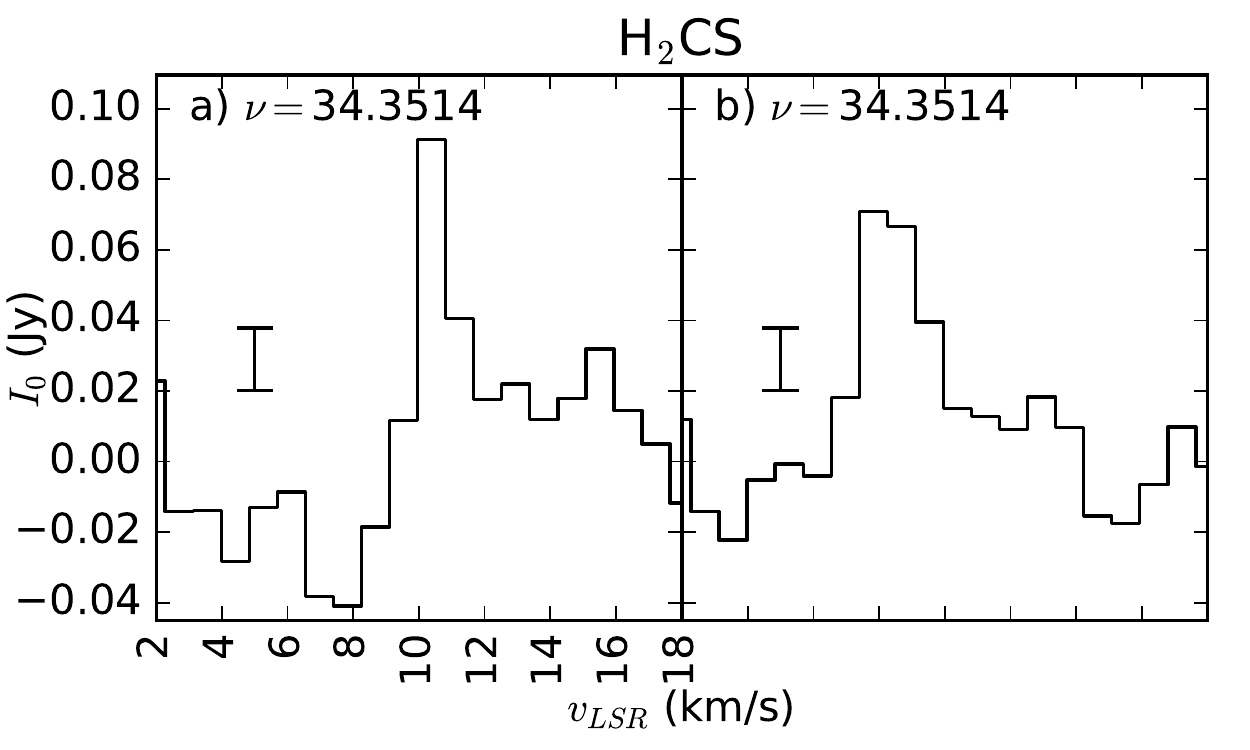}
\caption{Spectra of the two H$_2$CS peaks. a) is from the northern peak and b) is from 
the southern peak. The abscissa is \vlsr\ in \kms and the ordinate is intensity in \jbm\ and the 'I' bars denote the 
1$\sigma$ rms noise.\label{fig:h2cs_spec}}
\end{figure}

Figure~\ref{fig:h2cs_spec} shows the spectra 
from the northern (panel (a)) and southern (panel (b)) peaks, respectively. Each was best fit by a single Gaussian 
with \vlsr's of 8.4(2) and 10.6(4) \kms\ and FWHM of 2.3(6) and 1.4(10) \kms, respectively. Using a temperature 
of 30 K we find a beam averaged column density of 2.1\e{15} \cms\ for the CR-SW peak and 1.5\e{15} \cms\ for the CS1 peak. Given the column densities, there are no missing \tfmal\ transitions.

\clearpage

%--------------------------------------------------------------------------------------------------

\subsubsection{Sulphur Monoxide [SO and $^{34}$SO]}
\begin{figure}[!ht]
\includegraphics[scale=0.45]{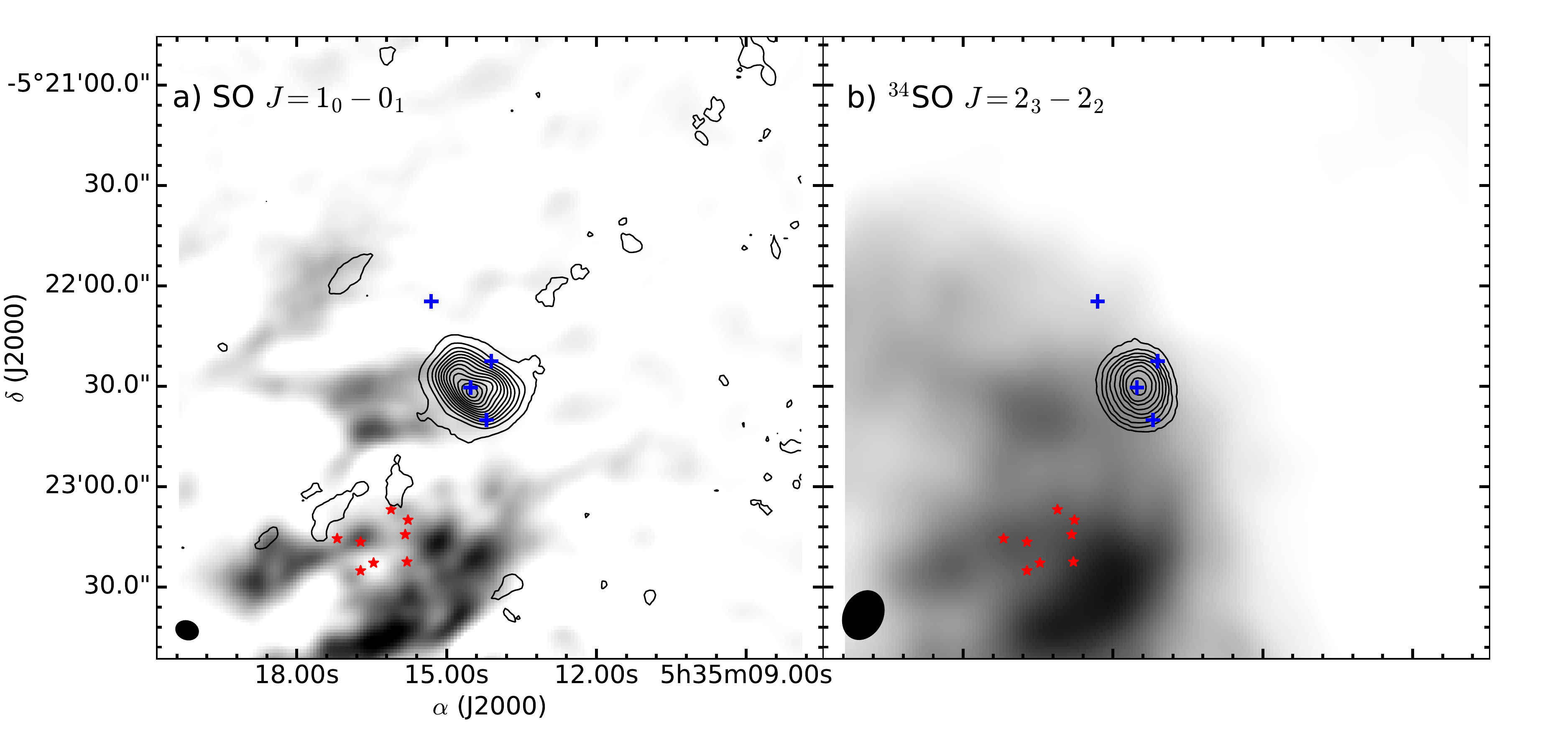}
\caption{Maps of SO and $^{34}$SO. a) shows the average intensity map of the detected SO transition. Contours are 
$\pm3\sigma, \pm21\sigma, \pm39\sigma, ...$, where $\sigma=1.0$ m\jbm. b) shows the average intensity map of 
$2_3-2_2$ transition of $^{34}$SO. The contours are $\pm3\sigma, \pm6\sigma, \pm9\sigma, ...$, where $\sigma=2.6$ 
m\jbm.\label{fig:so}}
\end{figure}

\begin{figure}[!ht]
\includegraphics[scale=.8]{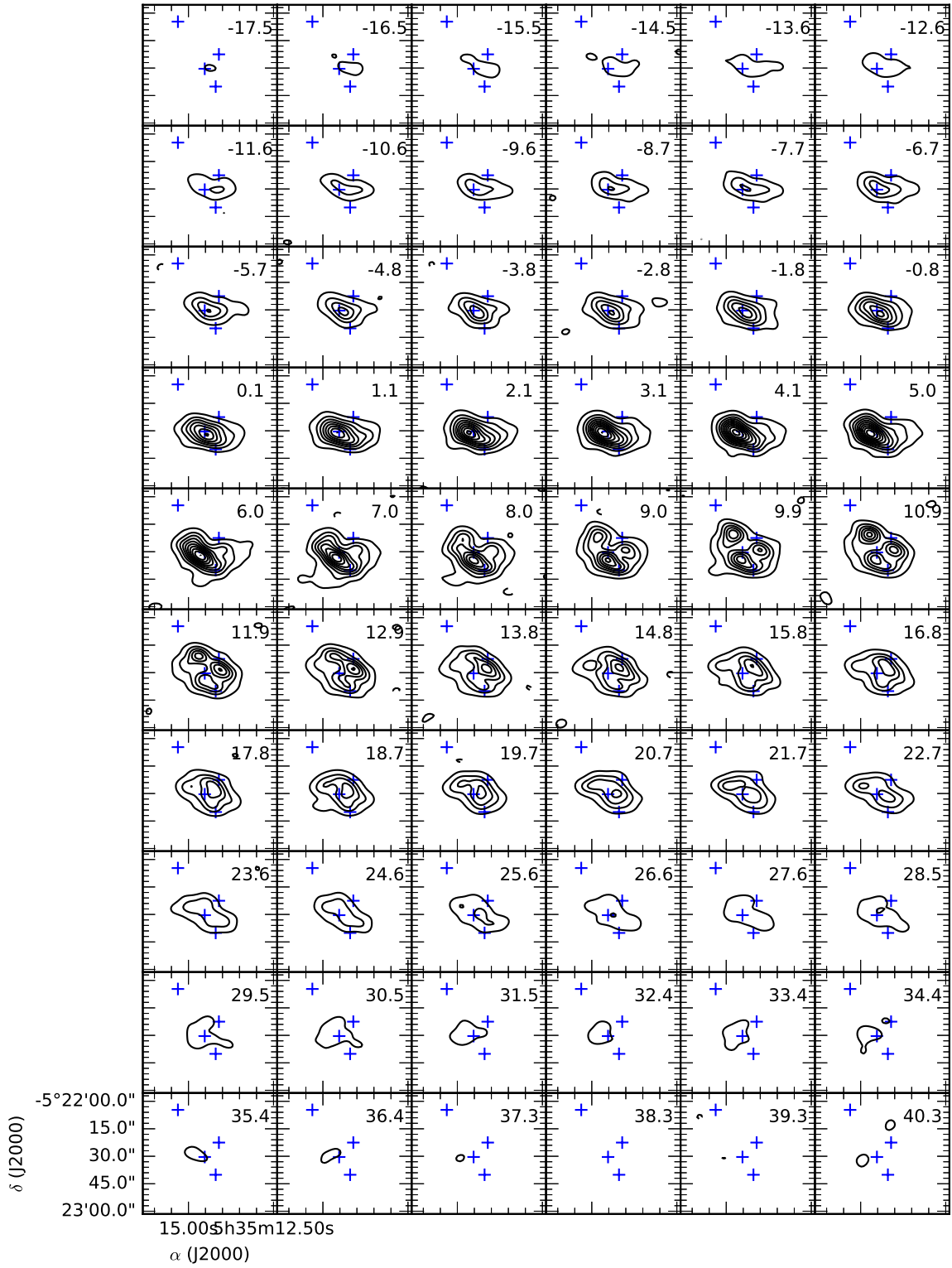}
\caption{Individual channels maps of SO. Contour levels are $\pm3\sigma, \pm11\sigma, \pm19\sigma ,...$ where 
$\sigma$=8.6 m\jbm. The velocity of each channel is given in the top right corner of each panel. \label{fig:so_chan}}
\end{figure}

All sulphur monoxide, SO and $^{34}$SO, transitions in our frequency range were detected. Figure~\ref{fig:so} shows the 
average intensity maps of the only SO transition (panel (a), C configuration) and the higher energy $^{34}$SO transition (panel (b), D configuration). 
From the average intensity maps it appears that SO has a rather extended distribution centered 
near the Hot Core. However, the channels maps in Figure~\ref{fig:so_chan} indicate a much more complex structure. The 
channel maps are shown from -17.5 to 40.3 \kms, with the \vlsr\ in the upper right corner of each panel. While the 
emission is constrained to a rather compact region, there are three distinct peaks, visible between 9.0 and 14.8 \kms, 
which surround the primary peak. Previous work by \citet{plambeck82}, \citet{wright96b}, and \citet{esp13}, of higher energy 
transitions near 86 and 219 GHz, indicated a clumpy SO structure. Our observations do indicate this clumpiness but also with an encasing shell like structure.

\begin{figure}[!ht]
\includegraphics[scale=.5]{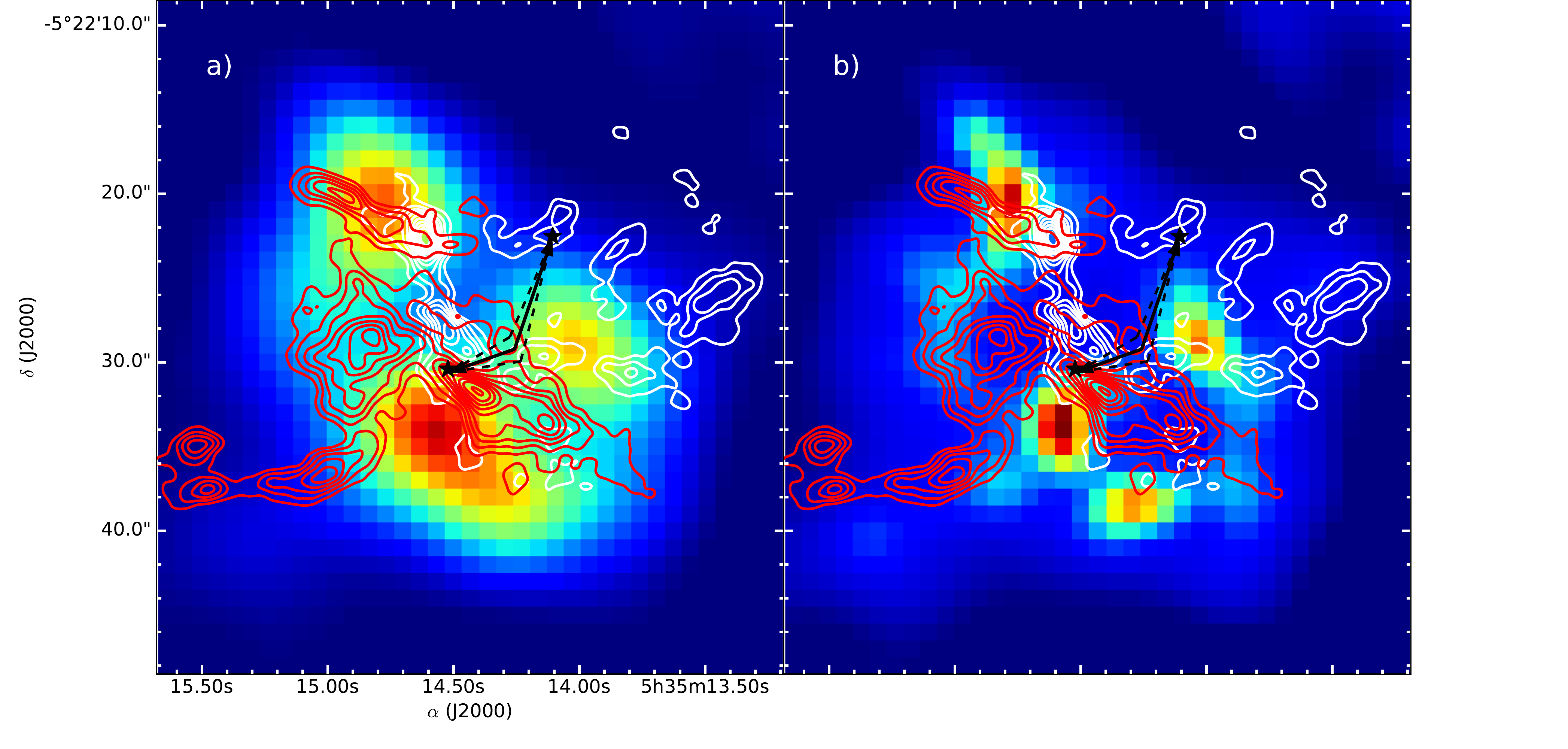}
\caption{Map of $\sim$10 \kms\ SO emission as a colorscale overlaid by SiO emission from \citet{plambeck09}. Both panels 
show the same data, but at different resolutions. a) was restored with a $6.8\arcsec\times5.5\arcsec$ 
beam while b) was restored with a $2.0\arcsec\times2.0\arcsec$ beam. 
The black stars indicate the current positions of Source I and BN, the
arrows indicate the most likely path of Source I and BN after their interaction $\sim$560 years ago,
and the dashed lines are the
uncertainties from proper motion, taken from \citet{goddi11}.\label{fig:so-sio}}
\end{figure}

\begin{figure}
\includegraphics[scale=0.8]{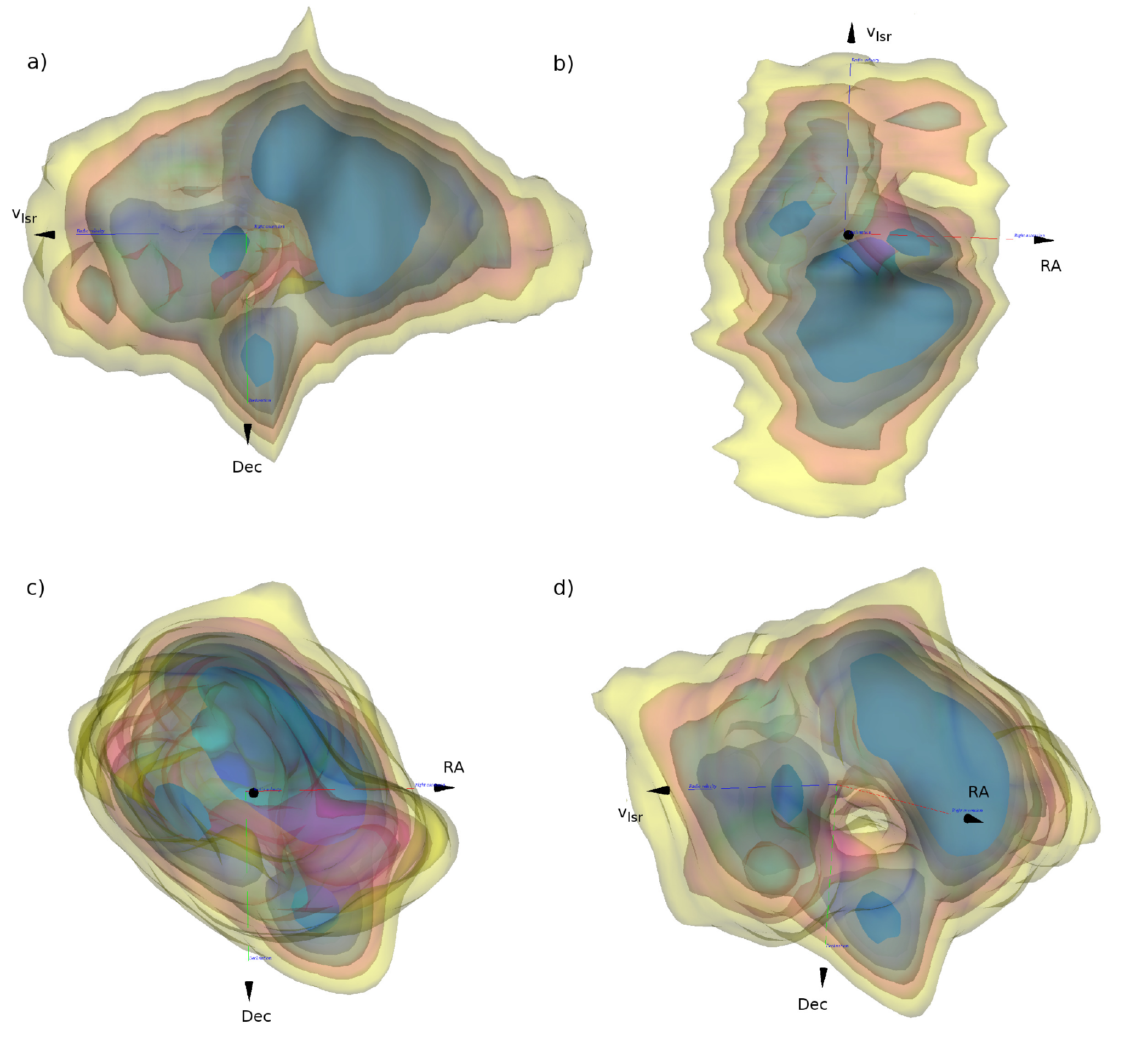}
\caption{The pseudo-3D SO structure of Orion-KL, using \vlsr\ as a proxy for distance. a) shows a view down the RA axis, b)
down the Dec axis, and c) down the velocity axis. d) shows the hole in 
SO emission which may have been created by BN escaping the region. \label{fig:so-mosaic}}
\end{figure}

Figure~\ref{fig:so-sio} shows the $\sim$10 \kms\ SO emission, from the center of the shell, as a colorscale overlaid by 
SiO emission from \citet{plambeck09}. The SiO emission indicates the extent of the outflow and shocked regions.
The SO emission is well correlated with the SiO 
emission edges and known structures. This indicates that the SO is being formed, or liberated from the grains, in the 
energetic regions as a shock moves interacts with the gas and dust. The northern most peak is coincident with the lower edge 
of \mae\ emission in the HC-NW. The western peak is coincident with IRc6, a region associated with emission of complex organic 
molecules \citep[e.g.,][]{friedel08,slww12,friedel12} and the third peak is near the Hot Core. Panel (b) is the same data as in panel (a), but restored with a $2\arcsec \times 2\arcsec$ beam and shows a nearly complete ring like 
structure, which is obscured in the lower resolution map, that is expected from the center  slice of a shell. The only 
notable gap is to the northwest, near BN. It is possible that after the interaction, BN created the hole as it left the 
region. The hole can be seen more clearly in Figure~\ref{fig:so-mosaic}(d). Figure~\ref{fig:so-mosaic} shows the pseudo-3D structure of SO, which uses \vlsr\ as a proxy for distance. The three SO clumps can be seen in all panels. Panel (a) looks down the RA axis, (b) looks down the DEC axis, (c) looks down the \vlsr\ axis, and (d) shows the hole in the SO structure

\begin{figure}[!ht]
\includegraphics[scale=1.0]{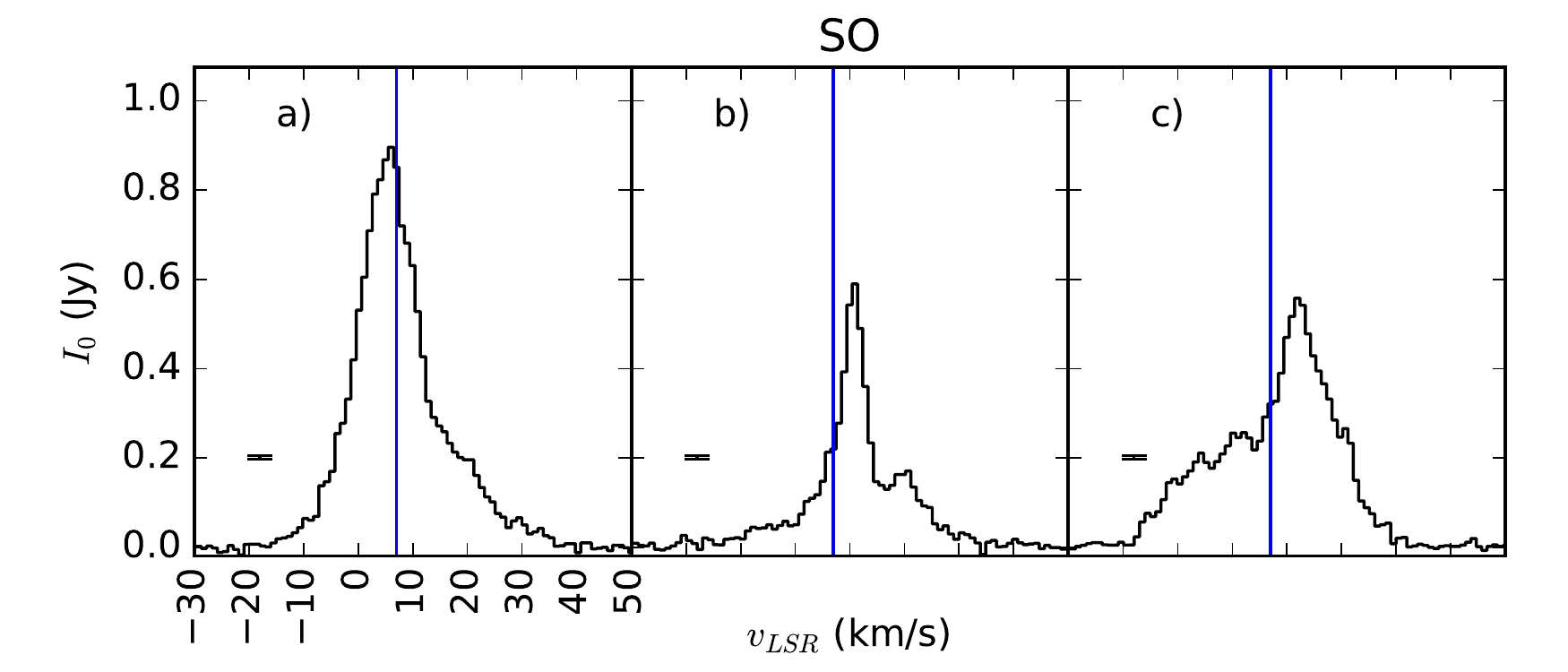}
\caption{Spectra from each of the SO peaks at $\nu$ = 30.0015 GHz. a) is from the southern peak, b) is from the 
northern peak and the rc) is from the western peak. The blue line is a reference set at a \vlsr\ of 7 
\kms. The abscissa is \vlsr\ in \kms and the ordinate is intensity in \jbm. The rest frequency for each transition is 
given in each panel and the 'I' bars denote the 1$\sigma$ rms noise. \label{fig:sospec}}
\end{figure}

Figure~\ref{fig:sospec} shows the spectra from the three main SO emission peaks. The blue line is a reference set at a 
\vlsr\ of 7.0 \kms. Panel (a), from the southern peak, was best fit by a pair of Gaussians, with \vlsr's of 5.3(0) 
and 20.9(3) \kms\ and FWHM of 12.4(1) and 11.5(8) \kms. Panel (b), from the northern peak, was best fit by three 
Gaussians, with \vlsr's of 9.2(81), 10.7(0), and 21.6(5) \kms\ and FWHM of 7.2(42), 4.0(2), and 8.1(12) \kms. Panel (c), from the western peak near IRc6, was best fit by at least three Gaussians, with \vlsr's of 1.6(5), 11.7(2), and 
18.5(87) \kms\ and FWHM of 25.0(9), 6.8(4), and 4.3(51) \kms. Since none of the peaks are coincident we calculated the 
column densities for each based on a range of temperatures (30-200 K). For the IRc6 peak we find column densities of 
0.3-1.9\e{17} \cms, 1.2-8\e{16} \cms, and 0.8-5.1\e{17} \cms, for the three Gaussians. For the northern peak we find 
column densities of 1.1-6.8\e{16} \cms, 0.2-1.1\e{17} \cms, and 0.2-1.4\e{17} \cms, for the three Gaussians. For the 
southern source we find column densities of 0.2-1.6\e{17} \cms\ and 1.4-8.9\e{17} \cms\ for the two Gaussians.

\begin{figure}[!ht]
\includegraphics[scale=1.0]{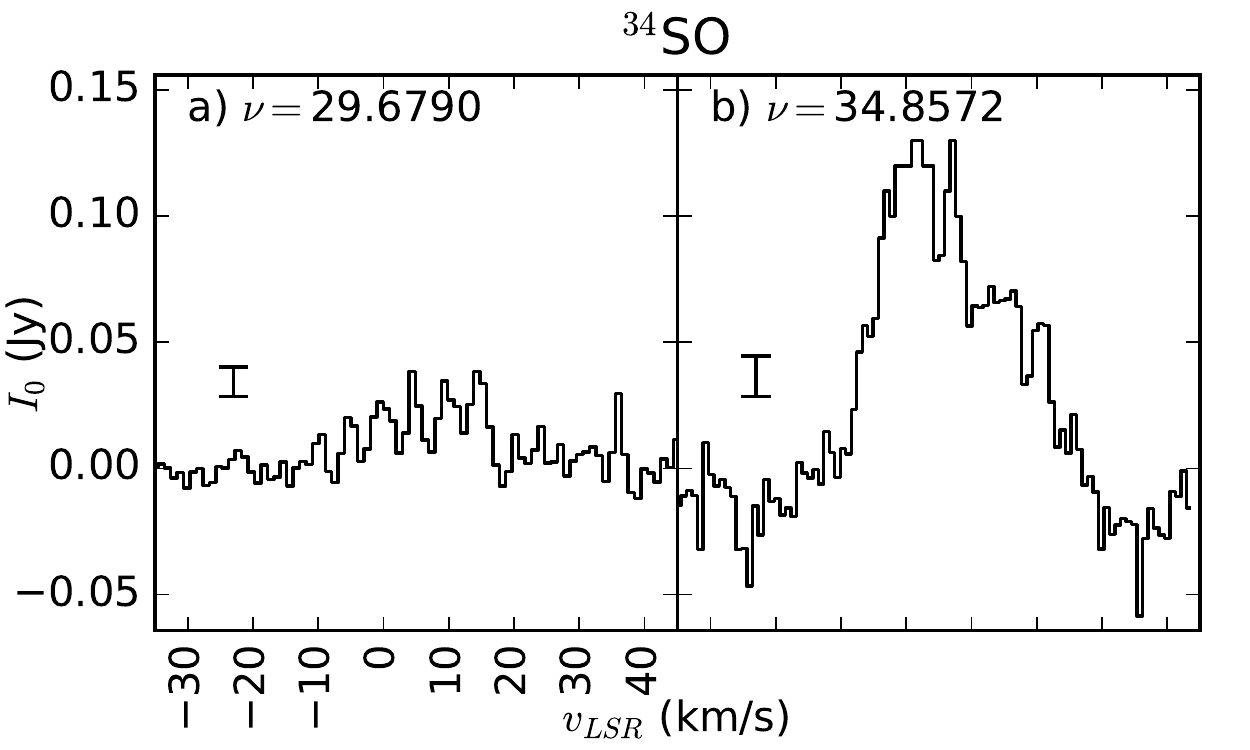}
\caption{Spectra from the two detected $^{34}$SO transitions, taken at the peak of emission. The abscissa is \vlsr\ in 
\kms and the ordinate is intensity in \jbm. The rest frequency for each transition is given in each panel and the 'I' 
bars denote the 1$\sigma$ rms noise. \label{fig:34so-spec}}
\end{figure}

Figure~\ref{fig:34so-spec} shows the spectra from the two detected $^{34}$SO transitions. Panel (a), from the 
lowest energy transition, was best fit by a single wide Gaussian, with a \vlsr\ of 8.0 \kms\ and FWHM of 22.8 \kms, 
although more components may be blended in. Panel (b), from the higher energy transition, was best fit by a 
pair of Gaussians, with \vlsr's of 2.4 and 17.6 \kms\ and FWHM of 15.5 and 10.5 \kms. For the lowest energy transition 
we find column densities of 0.7-4.2\e{16} \cms. For the higher energy transition we find column densities of 
0.9-3.4\e{16} \cms\ and 2.6-9.6\e{15} \cms, for the two Gaussians.

\begin{figure}[!ht]
\includegraphics[scale=0.5]{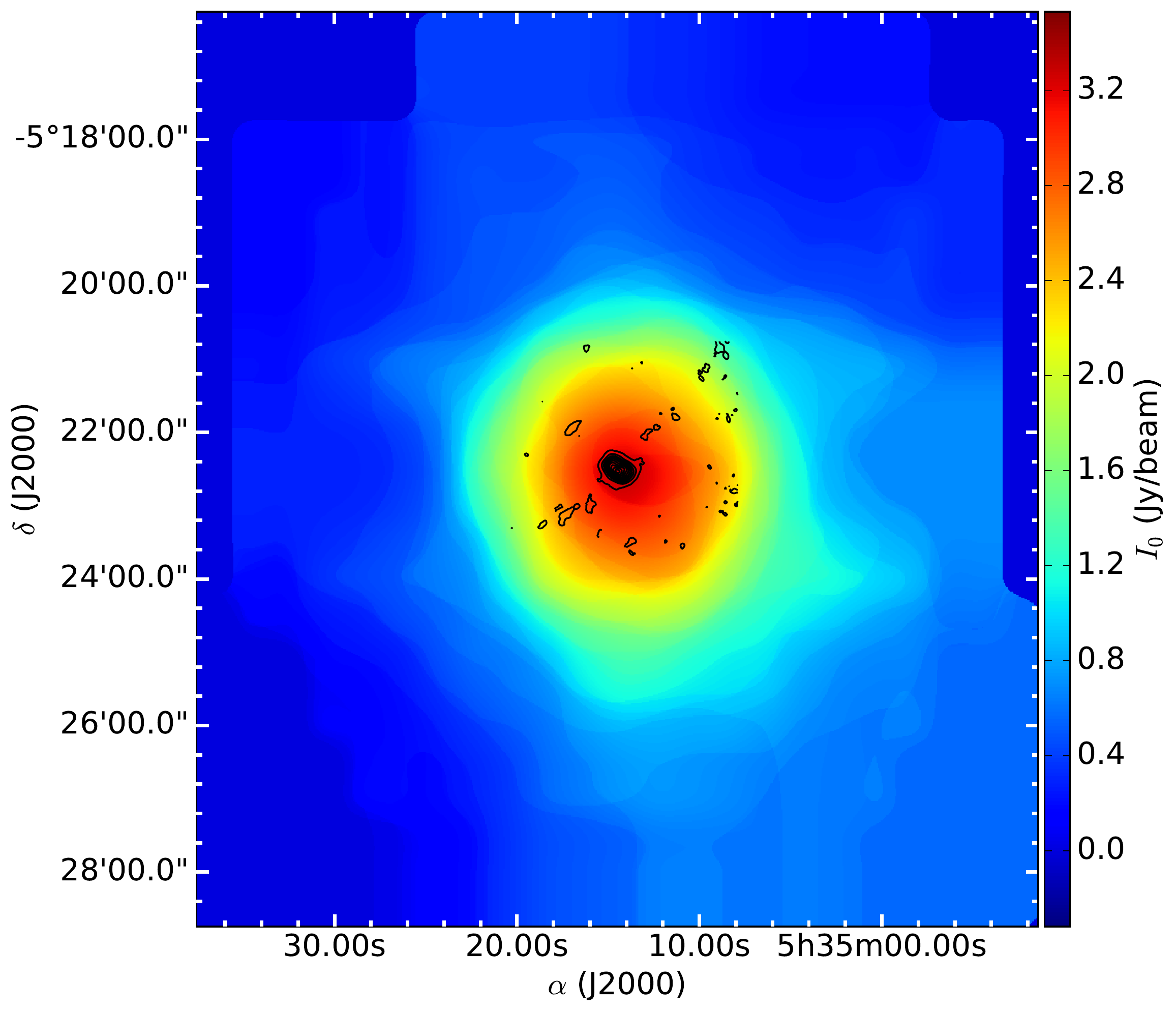}
\caption{Average intensity map of the CARMA single dish mode SO data as a colorscale overlaid with the contour average intensity map of the 
interferometric data. Contours are $\pm3\sigma, \pm27\sigma, \pm51\sigma, ...$, where $\sigma$ = 1.0 m\jbm.\label{fig:sosd}}
\end{figure}

\begin{figure}[!ht]
\includegraphics[scale=0.5]{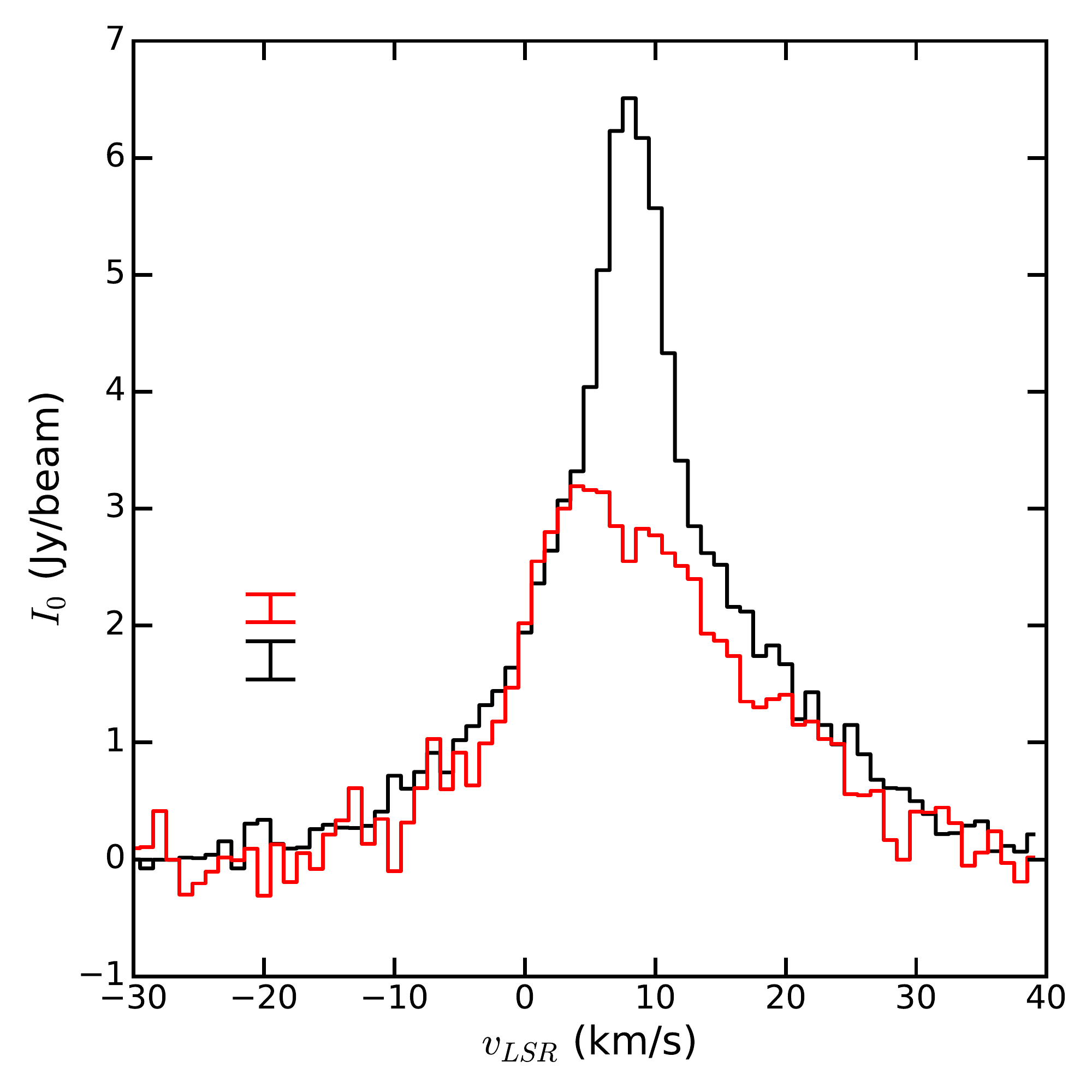}
\caption{Spectra from the single dish SO observations (black) and integrated spectra over the full emission region of the 
interferometric data (red). The abscissa is \vlsr\ in \kms and the ordinate is intensity in \jbm\ and the 'I' bars 
denote the 1$\sigma$ rms noise of the respective spectrum. \label{fig:sosdspec}}
\end{figure}

In order to understand the full extent of the SO emission the transition was observed in CARMA's single dish mode with 
a 32 point mosaic. Figure~\ref{fig:sosd} shows the resulting average intensity map in color overlaid with the interferometric 
observations as contours. As can be seen, the bulk of the flux is constrained to the Hot Core region. 
Figure~\ref{fig:sosdspec} shows a comparison of the spectra from the single dish (black) and interferometric (red) 
observations. The interferometric data were generated by convolving a 165\arcsec\ Gaussian with the output maps. Both 
spectra have similar structures, as the interferometric emission well traces the wide emission wings of the single dish 
data. However, there is a notable strong peak that is not detected by the interferometer. The interferometric data were 
best fit by three Gaussians with peaks of 0.7(3), 1.46(2), and 2.1(2) \jbm, FWHM of 4.6(23), 2.6(53), and 24.3(10) 
\kms, and \vlsr\ of 10.7(9), 3.7(53), and 8.1(0) \kms. Given the complexity of the SO emission there could be many more 
components which are lost in the blend. The single dish data were well fit with the same components, with the exception 
of the widest line which had an intensity of 2.6(4) \jbm, and a single component with a peak of 4.0(0) \jbm, FWHM of 
2.6(0) \kms, and \vlsr\ of 8.0(0) \kms.

\clearpage

%--------------------------------------------------------------------------------------------------

\subsubsection{Sulphur Dioxide [SO$_2$ and $^{34}$SO$_2$]}
\begin{figure}[!ht]
\includegraphics[scale=0.45]{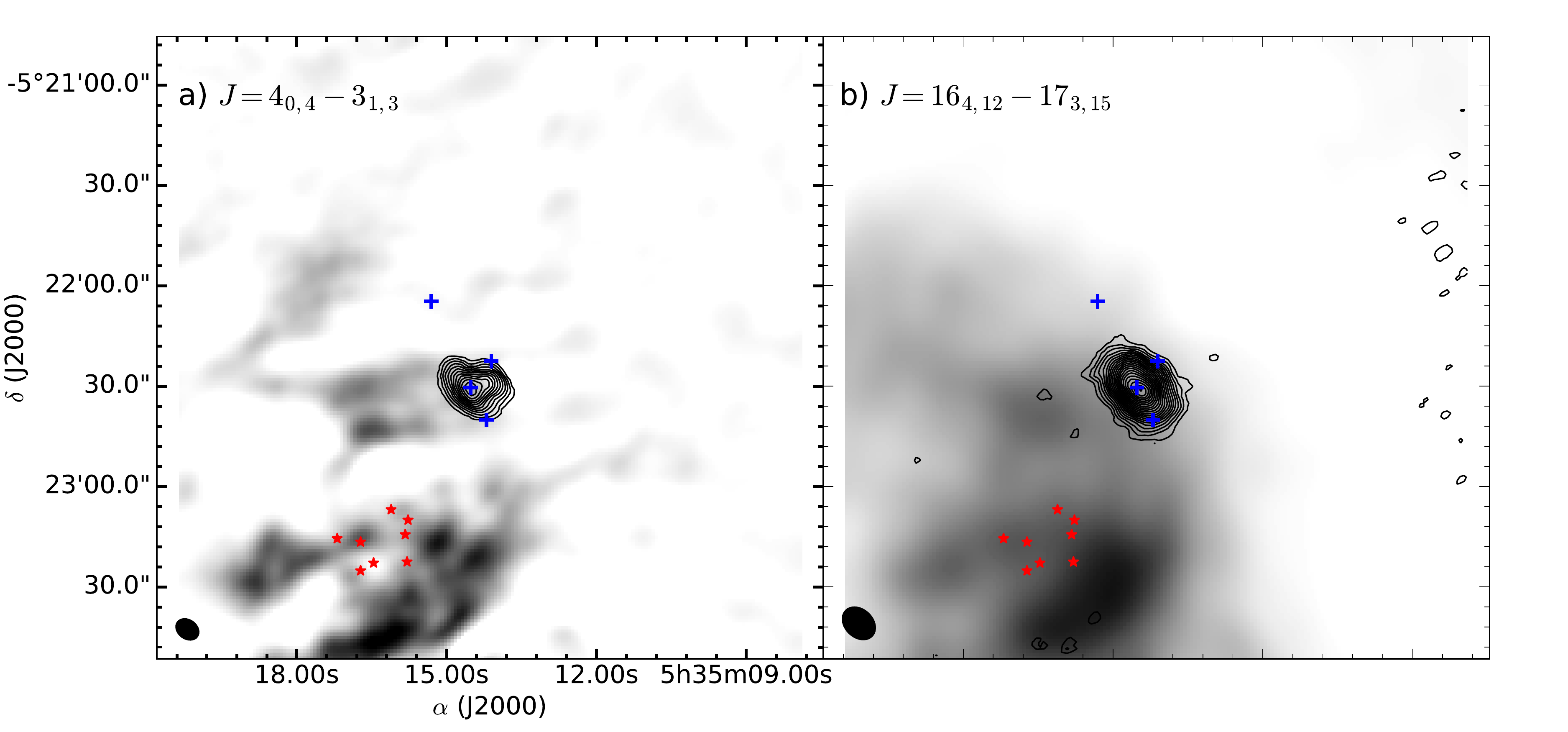}
\caption{Average intensity maps of \sot, overlaid on the appropriate continuum, from two of the transitions. a) 
is from the lowest energy transition which was observed in C configuration and b) is from a middle 
energy transition, which was observed in D configuration. Contours are $\pm3\sigma, \pm6\sigma, \pm9\sigma, ...$ where 
$\sigma$=4.4 and 2.0 m\jbm, respectively.\label{fig:so2}}
\end{figure}

A total of seven sulphur dioxide transitions (six \sot\ and one $^{34}$\sot) were detected with this survey. Like SO, 
\sot\ displays a shell like structure with three (or more) distinct peaks: HC-NE, IRc6, and HC/CR. Figure~\ref{fig:so2} shows average intensity maps, overlaid on the appropriate continuum, from two of the transitions. Panel (a) is from the lowest energy transition which was observed in C configuration and panel (b) is from 
a middle energy transition, which was observed with the combined C/D configurations. 

\begin{figure}[!ht]
\includegraphics[scale=.85]{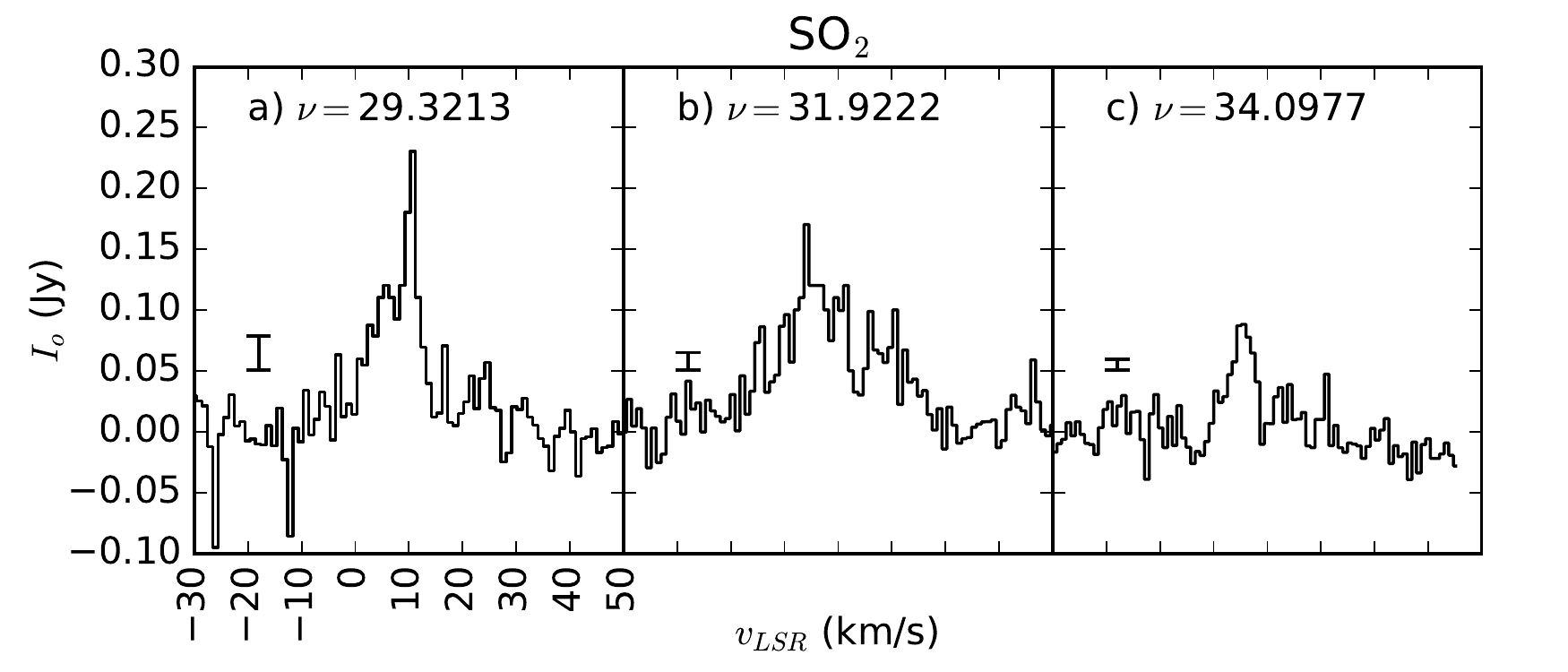}
\caption{Spectra of the three detected \sot\ transitions from the northern peak. The abscissa is \vlsr\ in \kms and the 
ordinate is intensity in \jbm. The rest frequency for each transition is given in each panel and the 'I' bars denote 
the 1$\sigma$ rms noise. \label{fig:so2_specN}}
\end{figure}

Not all transitions were detected from each peak and there were not enough transitions to construct a reliable rotation 
temperature diagram, therefore all column densities are calculated based on rotation temperatures of 30 - 200 K. 
Figure~\ref{fig:so2_specN} shows the spectra from the three detected transitions from the northern peak. The lowest two 
energy transitions (panels (a) and (b)) were best fit with pairs of Gaussians, with \vlsr's of 7.5(9) and 10.5(5) 
\kms\ (panel (a)) and 4.4(4) and 8.4(8) \kms\ (panel (b)) and FWHM of 17.8(21) and 1.9(13) \kms\ (panel (a)) and 
3.7(10) and 19.8(22) \kms\ (panel (b)). Given the overall low S/N of the spectra, it is possible that more components 
may be hidden in the profile. The highest energy transition (panel (c)) was best fit by a single Gaussian with a 
\vlsr\ of 4.9(3) \kms\ and FWHM of 5.4(8) \kms. These transitions yield a total column density of 3.6\e{16} - 8.6\e{20} \cms.

\begin{figure}[!ht]
\includegraphics[scale=.85]{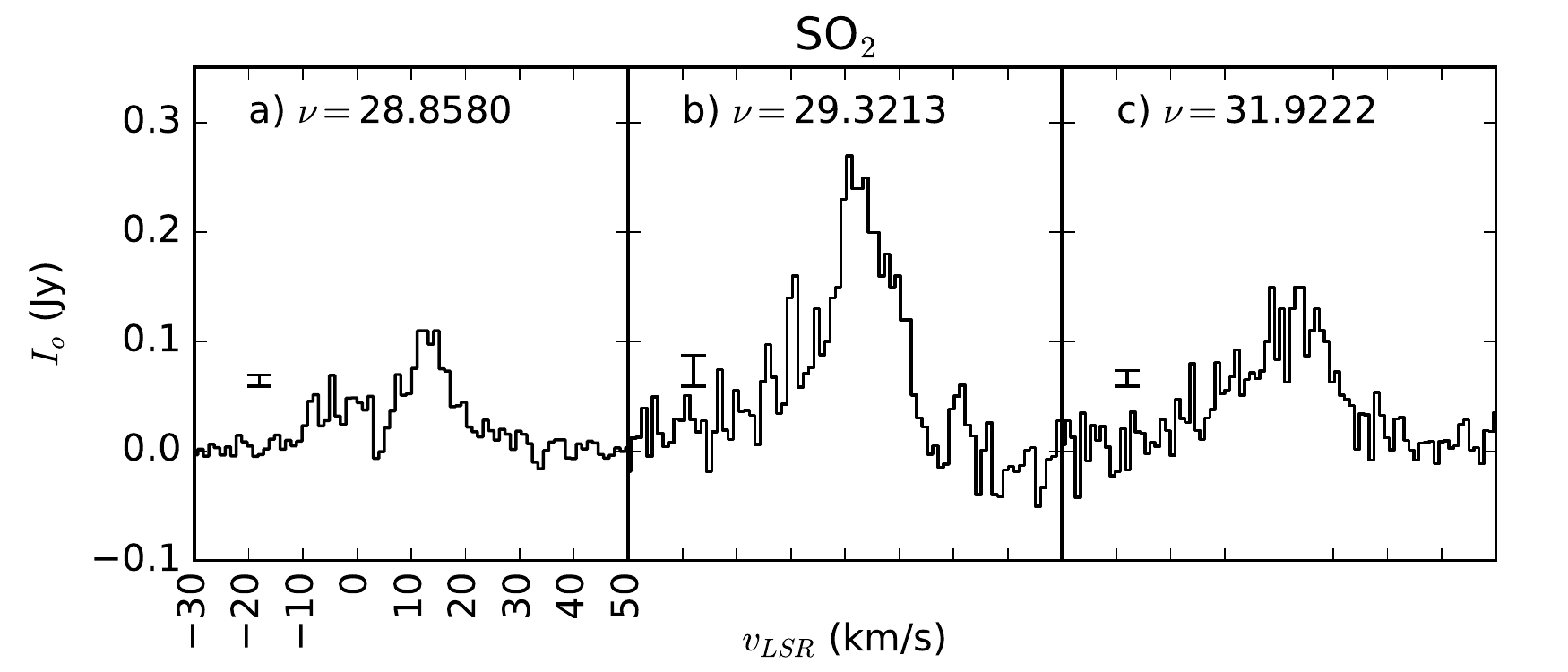}
\caption{Spectra of the three detected \sot\ transitions from the peak near IRc6.The abscissa is \vlsr\ in \kms and the 
ordinate is intensity in \jbm. The rest frequency for each transition is given in each panel and the 'I' bars denote 
the 1$\sigma$ rms noise.\label{fig:so2_specW}}
\end{figure}

Figure~\ref{fig:so2_specW} shows the three detected \sot\ transitions from the IRc6 peak. The lowest energy transition 
(panel (b)) was best fit by three Gaussians, with \vlsr's of 6.2(15), 12.0(8), and 18.6(16) \kms\ and FWHM of 
26.8(22), 5.6(17), and 6.1(30) \kms. The other transitions were best fit by a pair of Gaussians, with \vlsr's of 
-4.2(4) and 13.2(1) \kms\ (panel (a)) and 8.3(18) and 15.3(15) \kms\ (panel (c)) and FWHM of 9.9(11) and 10.2(5) 
\kms\ (panel (a)) and 26.1(26) and 36.0(47) \kms\ (panel (c)). The fit parameters give an overall column density of 
8.4\e{16} - 1.2\e{19} \cms.

\begin{figure}[!ht]
\includegraphics[scale=.85]{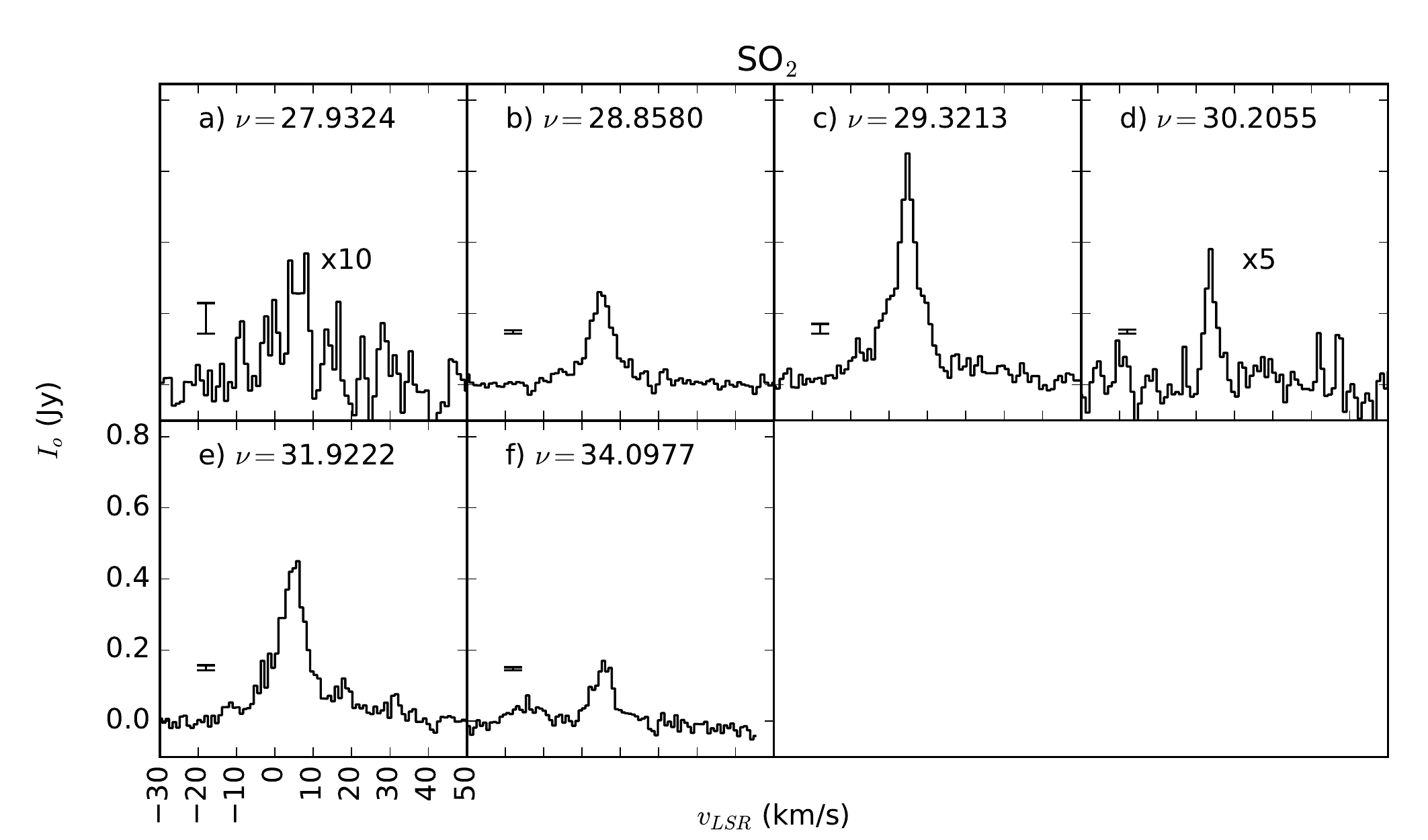}
\caption{Spectra of the six detected \sot\ transitions from the southern peak.The abscissa is \vlsr\ in \kms and the 
ordinate is intensity in \jbm. The rest frequency for each transition is given in each panel and the 'I' bars denote 
the 1$\sigma$ rms noise.\label{fig:so2_specS}}
\end{figure}

Figure~\ref{fig:so2_specS} shows all six detected \sot\ transitions from the southern peak. Three of the four lowest 
energy transitions (28.8, 29.3, and 31.9 GHz) were best fit by three Gaussians, with \vlsr's of -6.4(19), 5.0(2), and 
5.6(7) \kms, 2.9(9), 4.5(1), and 4.9(1) \kms, and -1.5(15), 4.9(2), and 11.4(13) \kms, respectively, and FWHM of 
2.8(47), 5.4(5), and 15.7(17) \kms, 40.4(33), 9.7(6), and 2.7(3) \kms, and 7.5(27), 5.9(5), and 32.9(22) \kms, 
respectively. The fourth of these lower energy transitions (30.2 GHz) was best fit by only a single Gaussian, with a 
\vlsr\ of 3.8(4) \kms\ and FWHM of 4.0(8) \kms. The other Gaussian components are likely lost in the noise as this 
transition has a linestrength that is a factor of four (or more) lower than the other transitions. The two higher 
energy transitions (27.9 and 34.1 GHz) were best fit by a single Gaussian, with \vlsr's of 4.5(9) and 5.5(2) \kms, 
respectively, and FWHM of 18.3(23) and 6.2(7) \kms, respectively. These give a total column density of 9.8\e{16} - 
1.8\e{21} \cms. While SO$_2$ appears to have a similar spatial distribution to SO, the spectra show that they likely 
occupy different gas in the northern core, due to the notably different \vlsr's and FWHM; toward IRc6 and the southern 
core the picture is less clear as there are similar velocity components for both molecules, but also ones that are 
distinctly different. This indicates that SO and SO$_2$ do not form under the same physical conditions; however, there 
is some overlap.

\begin{figure}
\includegraphics[scale=0.75]{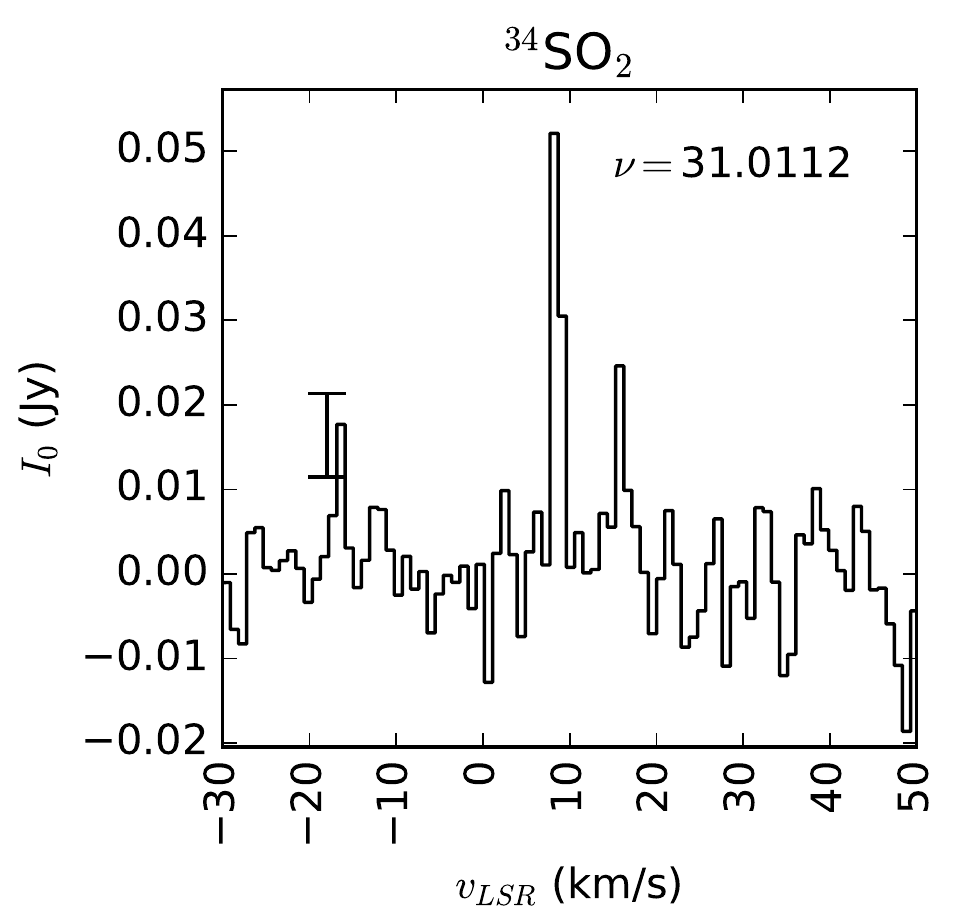}
\caption{The single detected $^{34}$\sot\ transition from the southern peak. The abscissa is \vlsr\ in \kms and the 
ordinate is intensity in \jbm. The rest frequency for each transition is given in each panel and the 'I' bars denote 
the 1$\sigma$ rms noise.\label{fig:34so2-spec}}
\end{figure}

Figure~\ref{fig:34so2-spec} shows the only detected $^{34}$\sot\ transition from this survey. This transition was 
detected toward the HC/CR region. It was best fit by a pair of narrow Gaussians, with \vlsr's of 8.5(9) and 15.9(3) 
\kms\ and FWHM of 1.2(22) and 2.0(8) \kms. Giving a total column density of 5.1\e{15} - 1.5\e{17} \cms. Based on the 
absence of any other detected $^{34}$\sot\ transitions we calculate that the rotation temperature must be no more than 
150 K, yielding an upper limit of 2.5\e{16} \cms\ for the total column density. Temperatures higher than this would 
have yielded a detection of the $17_{2,16}-16_{3,1}$ transition at 33.21281 GHz. Since \sot\ and $^{34}$\sot\ should 
share a common rotation temperature this yields an upper limit of 6.0\e{17} \cms\ for \sot.

\clearpage

%--------------------------------------------------------------------------------------------------

\subsubsection{Ammonia [NH$_3$]}
\begin{figure}[!ht]
\includegraphics[scale=0.75]{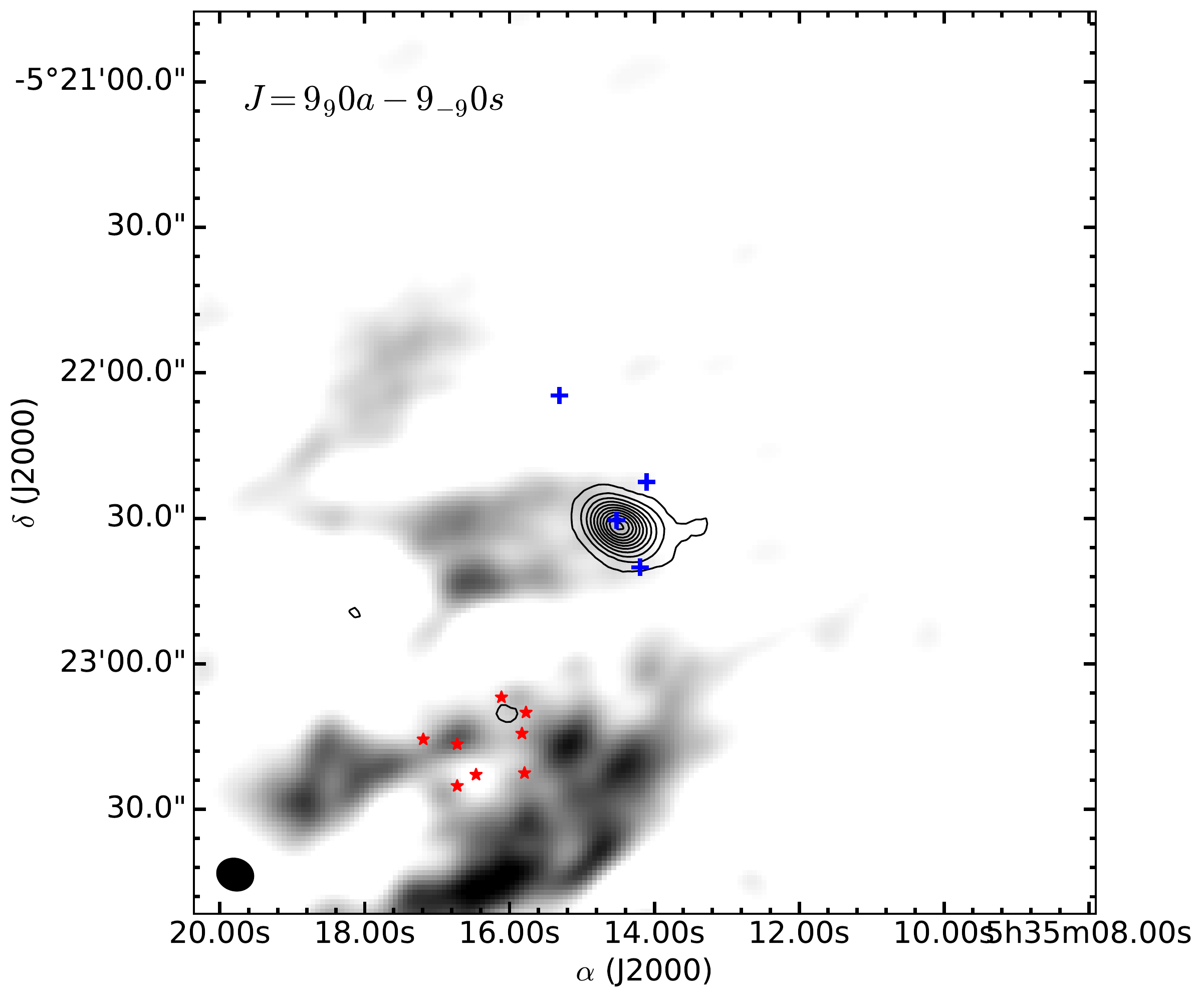}
\caption{Map of ammonia emission overlaid on the continuum. The contours are $\pm3\sigma$, $\pm15\sigma$, 
$\pm27\sigma$, ... where $\sigma$ = 2.0 m\jbm. The synthesized beam is shown in the lower left corner.\label{fig:nh3}}
\end{figure}

\begin{figure}[!ht]
\includegraphics[scale=0.65]{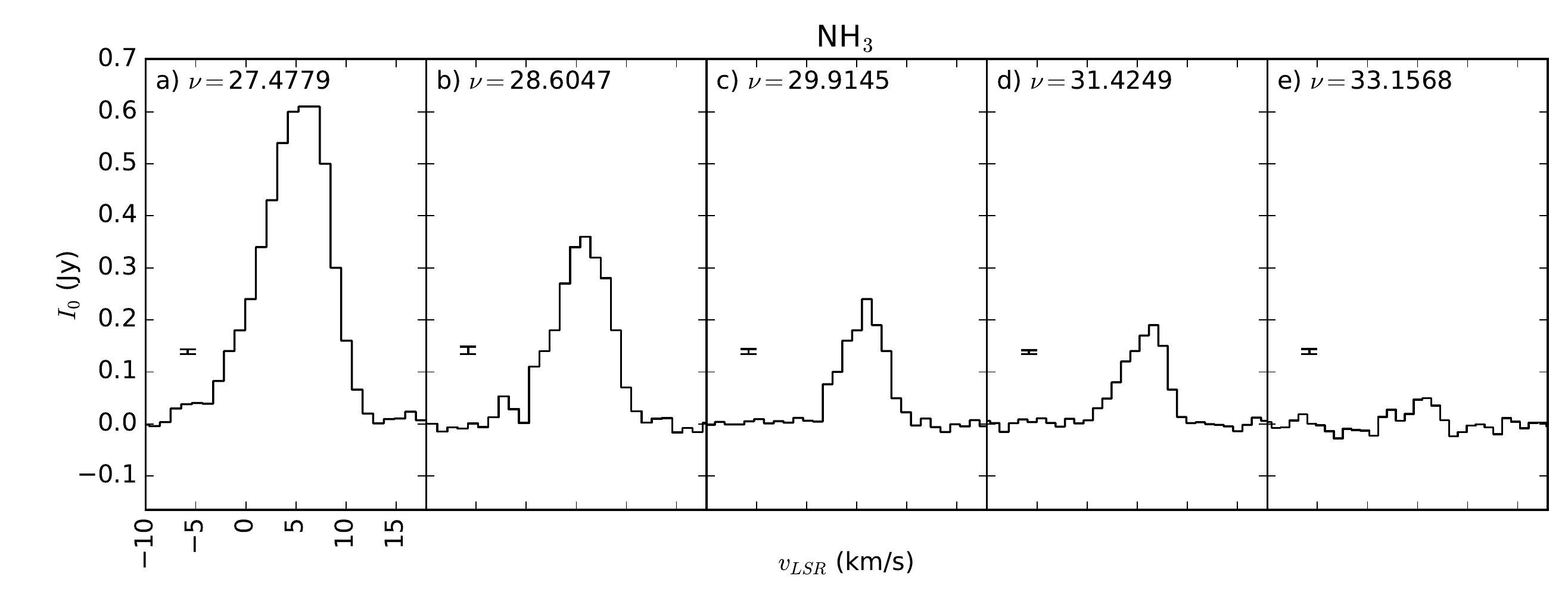}
\caption{Spectra of all detected ammonia lines. The abscissa is \vlsr\ in \kms and the ordinate is intensity in \jbm. 
The rest frequency for each transition is given in each panel and the 'I' bars denote the 1$\sigma$ rms noise. 
\label{fig:nh3_spec}}
\end{figure}

\begin{figure}
\includegraphics[scale=.2]{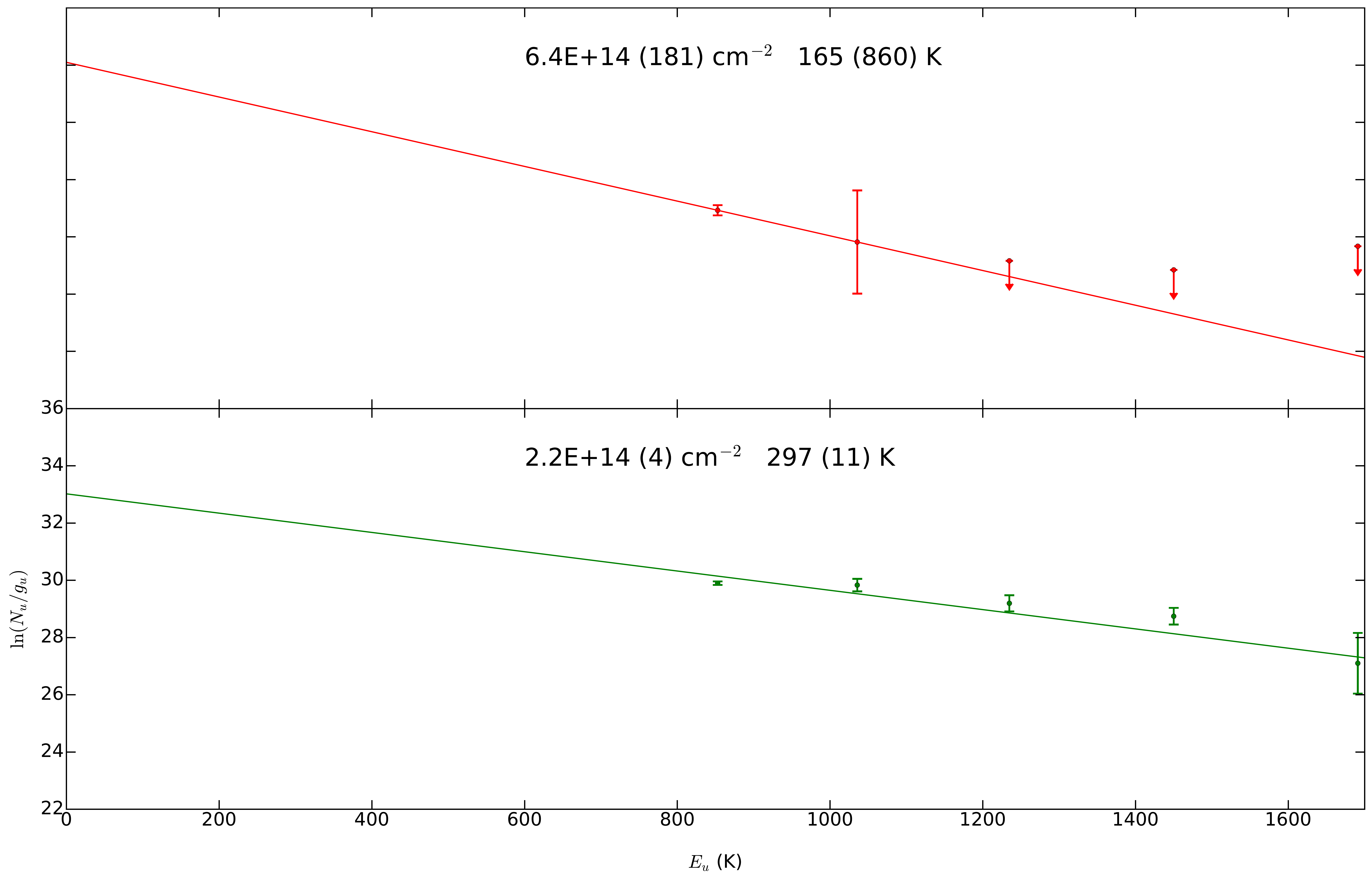}
\caption{Rotation temperature diagram from the ammonia transitions. a) shows the data and fit to the 
narrow component, three of the points are shown as upper limits based on the $1\sigma$ rms noise of the respective 
windows. b) shows the data and fits for the wide component.\label{fig:nh3-rtd}}
\end{figure}

A total of five ammonia lines were detected in this survey, mostly concentrated near the Hot Core, Figure~\ref{fig:nh3}.
Figure~\ref{fig:nh3_spec} show the spectrum from the emission peak of each detected line. All transitions 
were simultaneously fit with Gaussians by assuming a consistent FWHM and \vlsr. The transitions at 27.4779 and 28.6047 
GHz were best fit with a pair of Gaussians with \vlsr's of 1.0(0) and 6.0(0) \kms\ and FWHM of 6.3(0) and 5.7(0) \kms, 
respectively. The remaining three were best fit by a single Gaussian at 6.0(0) \kms\ and FWHM of 5.2(0) \kms. From 
these fits we constructed rotation temperature diagrams for each component (Figure~\ref{fig:nh3-rtd}).
The narrow component (panel (a)) only has two points, but does give reasonable results of 6.4\e{14} \cms\ and 165 K. 
The other three points are shown as upper limits based on the same line width and an intensity of $1\sigma$. The main 
component (panel (b)) is best fit with 2.2\e{14} \cms\ and 297 K. The temperature is consistent with previous 
observations of NH$_3$ by \citet{goddi11}, who found a temperature range of 160-490 K, at higher resolution. The total 
column density reported by \citet{goddi11} is three orders of magnitude larger than what we find in this work. This 
could be due to us sampling different gas that was resolved out by their smaller beam and/or a very small source size. 
Given our column densities and rotation temperatures there are no missing transitions.

\begin{figure}
\includegraphics[scale=.5]{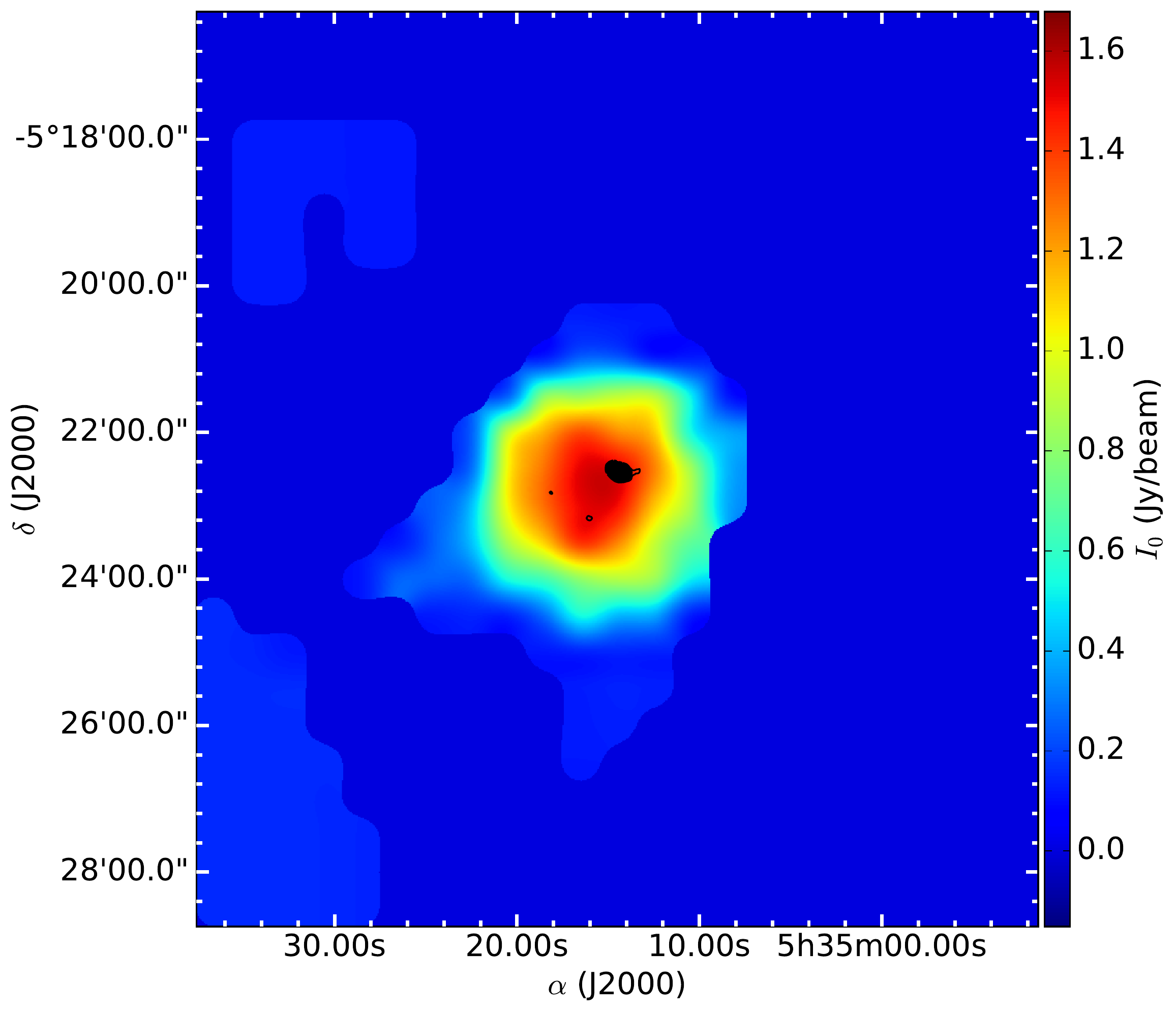}
\caption{Array average intensity map of the $9_90a-9_90s$ transition of NH$_3$ (contours) overlaid on the single dish average intensity map 
of the same. Contours are $\pm3\sigma, \pm6\sigma, \pm9\sigma, ...$ where $\sigma$=8.9 m\jbm. \label{fig:nh3-sd-map}}
\end{figure}

\begin{figure}
\includegraphics[scale=0.5]{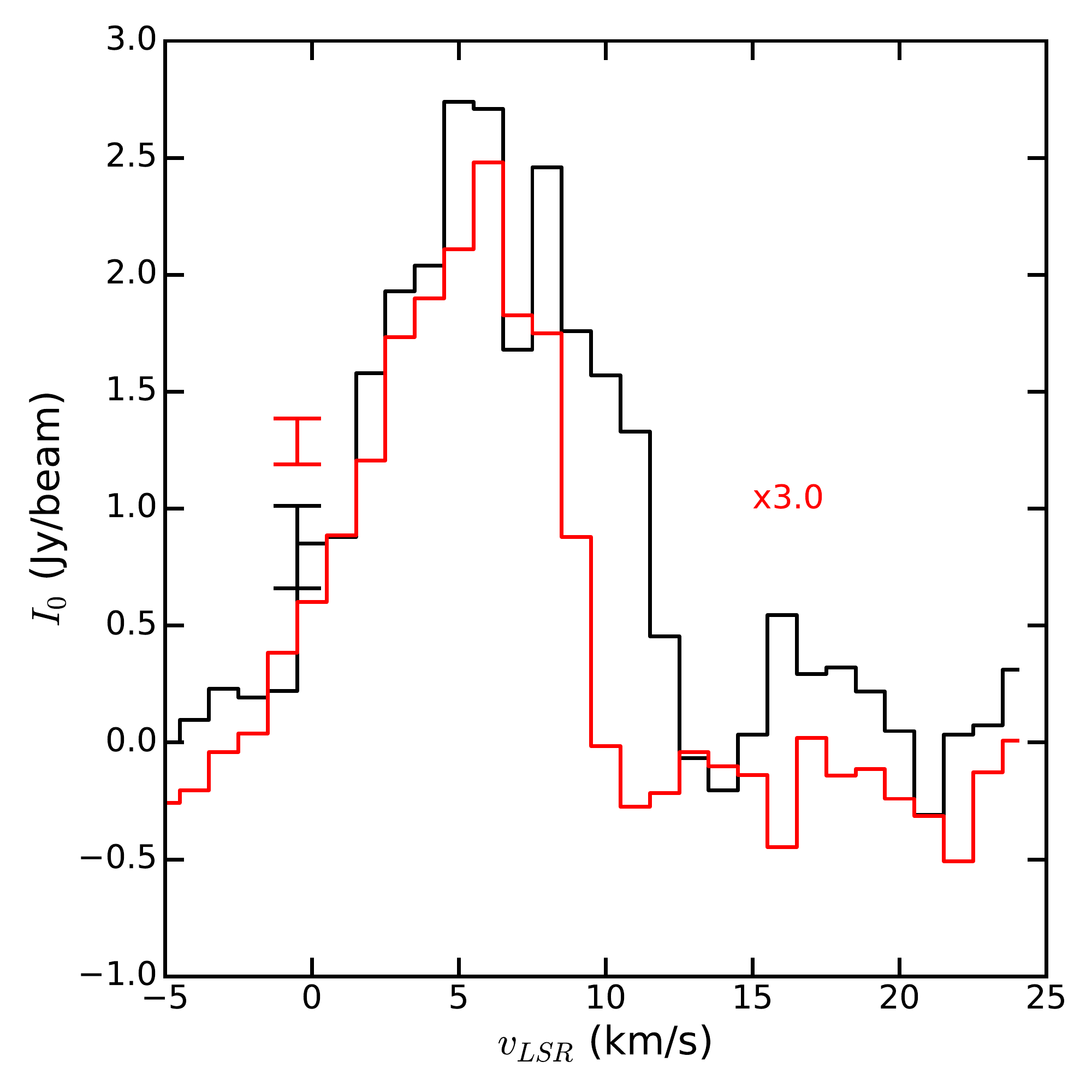}
\caption{Spectra of the single dish (black) and array (red) observations of the $9_90a-9_90s$ transition of NH$_3$. The 
array data have been scaled by a factor of 3.0. The abscissa is \vlsr\ in \kms and the ordinate is intensity in \jbm\ 
and the 'I' bars denote the 1$\sigma$ rms noise.\label{fig:nh3-sd-spec}}
\end{figure}

The $9_90a-9_90s$ transition of NH$_3$ was observed in single dish mode. Figure~\ref{fig:nh3-sd-map} shows the single 
dish average intensity map as colorscale overlaid by the array average intensity map as contours. Figure~\ref{fig:nh3-sd-spec} shows the 
single dish spectrum (black) and the array spectrum (red). The peaks of the average intensity maps do not perfectly correspond, 
nor does the NH$_3$ array spectrum match that of the single dish as it has to be scaled by a factor of 3.0 in order to 
match. For a transition of such high energy (850 K) to have so much flux resolved out is not usual, but not unprecedented. \citet{wilson93} found there are two temperature components to the NH$_3$ emission toward Orion-KL. According to Figure~3 of \citet{wilson93}, a transition at 850 K would have significant flux from both components. \citet{wilson00} noted a significant amount of flux was resolved out of array observations of a lower energy NH$_3$ transition, when compared to single dish observations. Thus the missing flux could be due to our observations resolving out the lower temperature component. The offset between the single dish and array peaks is likely due to the two temperature components being offset. Figure~3 of \citet{murata90} shows an elongation of the NH$_3$ emission to the southeast from Source I, and CO emission has a small peak in this direction as well.

\clearpage

%--------------------------------------------------------------------------------------------------

\subsubsection{Methyl Formate [\mef]}
\begin{figure}[!ht]
\includegraphics[scale=0.45]{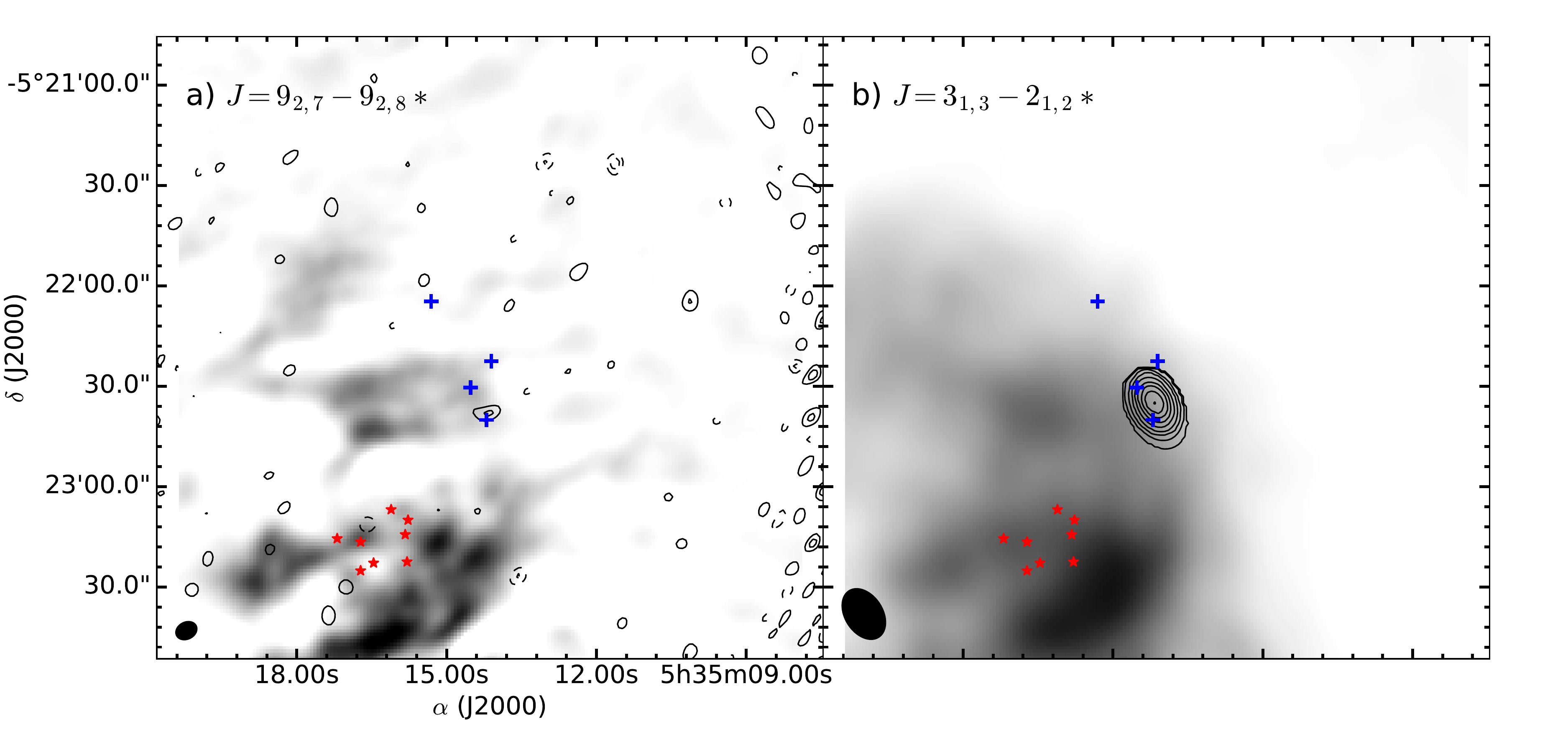}
\caption{Methyl formate maps, one each from a) C and b) D configurations. The emission contours 
are overlaid on the respective continuum. The contours are $\pm3\sigma$, $\pm5\sigma$, $\pm7\sigma$,... for panel (a)
 and $\pm3\sigma$, $\pm6\sigma$, $\pm9\sigma$,... for panel (b), where $\sigma$ = 6.0 and 11.7 m\jbm, 
respectively. The respective synthesized beams are in the lower left corner of each panel. \label{fig:mef}}
\end{figure}

\begin{figure}[!ht]
\includegraphics[scale=0.65]{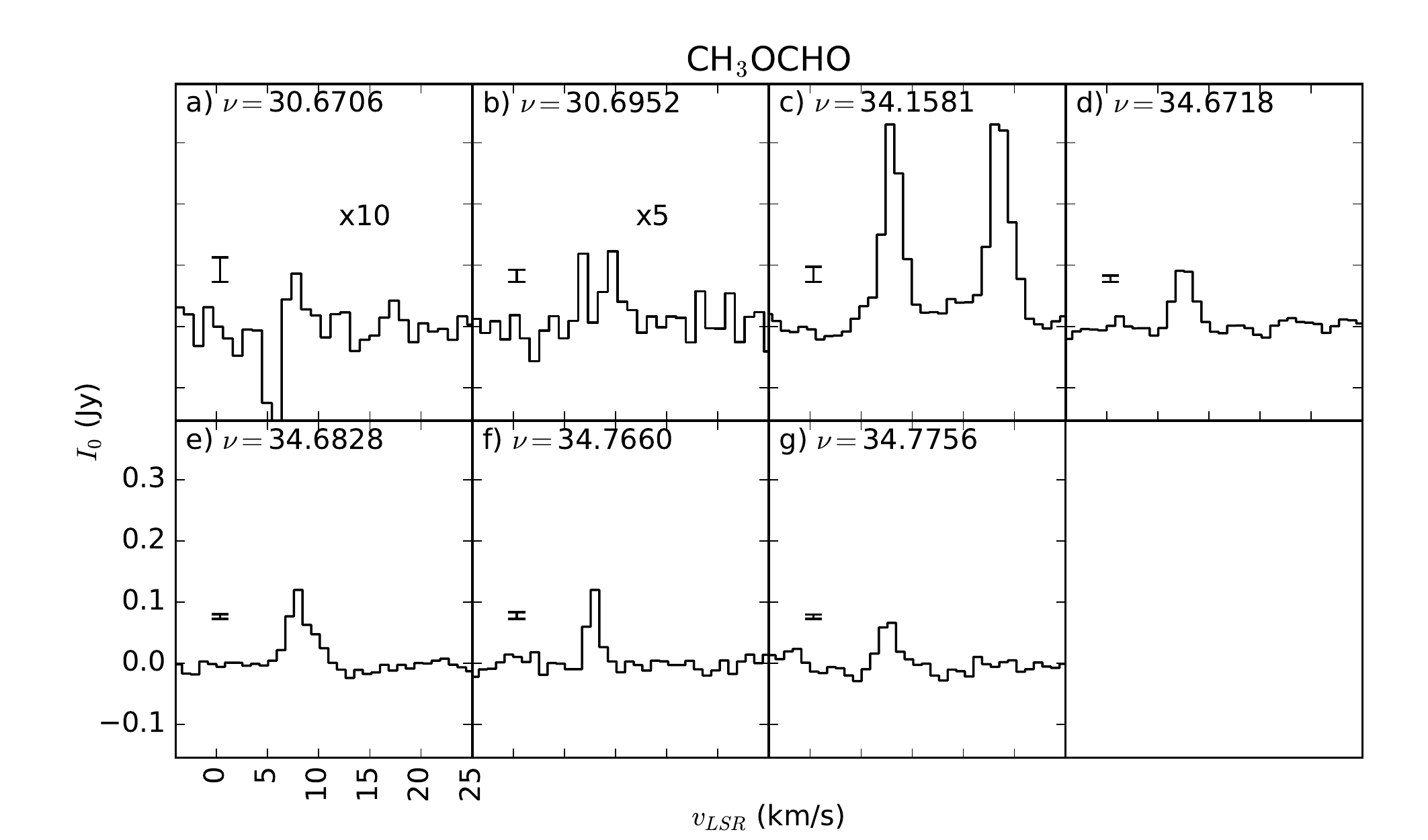}
\caption{Methyl formate spectra from each detected transition. Each transition is labeled by its rest frequency. The 
spectrum from 34.1581 GHz shows two \mef\ transitions. The abscissa is \vlsr\ in \kms and the ordinate is intensity in 
\jbm\ and the 'I' bars denote the 1$\sigma$ rms noise.\label{fig:mef_spec}}
\end{figure}

A total of eight \mef\ transitions were detected in this survey. Two of these transitions were detected with the C 
configuration, but were poorly fit by Gaussians, and were left out of the calculations. Figure~\ref{fig:mef} shows the 
maps of two transitions, one from C configuration (panel (a)) and one from D configuration peaking between
the Compact Ridge and the BN (panel (b)). The contours are  $\pm3\sigma$, $\pm5\sigma$, $\pm7\sigma$,... for panel (a) and $\pm3\sigma$, $\pm6\sigma$,
$\pm9\sigma$,...  for panel(b), where $\sigma$ = 6.0 and 11.7 m\jbm, respectively. 
Figure~\ref{fig:mef_spec} shows the spectra of 
each detected transition. The spectra were best fit with a \vlsr\ of $\sim$8.0 \kms\ and FWHM of $\sim$2.0 \kms, 
similar to that of \dme. Each transition was individually fit. The spectra from 34.1581 GHz shows two \mef\ 
transitions.

All detected transitions were very low energy ($<$16 K) and could not be fit by a rotation temperature diagram, due to 
the small spread in energies. \citet{friedelphd} found a total column density of 3.3(1)\e{17} \cms\ and a rotation 
temperature of 344(55) K. Using this temperature yields a column density nearly an order of magnitude higher. Given 
that these are very low energy lines, and that the rotation temperature in \citet{friedelphd} is dominated by high 
energy lines, it is likely that the emission region detected by this survey is dominated by cooler, more extended gas. 
Using a rotation temperature of 200 K yields a column density of $\sim$4\e{17} \cms. Using this temperature and column 
density there are no missing lines.

\begin{figure}
\includegraphics[scale=.5]{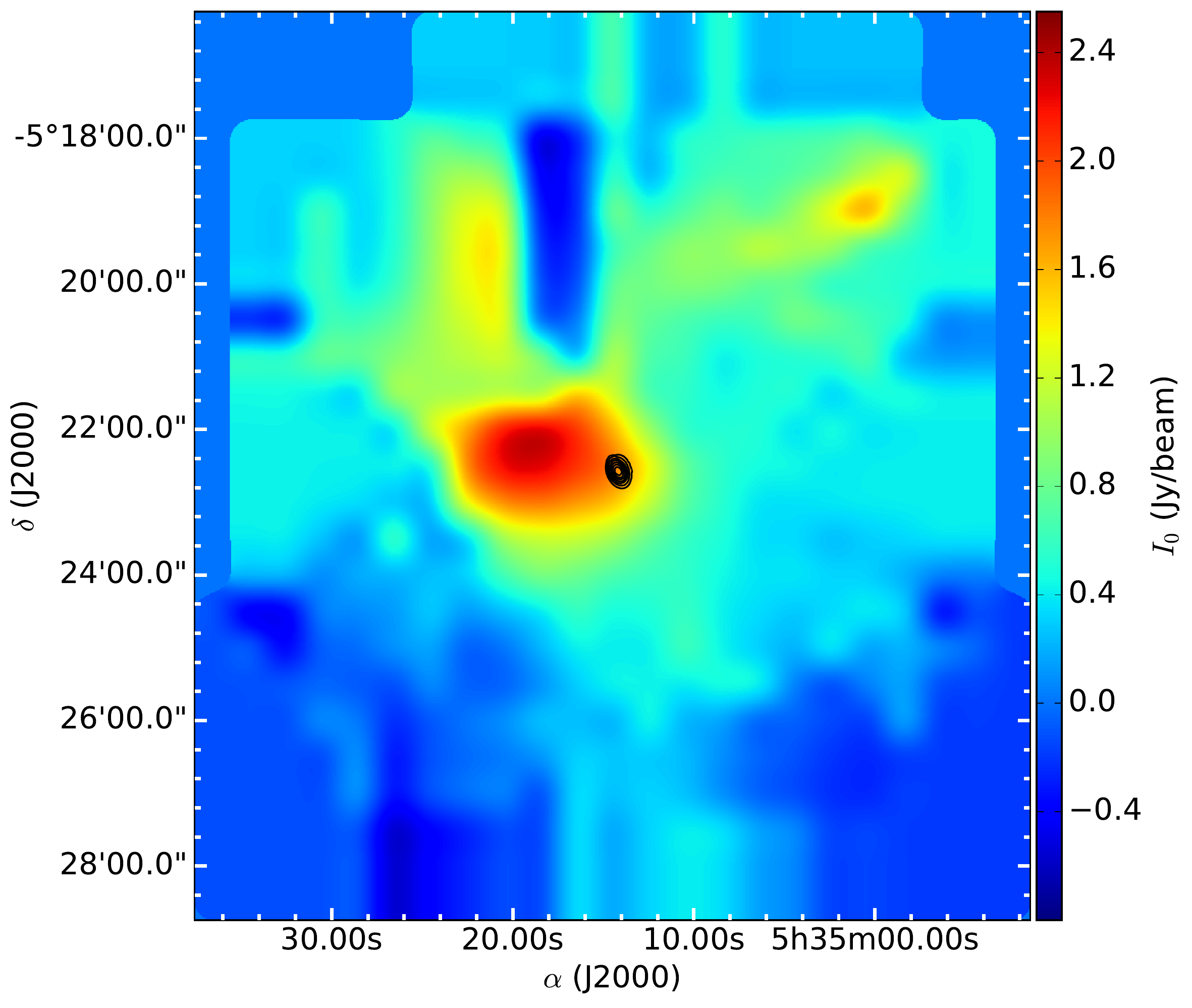}
\caption{Array average intensity map of the $3_{1,3}-2_{1,2}E$ transition of \mef\ (contours) overlaid on the single dish average intensity
map of the same transition. Contours are $\pm3\sigma, \pm6\sigma, \pm9\sigma, ...$ where $\sigma$=17.1 m\jbm.
\label{fig:mef-sd-map}}
\end{figure}

\begin{figure}
\includegraphics[scale=0.5]{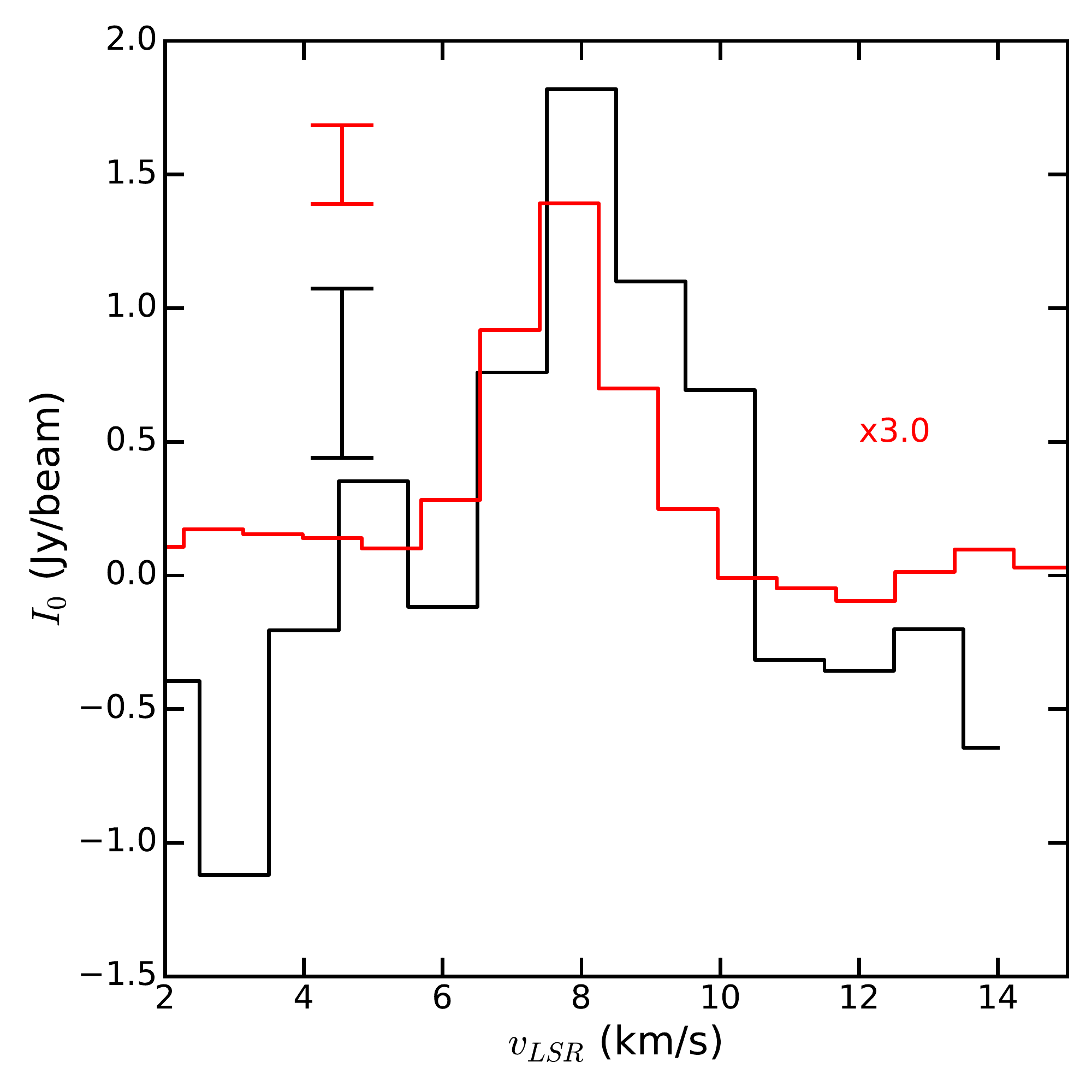}
\caption{Spectra of the single dish (black) and array (red) observations of the $3_{1,3}-2_{1,2}E$ transition of \mef. 
The array data have been scaled by 3.0. The abscissa is \vlsr\ in \kms and the ordinate is intensity in \jbm\ and the 
'I' bars denote the 1$\sigma$ rms noise.\label{fig:mef-sd-spec}}
\end{figure}

In order to investigate the full extent of the \mef\ emission the $3_{1,3}-2_{1,2}E$ transition was observed in single 
dish mode. Figure~\ref{fig:mef-sd-map} shows the single dish average intensity map as colorscale overlaid with the array map as 
contours and Figure~\ref{fig:mef-sd-spec} shows the spectra from the single dish observations (black) and array 
observations (red). The array data have been scaled by a factor of 3. Given the low signal-to-noise there is no 
convincing evidence of any extended structure.

\clearpage

%--------------------------------------------------------------------------------------------------

\subsubsection{Dimethyl Ether [\dme]}
A total of 17 dimethyl ether lines were detected in this survey, 14 with the C configuration and 3 with combined C and 
D configuration data. Typically, each transition is composed of four torsional states: {\it EE}, {\it AA}, {\it AE}, 
and {\it EA}, where {\it EE} is at the central frequency and {\it AA} and a combined {\it AE/EA} are equally spaced on 
either side. Three of the transitions (12 lines) in this work fit this description. However, the lines near 31.999 GHz 
and 33.943 GHz do not. In these transitions only a partial set of the four lines are detected. In the case of the 
31.999 GHz lines, the {\it AE} and {\it EA} components are not stacked and are below the detection threshold. In the 
case of the 33.943 GHz lines, the {\it AA} and {\it EE} are blended, the {\it EA} is detected but the separate {\it AE} 
component is below the detection threshold. \citet{friedelphd} found a rotation temperature of $\sim$84 K and a total 
column density of 7.2\e{16} \cms. Using this temperature we find a total column density of 0.3-2\e{16} \cms. This could 
be reconciled with the \citep{friedelphd} column density by using either a higher temperature or a source size of 
$\sim$5\arcsec$^2$. Given the column densities and temperatures, there are no missing transitions.

\begin{figure}[!ht]
\includegraphics[scale=0.45]{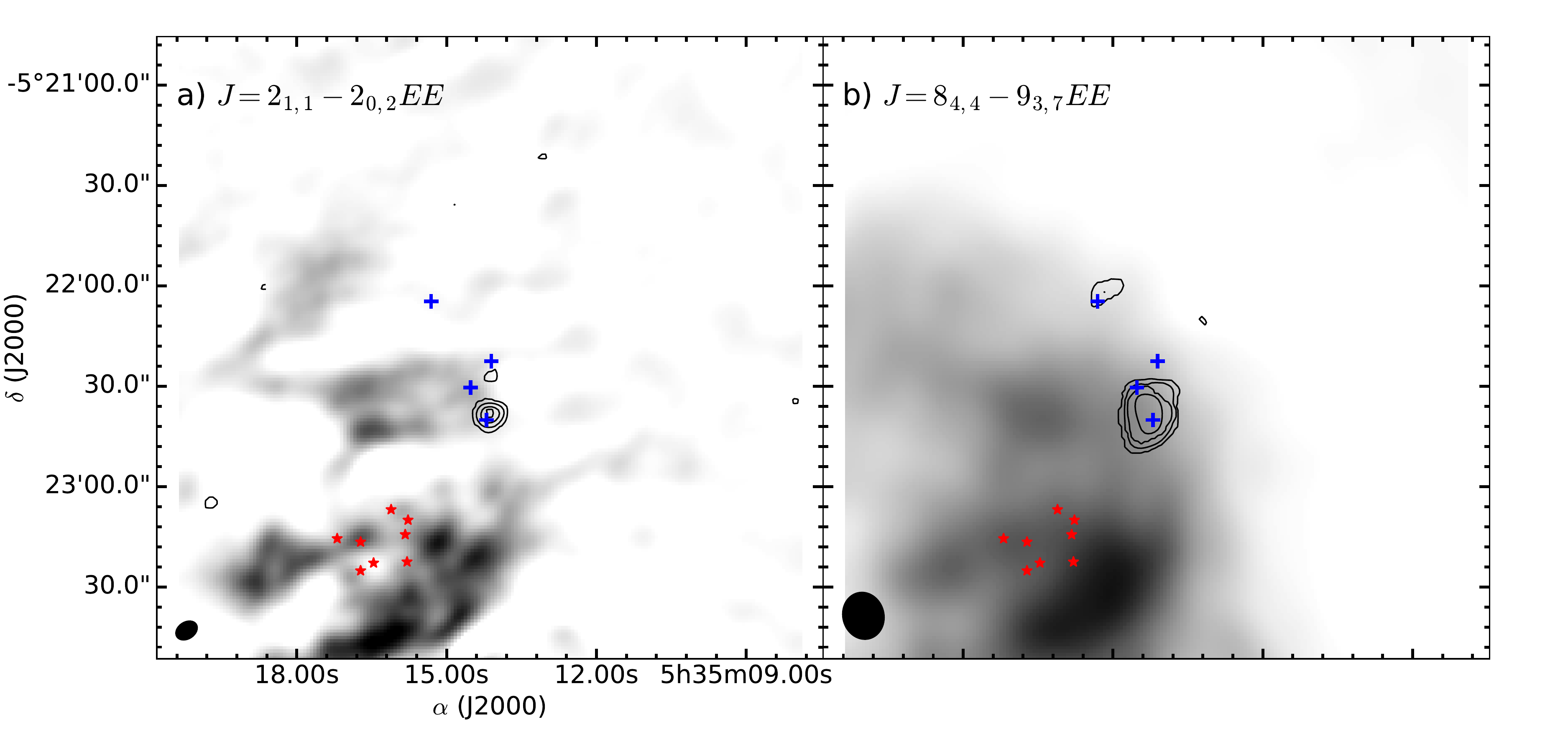}
\caption{Average intensity maps of \dme\ from the a) C configuration and b) both configurations. The contours 
are $\pm3\sigma$, $\pm6\sigma$, $\pm9\sigma$, ..., where $\sigma$ = 4.6 and 7.7 m\jbm, respectively.\label{fig:dme}}
\end{figure}

\begin{figure}[!ht]
\includegraphics[scale=0.75]{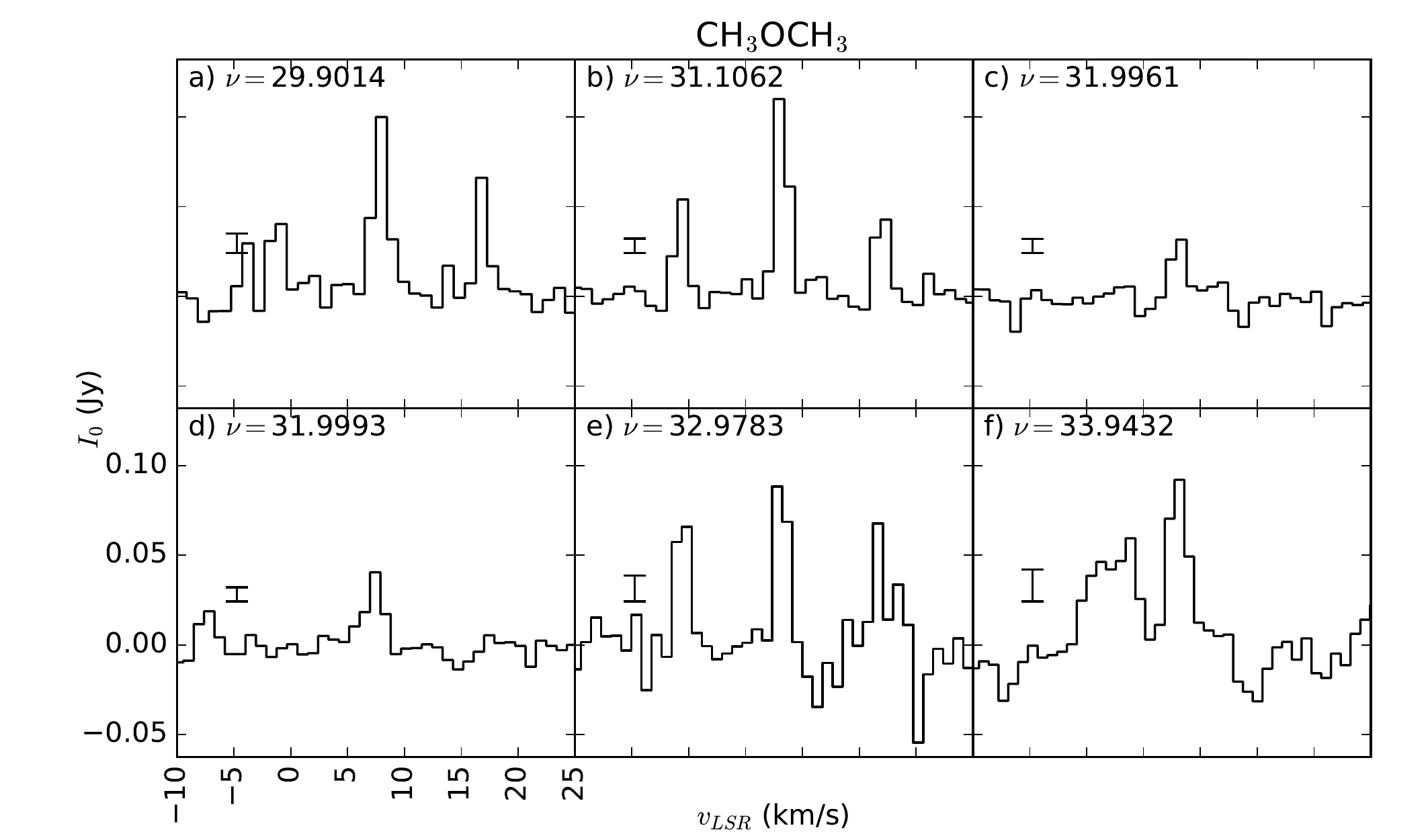}
\caption{Spectra of the detected \dme\ transitions. The abscissa is \vlsr\ in \kms and the ordinate is intensity in 
\jbm. The rest frequency for each transition is given in each panel and the 'I' bars denote the 1$\sigma$ rms noise.
\label{fig:dme-spec}}
\end{figure}

Figure~\ref{fig:dme} shows the C and D configuration maps of \dme. 
In the C configuration map (panel (a)) emission 
is detected near the Compact Ridge and near IRc 6, similar to \citet{slww12, friedel08}. In all instances the emission 
is unresolved. Figure~\ref{fig:dme-spec} shows the spectra from all detected \dme\ lines. They were best fit with 
single Gaussians with a \vlsr\ of $\sim$8.0 \kms\ and FWHM of $\sim$1.6 \kms. Each group of transitions ($EE$, $AA$, 
$AE$, and $EA$) was fit together, with a common \vlsr\ and FWHM for each group. In panels (a), (b), (e), and (f) there are 
multiple \dme\ lines, and the $v_{LSR}$ references the strongest ($EE$) rest frequency.

\clearpage

%--------------------------------------------------------------------------------------------------

\subsubsection{Ethyl Cyanide [\etcn]}
\begin{figure}[!ht]
\includegraphics[scale=0.8]{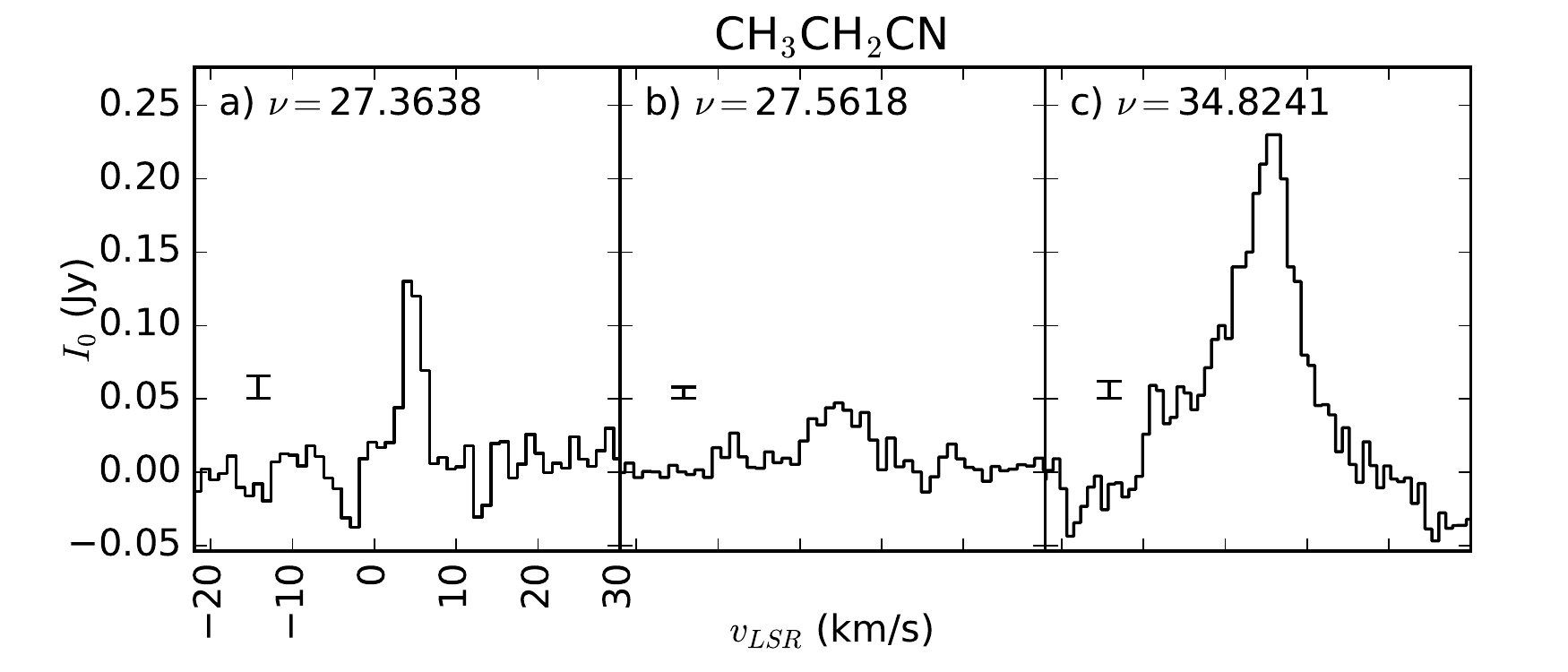}
\caption{The three detected \etcn\ transitions. a) is the blended \etcn/$^{13}$\mtoh\ line, b) 
is the $3_{1,2} - 2_{1,1}$ transition, and c) is the $4_{1,4} - 3_{1,3}$ transition. The abscissa is 
\vlsr\ in \kms and the ordinate is intensity in \jbm. The rest frequency for each transition is given in each panel and 
the 'I' bars denote the 1$\sigma$ rms noise.\label{fig:etcn_spec}}
\end{figure}

A total of three transitions of \etcn\ were detected in our survey, one of which is blended with a transition of 
$^{13}$\mtoh. While \etcn\ is one of the most ubiquitous emitters at higher frequencies \citep[e.g.,][]{friedelphd}, it 
is a much poorer emitter at lower frequencies due to notably lower line strengths. Figure~\ref{fig:etcn_spec} shows the 
three detected \etcn\ lines. Panel (a) shows the blended \etcn/$^{13}$\mtoh\ line, panel (b) shows the $3_{1,2} - 
2_{1,1}$ transition, and panel (c) shows the $4_{1,4} - 3_{1,3}$ transition. Both unblended \etcn\ transitions 
were best fit by two Gaussian components, a narrower one near 5 \kms, and a wider one near -6 \kms. \citet{friedelphd} 
also found that most \etcn\ lines at $\lambda$=3mm were best fit with two components. 
Based on the maps of \citet{slww12},  
we find this work fully encompasses the \etcn\ emission 
and are thus sampling the same gas. Based on the rotation temperatures from \citet{friedelphd} of 314 K, for the narrow 
component, and 118 K, for the wide component, we find total column densities of 1.5 - 2.9\e{17} and 8.4\e{15} - 
4.5\e{16} \cms, for the narrow and wide components respectively. The column density for the narrow component is a 
factor of 2-3 above that reported in \citet{friedelphd}, but the column density for the wide component agrees well with 
the value from \citet{friedelphd}. Given these column densities and rotation temperatures, no \etcn\ transitions are 
missing from our survey. Also given these values we determine that the blended \etcn/$^{13}$\mtoh\ line is at most 10\% 
\etcn.

\begin{figure}[!ht]
\includegraphics[scale=0.35]{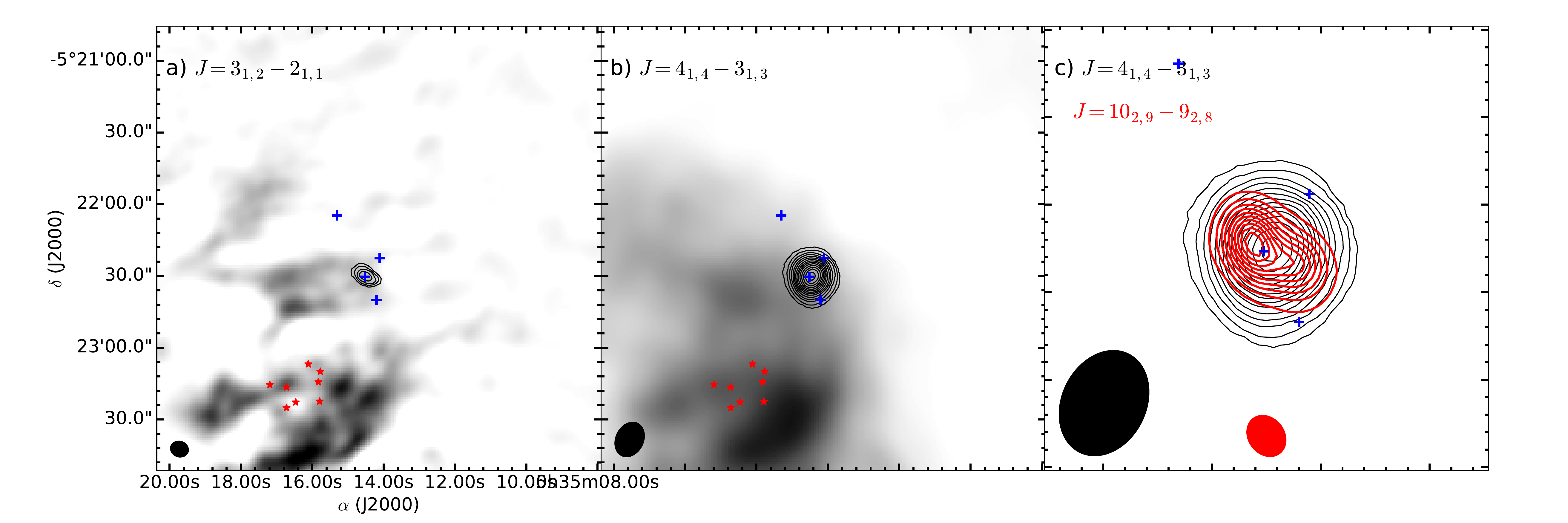}
\caption{Ethyl cyanide maps. a) shows the $3_{1,2} - 2_{1,1}$ transition and b) shows the 
$4_{1,4} - 3_{1,3}$ transition. Contours are $\pm3\sigma, \pm6\sigma, \pm9\sigma, ...$ where $\sigma=2.7, and 2.1$ 
m\jbm\ respectively. c) show the $4_{1,4} - 3_{1,3}$ in black contours overlaid with a $\lambda = 
3$mm \etcn transition from \citet{slww12} in red.\label{fig:etcn}}
\end{figure}

Figure~\ref{fig:etcn} shows maps of the two unblended detected \etcn\ transitions, peaking near the Hot Core. Panel (a) shows the 
$3_{1,2} - 2_{1,1}$ transition and panel (b) shows the $4_{1,4} - 3_{1,3}$ transition. 
Panel (c) shows the $4_{1,4} 
- 3_{1,3}$ in black contours overlaid with a $\lambda = 3$mm \etcn\ transition from \citet{slww12} in red. While the 
3mm emission was observed with a notably smaller beam, the $3_{1,2} - 2_{1,1}$ transition of \etcn, observed in single 
dish mode, was not detected above our 3$\sigma$ threshold (1$\sigma$=370 m\jbm). Combining this non-detection with the 
fact that the 1cm contours only show emission from the same region as the 3mm emission we can conclude that \etcn\ has 
little or no extended structure. This leads to the conclusion that \etcn\ is formed/liberated from the grains only in 
the hottest, densest, and most turbulent regions.

\clearpage

%--------------------------------------------------------------------------------------------------

\subsubsection{Cyanoacetylene [HC$_3$N]}
\begin{figure}[!ht]
\includegraphics[scale=0.7]{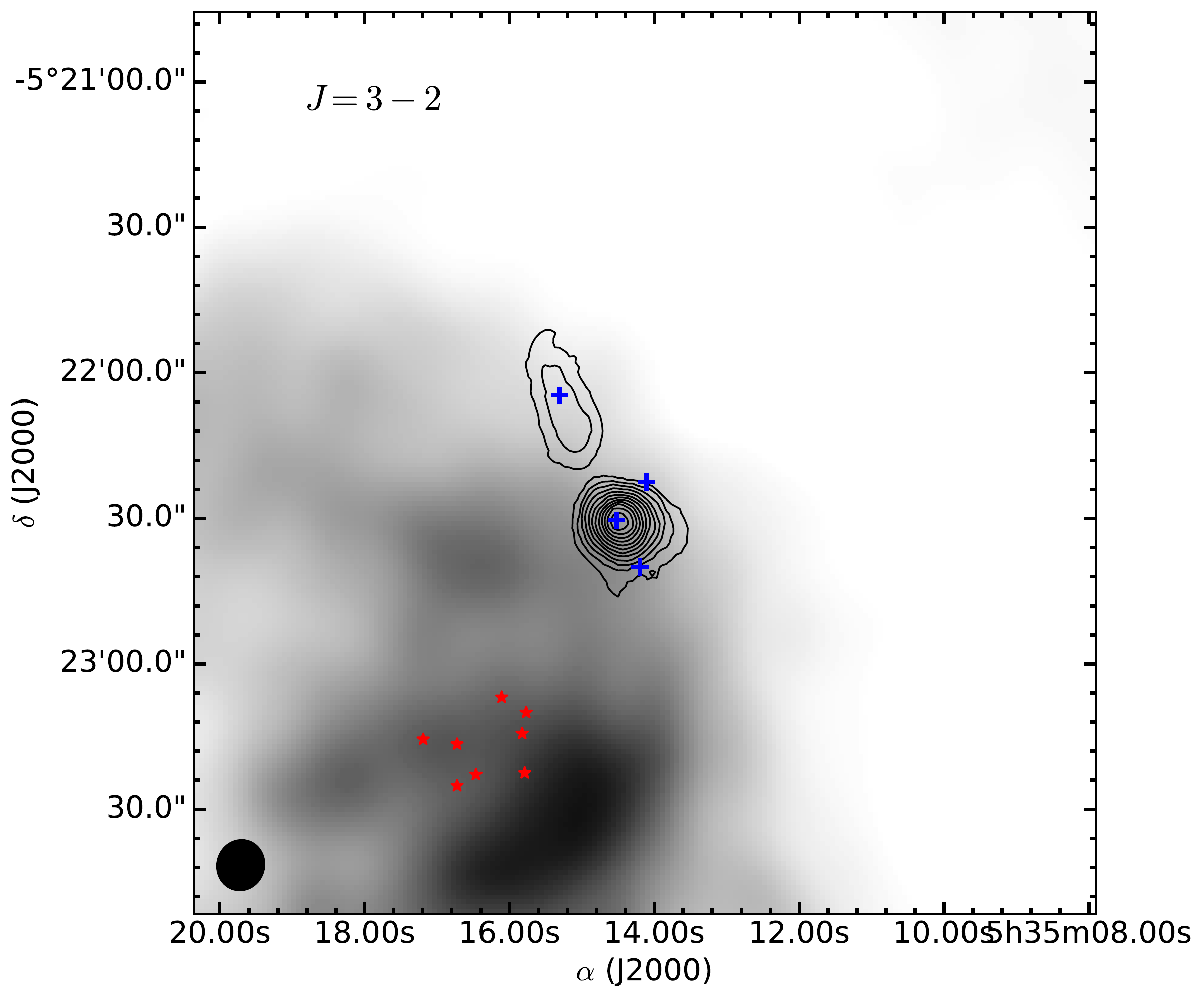}
\caption{Map of the HC$_3$N transition overlaid on the associated continuum. The contours are $\pm3\sigma$, 
$\pm6\sigma$, $\pm9\sigma$ ..., where $\sigma$ = 17.6 m\jbm.\label{fig:hc3n}}
\end{figure}

\begin{figure}[!ht]
\includegraphics[scale=1.0]{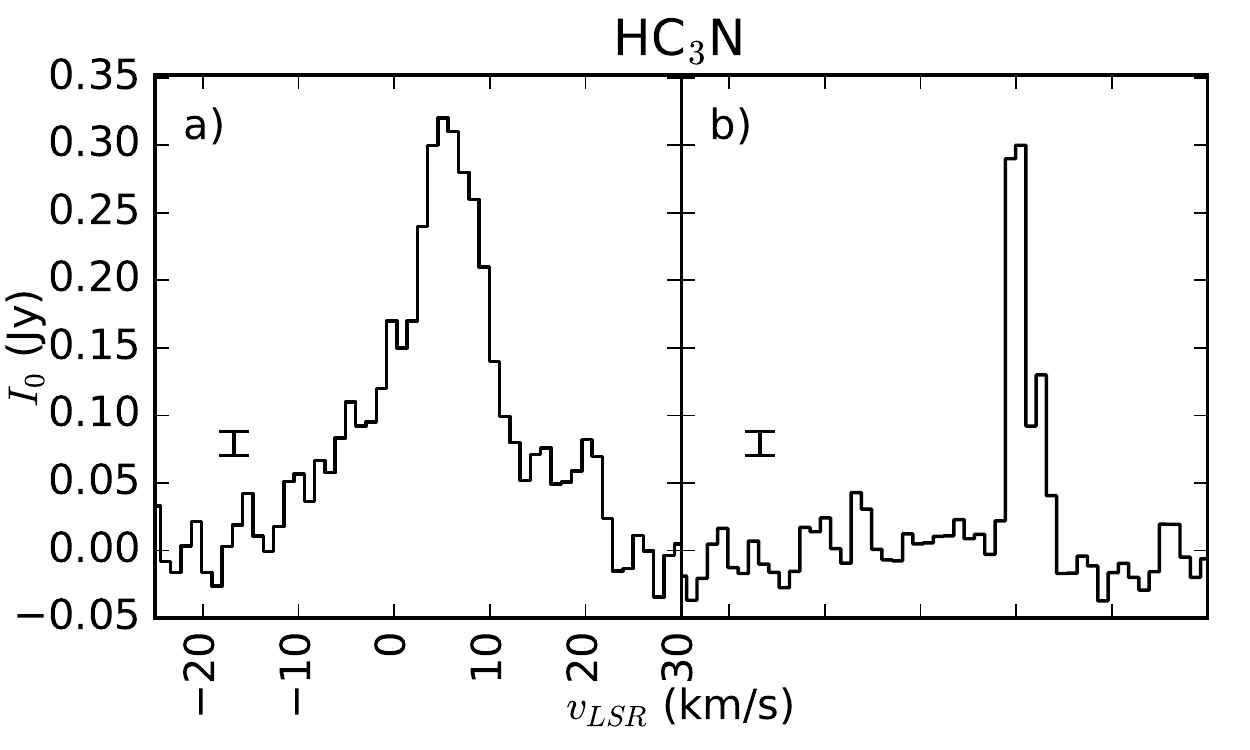}
\caption{Spectrum of the detected HC$_3$N transition ($\nu$ = 27.2943 GHz). The abscissa is \vlsr\ in \kms and the ordinate is intensity in 
\jbm and the 'I' bars denote the 1$\sigma$ rms noise. a) shows the spectrum toward the hot core and b) shows the spectrum toward the northwest peak.
\label{fig:hc3n_spec}}
\end{figure}

The only ground state HC$_3$N transition in our frequency coverage was detected, from two distinct sources. 
Figure~\ref{fig:hc3n} shows the map of the emission overlaid on the continuum. The strongest peak is centered near 
Source I and the weaker peak extends toward CS1. This structure is similar to other molecules like \fmal, 
H$_2$CS, and \mae, but does not quite match any of them. The only molecule that has notable peaks in both locations is 
H$_2$CS, but its strongest peak is near CS1, unlike HC$_3$N which has its strongest peak near source I. 
Figure~\ref{fig:hc3n_spec} shows the spectra from the two emission regions. The spectrum towards Source I is in  
panel (a) while the spectrum from the CS1 region is in the panel (b). The Source I spectrum was best fit 
with three Gaussian components, with \vlsr's of 2.2(15), 5.8(2), and 18.2(8) \kms\ and FWHM of 16.8(16), 5.5(9), and 
4.7(21) \kms. The CS1 peak was best fit by a single Gaussian with a \vlsr\ of 10.0(0) \kms\ and a FWHM of 1.8(0) 
\kms. The fits were performed by assuming the expected, optically thin, ratio of hyperfine components\footnote{Only 
five of the six hyperfine components were used in the fit as the sixth has a line strength over two orders of magnitude 
smaller than any other.}. Using a temperature range of 20 - 200 K we find column densities of 1.9\e{14}-7.7\e{15} \cms\ 
for Source I and 3.0\e{15}-3.3\e{16} for the northeast component.

\begin{figure}
\includegraphics[scale=.49]{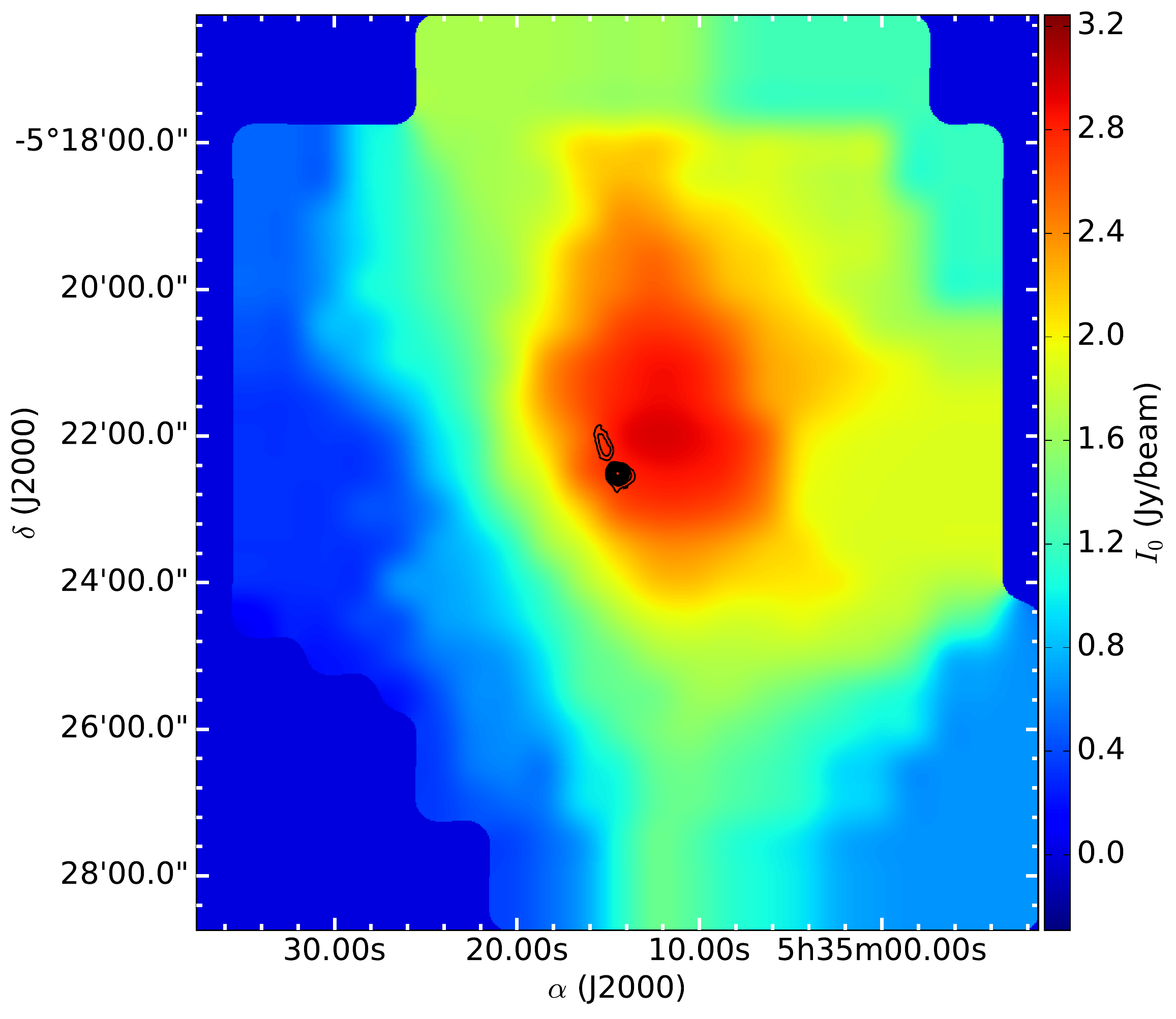}
\caption{Array average intensity map of the 3-2 transition of HC$_3$N (contours) overlaid on the color scale average intensity map of the 
same. Contours are $\pm3\sigma, \pm6\sigma, \pm9\sigma, ...$ where $\sigma$=17.6 m\jbm. \label{fig:hc3n-sd-map}}
\end{figure}

\begin{figure}
\includegraphics[scale=0.5]{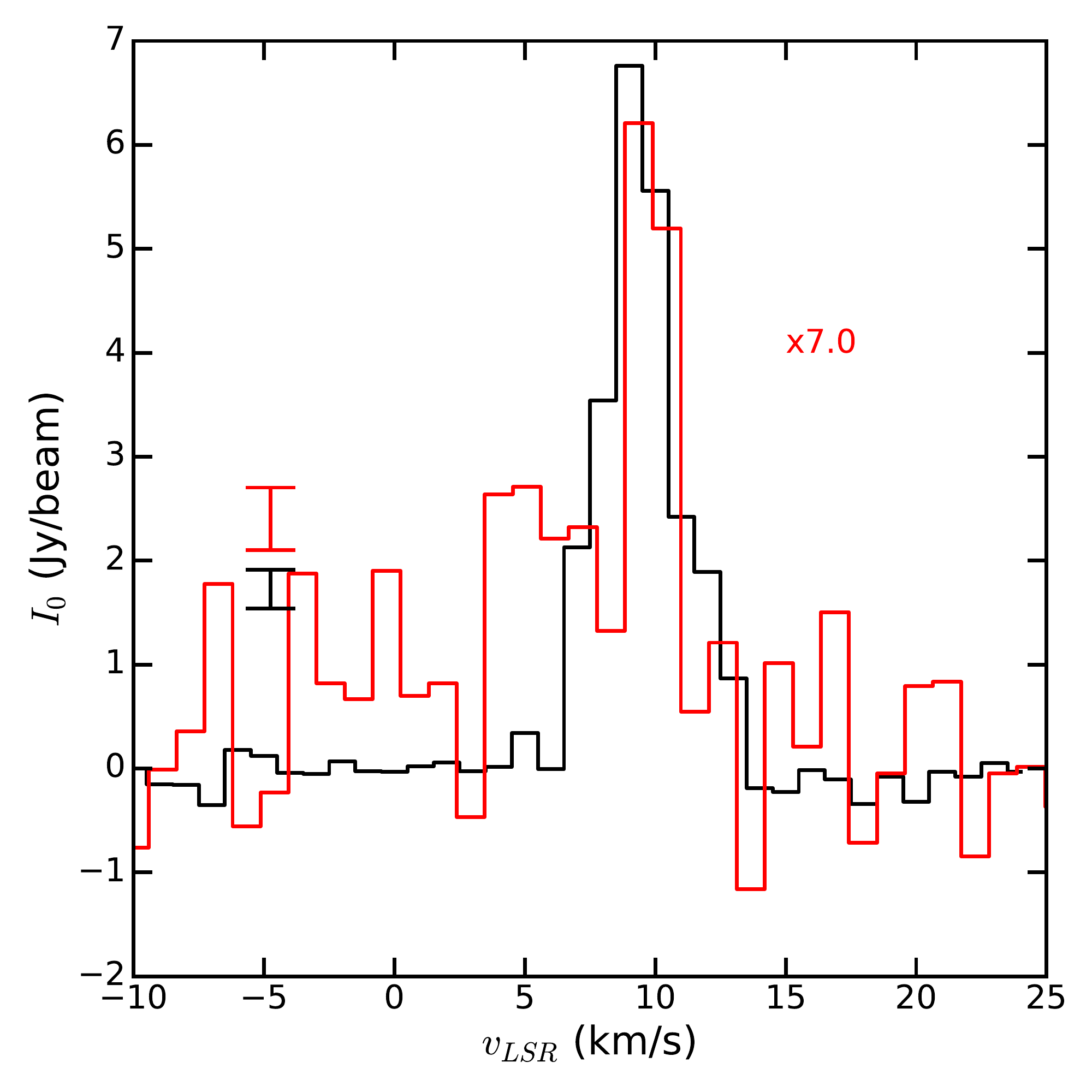}
\caption{Spectra of the single dish (black) and array observations (red) of the 3-2 transition of HC$_3$N. The array 
data have been scaled by a factor of 7.0. The abscissa is \vlsr\ in \kms and the ordinate is intensity in \jbm\ and the 
'I' bars denote the 1$\sigma$ rms noise.\label{fig:hc3n-sd-spec}}
\end{figure}

This transition of HC$_3$N was observed in single dish mode. Figure ~\ref{fig:hc3n-sd-map} shows the single dish average intensity 
colorscale map overlaid by the array average intensity map as contours, and Figure~\ref{fig:hc3n-sd-spec} shows the single dish 
(black) and array (red) spectra. Like NH$_3$ the peaks do not fully correspond, and the array spectrum has to be scaled 
by a factor of 7.0. The array data were best fit by a pair of Gaussians with \vlsr's of 5.4(5) and 9.8(1) \kms\ and 
FWHM of 3.1(13) and 1.4(6) \kms. The single dish data were best fit by a single Gaussian with a \vlsr\ of 9.1(0) \kms\ 
and FWHM of 2.8(0) \kms. This indicates that HC$_3$N has a notable extended component, which is not surprising as it is 
easily detected in many sources from hot cores to cold dark clouds.

%--------------------------------------------------------------------------------------------------

\section{Bringing it All Together}
In the previous sections we have described the distribution of the molecular species detected in this survey. These data paint a picture of complex molecules (\etcn, \dme, and \mef) and \mtoh\ being confined to the warmest, densest part of the region near the SiO outflow and highly extinguished infrared sources in Orion-KL. \fmal\ and NH$_3$ also show significant compact emission, but have both large scale features and emission from more distant regions (e.g.\ HC-NW) which should also be cooler. SO and SO$_2$ appear to have emission that is clumpy, but still shell like around the hottest part of the region. H$_2$CS is the next most extended followed by \mae, both of which occupy regions which are much cooler $\sim$30 K. \mae\ also has significant extended structure on the order of arcminutes. HC$_3$N is the only molecular species in this survey which occupies both the hottest and densest region (HC) and the more extended cooler regions (e.g. CS1).

\begin{figure}
\includegraphics[scale=0.3]{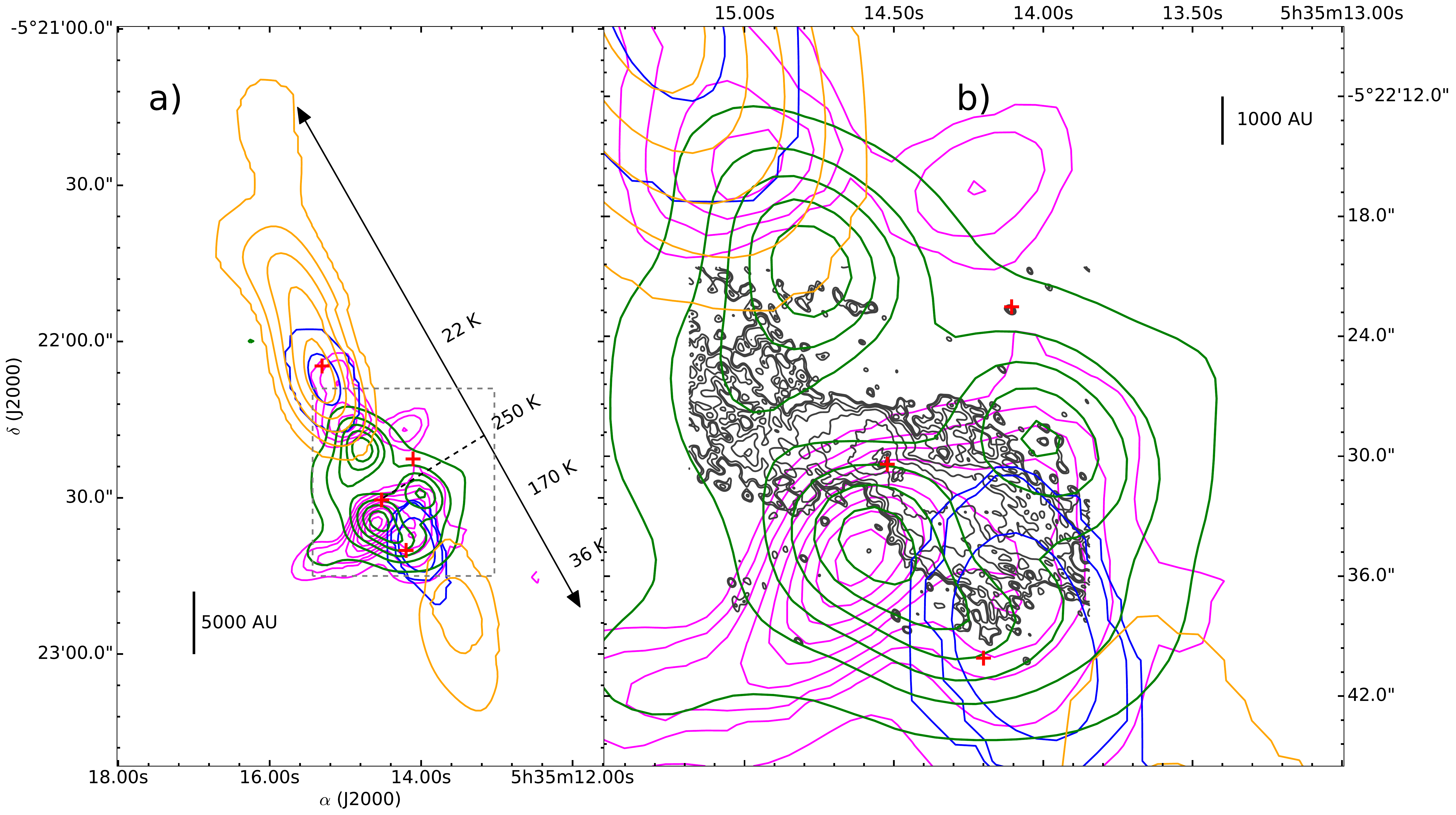}
\caption{A schematic of the temperature gradient, from the Hot Core where it is warmest ($\sim$250 K) moving outward to the coolest region, CS ($\sim$30 K). The contours are from \fmal\ (magenta), SO (green), H$_2$CS (blue), and \mae\ (orange). The arrows denote the temperature gradient and are labelled with temperatures derived from this work and \citet{friedelphd}. The dashed line connects the center of the gradient line with the Hot Core and the red + indicate CS1, BN, I, and CR positions. Figure~\ref{fig:temperature}(b) shows a zoom in of panel (a) with SiO contours (gray) added in.\label{fig:temperature}}
\end{figure}

Figure~\ref{fig:temperature}(a) illustrates the temperature gradient from the warmest region of the Hot Core ($\sim$250 K) moving outward to the coolest region, CS1 ($\sim$30 K). 
Note that none of the displayed molecules peak on top of the outflow, but rather at its edges or extremes. This shows how the outflow is energizing the region and driving the chemistry, with molecules formed dominantly in the gas phase occupying the coldest parts, through to the complex molecules which require the higher temperatures and densities to form and/or be desorbed from icy dust grain mantles.

%--------------------------------------------------------------------------------------------------

\section{Conclusion}
We have presented the results of the first $\lambda$=1 cm spectral line survey of Orion-KL. We detected a total of 89 
transitions from 14 molecular species, and no unidentified lines. The observations were conducted with Combined Array for Research in Millimeter-wave Astronomy (CARMA) in both array and single dish modes. 

SO and SO$_2$ (and their isotopomers) were found to have a shell like structure near Source I, with three distinct sub-cores on the edge. There 
also appears to be a hole in the SO emission that was created by BN as it exited the main core. This structure is not seen in any other detected molecular species because SO and SO$_2$ 
occupy a different physical region from the other detected species. Each of the SO/SO$_2$ clumps borders on one or more 
of the primary emission regions of complex organic molecules (Hot Core, Compact Ridge, IRc6, etc.), giving a picture of SO/SO$_2$ 
encasing the compact emission of these species. A majority of the 
emission comes from the compact source detected by the array, but there is a significant component that is more 
extended but $<$165\arcsec\ in total extent.

Methanol had the most detected transitions of any species and exhibits a primarily 
compact emission, with three distinct velocity components. Each of these components comes from a different set of 
physical conditions. The highest temperature component comes from the smallest structure and the lowest temperature 
comes from the largest structure. This may indicate either a shell like structure or multiple individual cores

Ammonia [NH$_3$] is also primarily compact, with two detected components, but like SO, has 
an extended component which is $<$165\arcsec. Both H and He recombination lines were detected, but were very dependent 
on the size of the synthesized beam. While there were notable knots of emission detected, the primary component was 
found to be extended.

Several molecular species, specifically formaldehyde [\fmal], thioformaldehyde [H$_2$CS], methyl acetate [\mae], and HC$_3$N 
show several regions of emission which are not compact. \mae\ has compact and extended emission regions which appear to be at the extreme 
ends of the SiO outflow and \fmal\ and HC$_3$N emission are concentrated near Source I, but also has emission at the 
northeast edge of the SiO emission, while H$_2$CS has its strongest peak at the northeast edge of the SiO outflow and 
weak emission near the Compact Ridge.
Ethyl Cyanide [\etcn], methyl formate [\mef], and, by proxy, dimethyl ether [\dme] showed only compact emission from 
the central core.

Based on data from this work and those of \citet{friedelphd,plambeck09,fav11,friedel11,goddi11,friedel12,slww12,brou13} 
we suggest the following general picture: SO and SO$_2$ were formed or liberated from grains by the initial shock from 
the Source I/BN interaction. Either at the same time or shortly after the interaction the SiO outflow from Source I 
began, which in turn heated up the region and sent further, more powerful shocks through the gas and dust. At the 
extreme ends of the outflow several molecular species were formed/liberated into the gas phase (e.g. \mae, \fa, \dme, 
\mef, and H$_2$CS). Further in, where the outflow is interacting with slightly denser material, molecules such \acetone\ 
are formed/liberated. In the densest region of the outflow molecules such as \etcn, NH$_3$, and \fmal\ are 
formed/liberated. \mtoh\ is the only molecule which is detected throughout the outflow, from the extremes to the 
densest parts. The outflow and BN formed "holes" in the SO shell as they escaped the inner region.

\acknowledgements
We thank the anonymous referee for comments and suggestions which made this work stronger.
This work was partially funded by NSF grant 
AST-0540459 and the University of Illinois. Support for CARMA construction was derived from the states of Illinois, 
California, and Maryland, the Gordon and Betty Moore Foundation, the Kenneth T. and Eileen L. Norris Foundation, the 
Associates of the California Institute of Technology, and the National Science Foundation.
\clearpage

\bibliography{refs}{}
\bibliographystyle{aasjournal}

\end{document}